\documentclass[prx,twocolumn,showpacs,superscriptaddress,preprintnumbers,amssymb,nofootinbib]{revtex4-2} 
\usepackage{graphicx}
\usepackage{amsthm}
\usepackage{thmtools}
\usepackage{url}
\usepackage{tikz}
\usepackage{relsize}
\usepackage{latexsym}
\usepackage{amssymb}
\usepackage{amsmath}
\usepackage{mathtools}
\usepackage{amsmath} 
\usepackage{amsfonts} 
\usepackage{upgreek}
\usepackage{bm}
\usepackage{multirow}
\usepackage{enumitem}
\usepackage{color}
\usepackage[colorlinks, citecolor=blue, linkcolor=blue]{hyperref}

\usepackage{physics}
\usepackage{tcolorbox}
\usepackage{manfnt}
\usepackage{lipsum}
\usepackage[ruled,linesnumbered]{algorithm2e}
\usepackage{tabularx}
\usepackage{nicematrix}
\usepackage{braket}
\usepackage[normalem]{ulem} 
\usepackage{enumitem}

\newcommand{\cP}{ {\cal P} }

\renewcommand{\O}{\mathrm{O}}

\newcommand{\secref}[1]{Sec.~\ref{#1}}
\newcommand{\appref}[1]{Appendix.\,\ref{#1}}
\newcommand{\eqnref}[1]{Eq.\,\eqref{#1}}

\newcommand{\figref}[1]{Fig.\,\ref{#1}}

\newcommand{\cJ}{\mathcal{J}}

\theoremstyle{plain} 

\newtheoremstyle{axiomwithbreak} 
  {3pt}   
  {3pt}   
  {} 
  {}      
  {\bfseries} 
  {.}     
  {\newline }   
  {}      

\theoremstyle{axiomwithbreak}

\theoremstyle{definition}               

\makeatother

\theoremstyle{plain}

\newtheorem*{theorem*}{Theorem}

\begin{document}

\title{Short-Range Vector Spin Glasses: \\ Parisi Hierarchies and Real-Space Ultrametricity}

\author{Akash Vijay}
\affiliation{Department of Physics and Institute for Condensed Matter Theory, University of Illinois Urbana-Champaign, Urbana, Illinois 61801, USA}
\author{Jaewon Kim}
\affiliation{Department of Physics and Institute for Condensed Matter Theory, University of Illinois Urbana-Champaign, Urbana, Illinois 61801, USA}
\author{Jong Yeon Lee}
\affiliation{Department of Physics and Institute for Condensed Matter Theory, University of Illinois Urbana-Champaign, Urbana, Illinois 61801, USA}
\affiliation{Korea Institute for Advanced Study, Seoul 02455, South Korea}

\date{\today}
\begin{abstract} 
Replica symmetry breaking (RSB) is the organizing principle of mean-field spin-glass theory, yet whether a continuous Parisi hierarchy can arise in a genuinely short-range system in finite spatial dimension remains unresolved.
Here we introduce a disordered model of $N$-component vector spins on a $D$-dimensional hypercubic lattice, with random matrix couplings on nearest-neighbor links and random tensor couplings on elementary plaquettes.
In the large-$N$ limit, the model supports thermodynamically stable one-step, full, and one-full RSB phases, alongside replica-symmetric paramagnetic and ferromagnetic phases.
We find that spatial locality constrains these ordered phases through two distinct infrared mechanisms.
($i$) Quenched random anisotropy destroys uniform ferromagnetism for $D\leq4$ through an Imry-Ma mechanism. ($ii$) A continuous Parisi hierarchy gives rise to replica Goldstone modes whose gaplessness is protected by a Ward identity. At leading order in $1/N$, these modes behave as a band of free spin waves, producing infrared-divergent fluctuations in $D\leq2$ and a Mermin--Wagner instability of the continuously broken phase.

Remarkably, the Parisi hierarchy also acquires a direct real-space interpretation. For two pure states with mutual overlap $q$, a characteristic length $\xi(q)$ describes how far the differences between their frozen magnetization patterns remain correlated.
Throughout a continuous sector of the Parisi hierarchy, $\xi(q)$ is independent of temperature and decreases monotonically with $q$. 
The ultrametric organization of pure states is therefore accompanied by an ultrametric hierarchy of real-space correlation lengths: states that separate near the root of the Parisi tree differ through long-wavelength frozen patterns, whereas states that separate farther from the root differ only on progressively shorter scales.

\end{abstract} 
\maketitle

\twocolumngrid 

\begin{figure*}[!t]
\centering
\includegraphics[width=\textwidth]{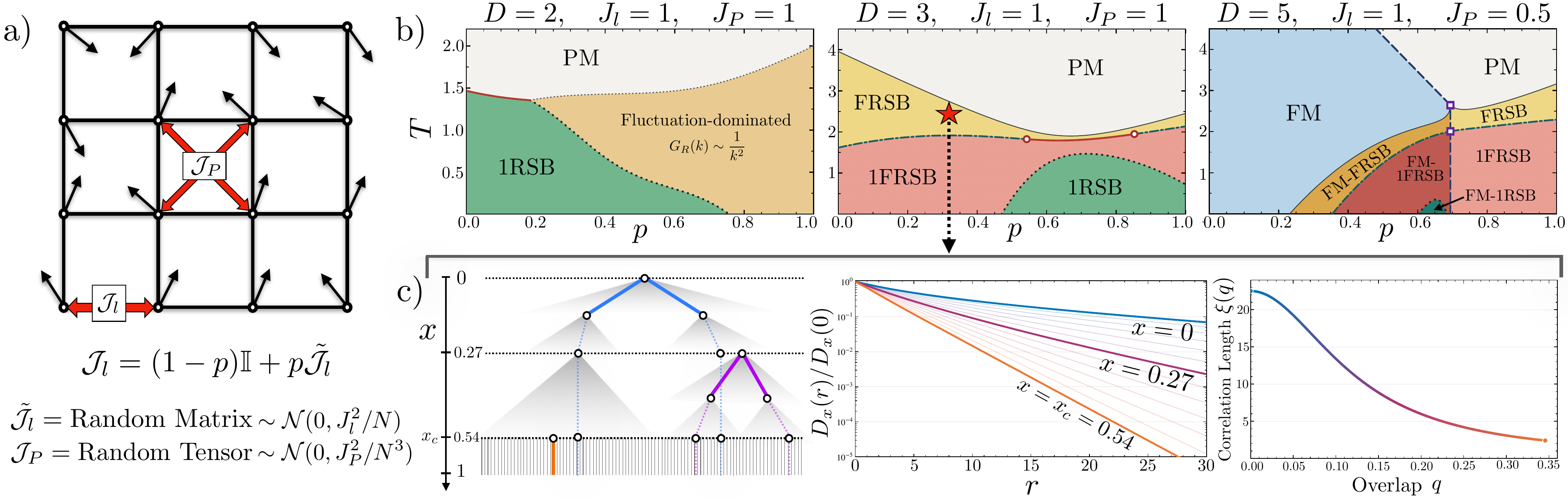}
\caption{ 
\textbf{(a) Short-range vector spin glass.} $N$-component spins reside on the vertices of a $D$-dimensional hypercubic lattice. Each nearest-neighbor link $l$ carries a matrix coupling $\mathcal{J}_{l}=(1-p)\mathbb{I}+p\widetilde{\mathcal{J}}_{l}$, which interpolates between clean ferromagnetic exchange at $p=0$ and a Gaussian random matrix at $p=1$. Each elementary plaquette couples its four corner spins through an independent random rank-four tensor $\mathcal{J}_{P}$. The disorder variances scale with $N$ as indicated, yielding a well-defined large-$N$ limit with free energy proportional to $NV$. 
\textbf{(b) Phase structure.}~Representative phase diagrams at $J_l=1$ for
$(D,J_P)=(2,1)$, $(3,1)$, and $(5,0.5)$.
Resolved regions show the thermodynamically dominant, replicon-stable saddles among the families considered. The prefix FM indicates coexistence with nonzero magnetization. 
In $D=2$, the dominant saddle in the amber ``fluctuation-dominated'' region has a continuous Parisi hierarchy. Its gapless replicons generate infrared-divergent $1/N$ fluctuations, rendering the saddle unstable. By contrast, the gapped 1RSB region is unaffected by this mechanism at the same order.
No ferromagnetic phase appears in $D=2$ or $3$, consistent with the Imry--Ma mechanism, which excludes uniform ferromagnetic order for $D\leq4$ in the presence of nonzero random anisotropy. Purple squares mark multicritical points, red circles denote glass--glass critical points, and the red star marks the $D=3$ FRSB state at $(p,T)=(0.30,2.10)$ analyzed in panel (c).
\textbf{(c) Hierarchy of spatial correlation scales.}
Left: schematic Parisi tree with shallow, intermediate, and deep levels highlighted. Unbranched terminal stems represent the Edwards--Anderson plateau $q(x)=q_{\rm EA}\simeq0.346$ for $x_c\leq x\leq1$, with $x_c\simeq0.543$.
Center: difference-field correlators $D^{x}(r)$ [Eq.~\eqref{eq:difference_correlator}], normalized by $D^{x}(0)=q_{\rm EA}-q(x)$. Blue and purple highlight $x=0$ and $x=x_c/2$, orange denotes the normalized limit $x\to x_c^{-}$, and faint curves sample additional intermediate levels.
Right: the correlation length $\xi(q)$ governing the asymptotic decay of $D^{x}(r)$ decreases with overlap. States that separate at shallow levels of the tree therefore differ through long-wavelength spatial patterns, whereas states that separate at deeper levels decorrelate over progressively shorter distances. 
Combined with overlap ultrametricity, $q_{ab}\geq\min(q_{ac},q_{bc})$, this gives $\xi(q_{ab})\leq\max\{\xi(q_{ac}),\xi(q_{bc})\}$: the Parisi hierarchy is inherited by the real-space correlation scales.}
\label{fig:panel_figure}
\end{figure*}

\makeatletter
\newcommand{\hideappendixsubsections}{%
  \renewcommand{\l@subsection}[2]{}%
  \renewcommand{\l@subsubsection}[2]{}%
}
\makeatother

\makeatletter
\begingroup
\renewcommand{\l@subsubsection}[2]{}
\tableofcontents
\endgroup
\makeatother


\section{Introduction}

One of the central open questions in spin-glass theory is whether the Parisi picture of replica symmetry breaking (RSB) describes the low-temperature phase of finite-dimensional spin glasses with strictly short-range interactions~\cite{EdwardsAnderson1975,BinderYoung1986}. 
In the fully-connected Sherrington--Kirkpatrick (SK) model~\cite{SherringtonKirkpatrick1975,SherringtonKirkpatrick1978}, this picture is firmly established. 
Early studies of the mean-field
equations~\cite{ThoulessAndersonPalmer1977,BrayMoore1980,Plefka1982} exposed the limitations of the replica-symmetric solution, which predicts a negative zero-temperature entropy~\cite{SherringtonKirkpatrick1975} and is unstable below the de Almeida--Thouless (AT)
line~\cite{AlmeidaThouless1978,BrayMoore1978}.
Parisi's RSB ansatz~\cite{Parisi1979,Parisi1980a,Parisi1980b,ParisiToulouse1980,Parisi1983} resolves these pathologies by replacing the unstable replica-symmetric saddle by a marginally stable hierarchical solution~\cite{DominicsKonder1983}. It is now proven to give the exact SK free energy~\cite{Guerra2003,Talagrand2006}.
Moreover, its physical content is striking: in the spin-glass phase, the equilibrium Gibbs measure decomposes into infinitely many pure states whose mutual overlaps are organized ultrametrically~\cite{Young1983,MezardParisiSourlasToulouseVirasoro1984a,MezardParisiSourlasToulouseVirasoro1984b,MezardVirasoro1985,MézardParisiVirasoro1986,RammalToulouseVirasoro1986,MezardParisiVirasoro1986b,GhirlandaGuerra1998,ParisiTersenghi2000,Panchenko2015}.

Beyond the SK model, studies of the random-energy model~\cite{Derrida1980,Derrida1981,Derrida1985}, \(p\)-spin Ising glasses~\cite{GrossMezard1986,Gardner1985,KirkpatrickThirumalai1987b}, spherical spin glasses~\cite{deAlmeidaJonesKosterlitzThouless1978,CrisantiSommers1994,CrisantiHornerSommers1993,CastellaniCavagna2005,CrisantiLeuzzi2004,CrisantiLeuzzi2006,CrisantiLeuzzi2007}, Potts glasses~\cite{GrossKanterSompolinsky1985,KirkpatrickWolynes1987}, finite-connectivity glasses~\cite{KanterSompolinsky1987,MezardParisi2001}, and hard-sphere structural glasses~\cite{KurchanParisiZamponi2012,KurchanParisiUrbaniZamponi2013,CharbonneauKurchanParisiUrbaniZamponi2014,FranzParisi2016} have revealed a much broader landscape of mean-field phases, with one-step, full, and mixed patterns of RSB. The corresponding nonequilibrium dynamics is equally rich, exhibiting aging, dynamical arrest, and violations of fluctuation--dissipation relations~\cite{Sompolinsky1981,SompolinskyZippelius1982,KirkpatrickThirumalai1987,Bouchaud1992,CugliandoloKurchan1993,CugliandoloKurchan1994,Franz1994,FranzParisi1995,FranzParisi1997,Monasson1995,CugliandoloLeticiaKurchan1997,FranzMezardParisiPeliti1998,FranzMezardParisiPeliti1999,JohOrbachWoodHammannVincent1999,MontanariTersenghi2003,FolenaFranzTersenghi2020}.

In finite dimensions, however, spatial fluctuations can disrupt the mean-field organization of states, and exact analytical results are scarce.
The principal alternative to the Parisi picture is the droplet theory~\cite{SouthernYoung1977,BrayMoore1984,McMillan1984,FisherHuse1986,BrayMoore1987,FisherHuse1987,HuseFisher1987,FisherHuse1988}. In its canonical form, this picture holds that a zero-field short-range Ising spin glass supports, for each disorder realization, only a single pair of pure states related by a global spin flip, and that its low-energy excitations are compact droplets of linear size $L$ whose characteristic free-energy cost scales as $L^{\theta}$.
The theory further predicts that any nonzero magnetic field destroys spin-glass order, in contrast with the de Almeida--Thouless (AT) transition of mean-field theory~\cite{AlmeidaThouless1978,FisherHuse1988}. 
Independently, the self-averaging constraints of Newman and Stein rule out a direct transplantation of the standard SK state picture to short-range systems~\cite{NewmanStein1992,NewmanStein1996,
NewmanStein1997}, without excluding more general RSB descriptions formulated through metastates~\cite{Read2014}.
These competing descriptions agree on the presence of frozen order: within a pure state, the local magnetizations form a spatially random pattern with no conventional long-range order, but with a nonzero Edwards--Anderson order parameter $q_{\rm EA}$~\cite{EdwardsAnderson1975}.
However, they differ in how many pure states carry this order and how their mutual overlaps are organized.

Extensive numerical studies~\cite{BhattYoung1985,Ogielski1985,FranzParisiVirasoro1994,JanusCollab2013,HukushimaNemoto1996,KatzgraberKorner2006,HasenbuschPelissetto2008,WangMachtaKatzgraber2015,Kawamura1992,MarinariParisiTersenghiRuizLorenzoZuliani2000,VedulaMooreSharma2025,ParisiRitort1993,MarinariParisiRuizLorenzoRitort1996,MarinariParisiRuizLorenzo1998,Young1997,KrzakalaMartin2000,PalassiniYoung2000,HartmannYoung2001,AspelmeierBrayMoore2002,KatzgraberHelmutYoung2003,KatzgraberHamzZhueOchoaMunozBauza2015,YucesoyKatzgraberHelmut2012,WangMachtaKatzgraber2014,WangWallinLidmar2020,HedYoungDomany2004,ContucciGiardinaGibertiParisiVernia2007,KatzgraberHartmann2009,LarsonKatzgraberMooreYoung2010,YoungKatzgraber2004,KatzgraberLarsonYoung2009,LeuzziParisiTersenghiRuizLorenzo2008,SinghYoung2017,Fernandezetal2022,Aguilaretal2024,VedulaMooreSharma2024} and a complementary field-theoretic program~\cite{HarrisLubenskyChen1976,BrayRoberts1980,FisherSompolinsky1985,DominicisKondorTemesvari1993,DrosselBokilMoore2000,MooreDrossel2002,TemesváriDominicisPimentel2002,TemesvariDominicis2002,Temesvari2008,ParisiTemesvári2012,MooreBray2011,CharbonneauYaida2017,MooreRead2018,HollerRead2020,AngeliniLucibelloParisiPerrupatoTersenghiRizzo2022,AngeliniParisiTersenghi2013,Castellana2025} have clarified several aspects of spin-glass ordering. 
Simulations of the three-dimensional Edwards--Anderson (EA) model strongly support a continuous transition at finite temperature and zero field into a glassy phase with $q_{\rm EA}>0$~\cite{BhattYoung1985,Ogielski1985,HasenbuschPelissetto2008,JanusCollab2013}. Domain-wall and interface calculations place the lower critical dimension near $d_\ell\simeq2.5$~\cite{FranzParisiVirasoro1994,Boettcher2005}, while the replicated Landau--Ginzburg theory of the overlap field has upper critical dimension $d_u=6$~\cite{HarrisLubenskyChen1976,FisherSompolinsky1985}. These results do not, however, determine how the pure states are organized below the transition.

That organization remains disputed. Numerical studies of low-temperature overlap distributions and excitations have been interpreted in favor of both pictures~\cite{MarinariParisiTersenghiRuizLorenzoZuliani2000,KrzakalaMartin2000,PalassiniYoung2000,YucesoyKatzgraberHelmut2012,WangMachtaKatzgraber2014,WangWallinLidmar2020,VedulaMooreSharma2025}, and direct tests of ultrametricity have also reached opposing conclusions~\cite{HedYoungDomany2004,ContucciGiardinaGibertiParisiVernia2007,KatzgraberHartmann2009}. The fate of the AT line is equally unsettled. In a magnetic field, the one-loop renormalization group admits no stable perturbative fixed point below six dimensions~\cite{BrayRoberts1980}. This obstructs a weak-coupling description of the transition but does not establish its absence. Higher-order calculations, alternative expansions, and simulations continue to support different scenarios~\cite{CharbonneauYaida2017,HollerRead2020,AngeliniLucibelloParisiPerrupatoTersenghiRizzo2022,Fernandezetal2022,Aguilaretal2024,VedulaMooreSharma2024}.

This impasse motivates the search for analytically tractable models that retain essential features of short-range systems.
Early work emphasized the gauge-theoretic structure of frustration and led to exact identities for local spin-glass models on the Nishimori line~\cite{Mattis1976,Toulouse1977,VannimenusToulouse1977,Kirkpatrick1977,FradkinHubermanShenker1978,Hertz1978,Nishimori1980,Nishimori1981,GeorgesHanselBouchaud1985,Nishimori_book,GoldbartMartinFradkin1986}. Other approaches introduce controlled limits.
Kac models recover mean-field thermodynamics as the interaction range diverges~\cite{FranzToninelli2004,FranzToninelli2004b,FranzMontanari2007}. Bethe-lattice models retain finite connectivity but lack a finite-dimensional Euclidean geometry~\cite{MezardParisi2001,
AngeliniLucibelloParisiPerrupatoTersenghiRizzo2022}. 
Another route preserves the lattice and local interactions while enlarging the number of degrees of freedom per site. The short-range $p$-spin model of Campellone, Parisi, and Ranieri, for example, places $M$ spins at each site. Its site-factorized $M\to\infty$ saddle recovers the mean-field $p$-spin equations, with spatial structure entering through $1/M$ fluctuations. Their analysis focuses on the 1RSB phase, leaving the lower-temperature continuous-RSB regime unexamined~\cite{CampelloneParisiRanieri1999}.
Large-component vector glasses offer a related approach~\cite{Ginzburg1981,BrayMooreYoung1986,AspelmeierMoore2004,LeeYoung2005,LeeDharYoung2005,SharmaYoung2010,Moore2012}, although infinite-component solutions of scalar and anisotropic random-exchange models tend to remain replica symmetric\footnote{Related random-anisotropy models have been studied at large $N$~\cite{Goldschmidt1983}, with RSB obtained through $1/N$ corrections~\cite{Goldschmidt1984}.
Continuous RSB also appears in large-$N$ theories of elastic manifolds and generalized random anisotropy~\cite{MezardParisi1991,
LeDoussalWiese2003,MouhannaTarjus2016}.}. 

These approaches motivate two related questions for a microscopic lattice model with strictly local interactions. Can a Parisi hierarchy be thermodynamically stable? And if so, how do pure states that separate at different levels of the hierarchy differ in real space? The latter has so far been approached only numerically through clustering analyses of the EA model that suggested a hierarchy of states generated by correlated spin domains~\cite{DomanyHedPalassiniYoung2001}.

In this work, we introduce a model that gives analytic access to both questions. It consists of $N$-component spins on a $D$-dimensional hypercubic lattice, coupled through random matrices on nearest-neighbor links and random rank-four tensors on elementary plaquettes. A nonrandom ferromagnetic exchange is also present, with relative strength controlled by $p$ [\figref{fig:panel_figure}(a)]. For $p<1$, this clean exchange keeps the spin propagator spatially dispersive, so spatial correlations remain an essential part of the theory. In the $N\to\infty$ limit, the model is governed by an exact saddle point that gives direct access not only to the Parisi structure, but also to replica-resolved correlation functions in real space.
In the fully disordered limit $p=1$, the clean exchange vanishes and the spin propagator becomes local in position space. The large-$N$ saddle-point equations then reduce to those of the mixed $2+4$ spherical spin glass~\cite{CrisantiLeuzzi2004,CrisantiLeuzzi2006}, up to dimension-dependent effective couplings.


We find thermodynamically stable phases with a continuous Parisi hierarchy in this strictly local model [\figref{fig:panel_figure}(b)]. Strikingly, the ultrametric organization of overlaps is accompanied by a hierarchy of real-space correlation lengths, giving the Parisi structure a direct spatial interpretation [\figref{fig:panel_figure}(c)]. The model therefore demonstrates that strict locality and finite dimensionality are compatible with a continuous Parisi hierarchy in the large-$N$ limit, while also providing a systematic $1/N$ expansion for studying fluctuations about these saddles.

The model is defined in \secref{sec:model}, and the main results are summarized in \secref{sec:summary}. \secref{sec:largeN} derives the large-$N$ saddle-point equations and replica-resolved spin correlators. \secref{sec:RS} develops the replica-symmetric analysis, including the fate of ferromagnetic order and the replicon instabilities that motivate replica symmetry breaking. Finally, \secref{sec:RSB} constructs and classifies the RSB solutions, assembles the full phase diagram, and analyzes the fluctuations and spatial correlations of the RSB phases.

\begin{figure}[!t]
    \centering
    \includegraphics[width=0.98\columnwidth]{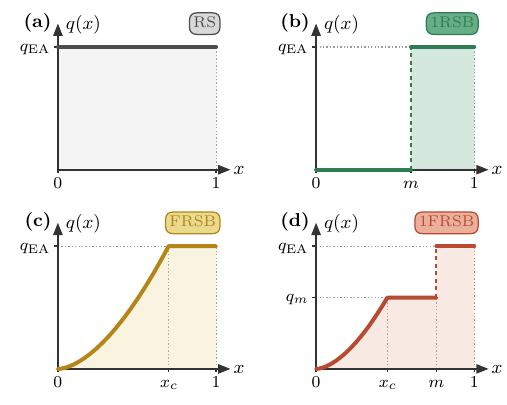}
    \caption{\textbf{Representative Parisi profiles for the RS, 1RSB, FRSB, and 1FRSB phases.}
    \textbf{(a)} In the RS phase, the Parisi profile is constant, $q(x)=q_{\rm EA}$.
    \textbf{(b)} In the 1RSB phase, $q(x)$ jumps from $0$ to the Edwards--Anderson overlap $q_{\rm EA}$ at the breaking point $m$.
    \textbf{(c)} In the FRSB phase, $q(x)$ varies continuously up to $x_c$, where it reaches $q_{\rm EA}$ and remains on the Edwards--Anderson plateau for $x>x_c$.
    \textbf{(d)} In the 1FRSB phase, the continuous branch reaches $q_m$ at $x_c$, followed by a plateau up to $m$ and a finite jump to $q_{\rm EA}$.  When $x_c=m$, the intermediate plateau has zero width.}
    \label{fig:1FRSBvsFRSB}
\end{figure}

\subsection{Model}
\label{sec:model}
We consider $N$-component real vector spins $S(v)$ on the vertices of a $D$-dimensional hypercubic lattice with $V=L^D$ sites and periodic boundary conditions. Each spin obeys the local constraint $\sum_{i=1}^{N}S(v)_i^2=N$.
Nearest-neighbor links $l$ carry matrix couplings, while elementary plaquettes $P$ carry rank-four tensor couplings
[\figref{fig:panel_figure}(a)]. The Hamiltonian is
\begin{equation}
    \begin{split}
    \label{eq:Ham}
    H &= -\hspace{-9pt}\sum_{l = \langle v_{1},v_{2}
    \rangle} \hspace{-8pt} 
    h_{l}[\mathcal{J}_{l}] - \hspace{-9pt} \sum_{P \ni (v_{1 \cdots 4})} \hspace{-9pt} h_{P}[\mathcal{J}_{P}],\\
    h_l[\mathcal{J}_{l}] &=
    (\mathcal{J}_{l})_{i_{1}i_{2}}S(v_{1})_{i_{1}} S(v_{2})_{i_{2}} \\
    h_{P}[\mathcal{J}_{P}] &= (\mathcal{J}_{P})_{i_{1}i_{2}i_{3}i_{4}}S(v_{1})_{i_{1}} S(v_{2})_{i_{2}} S(v_{3})_{i_{3}} S(v_{4})_{i_{4}},
    \end{split}
\end{equation}
where repeated component indices are summed.
The link matrices interpolate between clean
ferromagnetic exchange and random exchange,
\begin{align}
    \mathcal{J}_{l} &= (1-p) J_0 \mathbb{I} + p \tilde{\mathcal{J}}_{l}.
\end{align}
We set the ferromagnetic coupling strength $J_0  = 1$ for simplicity.
The entries of $\widetilde{\mathcal J}_l$ and $\mathcal J_P$ are independent Gaussian random variables with zero mean and variances $J_l^2/N$ and $J_P^2/N^3$, respectively. Couplings on different links and plaquettes are drawn independently. The corresponding powers of $N$ ensure a well-defined large-$N$ limit in which the free energy is extensive, scaling as $NV$. 
Although the disorder ensemble is invariant under global $O(N)$ rotations, a generic disorder realization explicitly breaks this symmetry, leaving only a global $\mathbb Z_2$ spin-flip symmetry $S(v)\to -S(v)$. 
The independent parameters of the model are the temperature $T$, the interpolation parameter $p$, and the disorder strengths $J_l$ and $J_P$. Throughout, we set $p<1$ so that the spin propagator remains spatially dispersive. We consider the fully disordered limit $p=1$ only as a reference when useful.

Since the Gaussian random couplings enter the Hamiltonian linearly, their disorder average can be performed exactly. In the replicated theory, this generates local interactions between the replica-overlap fields
\begin{equation}
    Q^{\alpha\beta}(v)
    =\frac1N S^\alpha(v)^T S^\beta(v),
\end{equation}
where $\alpha$ and $\beta$ label replicas. The random links generate quadratic overlap interactions, while the random plaquettes generate quartic ones. Integrating out the spin components then produces an effective action proportional to $N$, so the theory is governed by the minimum free energy, stable saddle point in the $N\to\infty$ limit (\secref{sec:largeN}).

At a translation-invariant Parisi saddle point, the off-diagonal overlaps are encoded in a nondecreasing function $q(x)$, where $x\in[0,1]$ labels the levels of the replica hierarchy at which two replicas separate and $q(x)$ is their overlap.
A $k$-step ansatz produces a $q(x)$ with $k$ finite jumps, one for each level at which replica symmetry is further broken. 
When the number of breaking levels becomes infinite, these jumps merge into a continuous rise, and replica symmetry is broken continuously over an interval of the hierarchy.
We call a phase \emph{continuously broken} when $q(x)$ is continuous and strictly increasing over a nonzero interval. Representative profiles for one-step (1RSB), full (FRSB), and one-full (1FRSB) glassy phases are shown in \figref{fig:1FRSBvsFRSB}. In this sense, 1RSB is discontinuous, since replica symmetry is broken through a single finite jump, whereas both FRSB and 1FRSB contain a continuously broken sector.

\subsection{Summary of Results}
\label{sec:summary}
 
The large-$N$ solution of our local model connects three aspects of glassy order: the hierarchy of equilibrium states, the spatial patterns that distinguish them, and the fluctuations that threaten their stability. The results below refer to the translation-invariant saddles of the $N\to\infty$ theory and, where stated, to their leading $1/N$ fluctuations.
 
\begin{itemize}[leftmargin=13pt]

\item \textbf{Stable hierarchical glasses with strictly local interactions.} The equilibrium phase diagram [\figref{fig:panel_figure}(b)] contains thermodynamically stable 1RSB, FRSB, and 1FRSB glassy phases. 
FRSB consists of a continuously broken Parisi sector that terminates at the Edwards--Anderson plateau, whereas 1FRSB contains a continuous sector followed by an intermediate plateau and a final finite jump to the Edwards--Anderson plateau.
For example, in $D=3$ at $J_P=J_l=1$, cooling drives the system from the paramagnet into FRSB and then 1FRSB, while a 1RSB dome appears at strong frustration and low temperature.
No other Parisi structure is possible since the saddle-point equations together with replicon stability permit at most one continuously broken sector, followed by at most one jump, in every integer dimension (see~\secref{sec:no-other-saddles}).

Plaquette disorder is essential for replica symmetry breaking. 
With link disorder alone, the glass remains replica symmetric and marginal, whereas any nonzero $J_P$ destabilizes this solution.
At fixed link-disorder strength $J_l$, the relative strength of plaquette disorder then controls the character of the broken phase: weak plaquette disorder favors continuously broken FRSB and 1FRSB phases, while stronger plaquette disorder favors one-step order.
Accordingly, the phase diagram evolves from predominantly FRSB and 1FRSB at weak $J_P$ to predominantly 1RSB at strong $J_P$.
In the latter regime, the paramagnet can transition directly into 1RSB through a random first-order transition, with a finite jump in the Edwards--Anderson overlap but no latent heat.

\item \textbf{Real-space ultrametricity.}
Locality gives the Parisi hierarchy a direct spatial interpretation, as illustrated in \figref{fig:panel_figure}(c). 
Let $\mu_a(v)\equiv N^{-1/2}\langle S(v)\rangle_a$ denote the normalized local magnetization in pure state $a$.
The thermodynamic overlap between two pure states $a$ and $b$ is $q_{ab}=V^{-1}\sum_v\mu_a(v)^{T}\mu_b(v)$, while their self-overlaps satisfy $q_{aa}=q_{bb}=q_{\rm EA}$.
To characterize how the hierarchical separation between two states is organized in real space, we define the difference field
$\delta\mu_{ab}(v)\equiv\mu_a(v)-\mu_b(v)$ and its two-point correlator at fixed overlap,
\begin{equation}
\label{eq:difference_correlator}
    D^{x}(r)\equiv\frac{1}{2V}\,\overline{\Bigl\langle\sum_{v}\delta\mu_{ab}(v)^{T}\,
    \delta\mu_{ab}(v+r)\Bigr\rangle_{q_{ab}=q(x)}} .
\end{equation}
The brackets denote the Gibbs-weighted average over pairs of pure states with mutual overlap $q(x)$ (see \appref{app:pure_state_decomposition}), and the overline denotes the disorder average.

In the replicated theory, $D^{x}(r)$ is the difference between two replica-off-diagonal spin correlators: one for replicas occupying the same pure state and one for replicas whose states branch at level $x$ of the Parisi tree. It can therefore be obtained directly from the saddle point.
At zero separation, $D^{x}(0)=q_{\rm EA}-q(x)$, so the overlap deficit measures the mean-square difference between the two frozen magnetization patterns. On the other hand, at large separation,
\begin{equation}
    D^{x}(r)\sim\frac{e^{-r/\xi(x)}}{r^{(D-1)/2}},
    \qquad r\gg1,
\end{equation}
which defines a level-dependent correlation length $\xi(x)$ governing the spatial scale over which these differences remain correlated. Thus, the overlap measures how strongly two states differ, while $\xi$ measures the spatial scale over which that difference is organized. Along a continuously broken sector, $\xi$ decreases monotonically with $q$. The ultrametric relation
$q_{ab}\geq\min(q_{ac},q_{bc})$ therefore implies a corresponding ultrametric relation for the correlation lengths,
\begin{equation}
    \xi_{ab}\leq\max\{\xi_{ac},\xi_{bc}\},\qquad \xi_{ab}\equiv\xi(q_{ab}).
\end{equation}
States that separate near the root of the Parisi tree differ through long-wavelength frozen patterns, whereas states that share a deeper common branch differ only on progressively shorter scales.

\item \textbf{Parisi--Toulouse scaling of the real-space hierarchy.}
Along every continuously broken sector, the saddle equations imply
\begin{align}
    x(q,T)=T f(q),
    \qquad
    \xi(q,T)=\xi(q),
\end{align}
where both $f(q)$ and $\xi(q)$ are independent of temperature.
Thus, a given overlap $q$ is associated with a fixed real-space correlation length, while changing the temperature only shifts the location $x$ at which that overlap appears in the Parisi hierarchy. The range of overlaps realized by the continuous sector may nevertheless vary with temperature as its endpoints move.
This provides a real-space analogue of Parisi--Toulouse scaling.

\item \textbf{Imry--Ma destruction of ferromagnetism for $D\leq4$.}
A generic realization of the random matrix and tensor couplings breaks the global $O(N)$ symmetry down to $\mathbb{Z}_{2}$ and locally favors different spin orientations.
The resulting quenched random anisotropy competes with ferromagnetic stiffness and destroys uniform alignment beyond a characteristic Larkin scale for $D\leq4$.
In the large-$N$ theory, the same physics appears as an infrared divergence of the relevant lattice Watson integral.
Moreover, the Larkin scale has the same scaling as the glass correlation length governing spatial correlations of the frozen magnetization pattern within a fixed pure state.
Replica symmetry breaking does not remove this infrared obstruction, so magnetized RS, 1RSB, FRSB, and 1FRSB phases are all excluded for $D\leq4$.

\item \textbf{Coexistence of ferromagnetism and hierarchical glass order.}
For $D>4$, where uniform ferromagnetism is allowed, the replica-symmetric ferromagnet loses replicon stability along a de Almeida--Thouless line while retaining a finite magnetization.
The equilibrium continuation is therefore a mixed phase in which ferromagnetic order coexists with a nontrivial Parisi hierarchy, realized as FM--1RSB, FM--FRSB, or FM--1FRSB depending on parameters.
Thus, hierarchical glass order can emerge within an already magnetized phase, rather than only after ferromagnetic order has disappeared.

\item \textbf{Gapless replicons and Mermin--Wagner instability in $D\leq2$.}
Every continuous Parisi hierarchy is accompanied by gapless replicon modes generated by ``small permutations'' of the Parisi tree. These are Goldstone modes whose gaplessness is protected by a Ward identity.
At leading order in $1/N$, they behave as free spin waves with propagator
\begin{align}
    G_R(k;x)\sim \frac{1}{\rho_R(x)k^2}, 
\end{align}
where $\rho_R(x)$ is a level-dependent stiffness.
Their long-wavelength fluctuations therefore diverge logarithmically in $D=2$ and as a power law for $D<2$.
Thus, FRSB and 1FRSB saddles containing a continuously broken sector suffer a Mermin--Wagner-type infrared instability in $D\leq2$, whereas this argument does not exclude a gapped 1RSB phase.

\end{itemize}

\section{Large-$N$ Saddle Point}
\label{sec:largeN}

In this section, we develop the large-$N$ saddle-point theory that forms the basis of our subsequent analysis. 
We first average over quenched disorder using the replica trick, which generates local interactions between replicas.
We then integrate out the $N$ spin components to obtain an effective action for the replica-overlap field $Q^{\alpha\beta}(v)$ and its conjugate Lagrange-multiplier field $\Lambda^{\alpha\beta}(v)$. 
Since this action is proportional to $N$, the functional integral is dominated by its saddle points as $N\to\infty$, with fluctuations organized in a systematic $1/N$ expansion.
The resulting saddle-point equations determine both the thermodynamics and the replica-resolved spin correlation functions.

\subsection{Replica Trick}
For a fixed realization of the link and plaquette disorder \(\mathcal{J}_{l}\) and \(\mathcal{J}_{P}\), the partition function at inverse temperature $\beta$ is given by
\begin{equation}
  Z = \int \prod_v \Big[ dS(v) \, \delta( |S(v)|^{2} - N ) \Big] e^{-\beta H[ S(v);\mathcal{J}_{l},\mathcal{J}_{P}]}.
\end{equation}
The quenched free energy is determined by $\overline{\log Z}$, where $\overline{(...)}$ denotes the disorder average.
We evaluate $\overline{\log Z}$ using the replica trick
\begin{align}
    \overline{\log Z} =\lim_{n \rightarrow 0^{+}} \frac{1}{n}  \log \overline{Z^n}.
\end{align}
For positive integer $n$, $Z^n$ describes $n$ copies of the system subject to the same disorder realization.
Averaging over their common disorder generates interactions between the replicas. We first evaluate $\overline{Z^n}$ at integer $n$ and then analytically continue to $n\to0^+$.

Since the link and plaquette couplings are independent random variables, the disorder average can be conveniently factorized as,
\begin{equation}
\begin{split}
    \overline{Z^n} = \int \prod_{v,\alpha} \Big[ dS^{\alpha}(v) \, \delta( |S^{\alpha}(v)|^{2} - N ) \Big]&  \\
    \times \prod_{l} \bigg[\int d\mathcal{J}_{l} p(\mathcal{J}_{l})e^{-\beta\sum_{\alpha = 1}^{n}H_{l}[S^{\alpha}(v);\mathcal{J}_{l}]}\bigg]&  \\
    \times \prod_{P}\bigg[ \int  d\mathcal{J}_{P} p(\mathcal{J}_{P})e^{-\beta\sum_{\alpha = 1}^{n}H_{P}[S^{\alpha}(v);\mathcal{J}_{P}]}\bigg]&,
    \end{split}
\end{equation}
where $\alpha = 1,\cdots, n$ labels the replicas. Furthermore, because the disorder variables are Gaussian distributed and enter the Hamiltonian linearly, each disorder integral can be evaluated exactly. In terms of the local overlap field $Q^{\alpha\beta}(v) = \frac{1}{N} S^{\alpha}(v)^{T}S^{\beta}(v)$, the random link couplings generate quadratic interactions and the random plaquette couplings generate quartic interactions. Thus, 
\begin{align}
    \overline{Z^n} = \int \prod_{v,\alpha} &\Big[ dS^{\alpha}(v) \, \delta( |S^{\alpha}(v)|^{2} - N ) \Big] e^{-H_{n}[S^{\alpha}]} \,, \textrm{ where }  \nonumber \\
    H_{n}[S^{\alpha}] = &-\frac{N\beta^{2}J^{2}_{l}p^{2}}{2}\sum_{\alpha , \beta}\sum_{l= \langle vv' \rangle}Q^{\alpha\beta}(v)Q^{\alpha\beta}(v') \nonumber \\
    &- \beta(1-p)\sum_{\alpha }\sum_{l= \langle vv' \rangle}S^{\alpha}(v)^{T}S^{\alpha}(v') \nonumber \\
    &- \frac{N\beta^{2}J^{2}_{P}}{2}\sum_{\alpha ,\beta}\sum_{P}\prod_{v\in P}Q^{\alpha\beta}(v) .
\end{align}
Note that $Q^{\alpha\beta}(v)$ is a symmetric matrix with diagonal components equal to 1 due to the local constraint $S^{\alpha}(v)^{T}S^{\alpha}(v)=N$.

We next treat the off-diagonal overlaps $Q^{\alpha\neq\beta}(v)$ as independent integration variables and enforce their definition in terms of the spins using Lagrange multipliers. For each site $v$ and replica pair $\alpha<\beta$, we insert
\begin{equation}
\begin{split}
    1={}&\int dQ^{\alpha\beta}(v)
    \int_{-i\infty}^{i\infty}
    \frac{N\,d\Lambda^{\alpha\beta}(v)}{2\pi i}
    \\
    &\times e^{-\Lambda^{\alpha\beta}(v)
    [S^\alpha(v)^T S^\beta(v)-NQ^{\alpha\beta}(v)]}.
    \label{eq:delta_insertion_offdiag}
\end{split}
\end{equation}
Similarly, we introduce a diagonal Lagrange multiplier to enforce the spin-length constraint at each site $v$ and in each replica $\alpha$,
\begin{equation}
\begin{split}
    \delta\bigl(|S^\alpha(v)|^2-N\bigr)
    ={}&\int_{-i\infty}^{i\infty}
    \frac{d\Lambda^{\alpha\alpha}(v)}{4\pi i}
    \\
    &\times e^{-\frac12\Lambda^{\alpha\alpha}(v)
    [|S^\alpha(v)|^2-N]}.
    \label{eq:delta_insertion_diag}
\end{split}
\end{equation}
Together, these multipliers form a symmetric matrix $\Lambda^{\alpha\beta}(v)$. In what follows, we omit field-independent normalization factors from the functional integrals.

For notational convenience, we now introduce the nearest-neighbor kernel $K(v-v')$, which rewrites a sum over links as an unrestricted sum over lattice vertices:
\begin{align}
    \sum_{l = \langle vv' \rangle} (\cdots)= \frac{1}{2}\sum_{v}\sum_{v'}K(v-v')(\cdots),
\end{align}
where the factor of $1/2$ avoids double counting of each link.
For the hypercubic lattice, with $r = v-v'$, this kernel is given by,
\begin{align}
  K(r) = \begin{cases}
    1, & r=\pm\hat{e}_{\mu},\\
    0, & \text{otherwise}.
  \end{cases}
\end{align}
Its Fourier transform is 
\begin{align}
    K(k)=2\sum_{\mu}\cos k_{\mu},
\end{align}
where $k_{\mu} = {2\pi n_\mu}/{L}$ and $n_{\mu} = 0,\cdots, L-1$.

With these definitions, the delta-function constraints introduced above render the remaining spin integral Gaussian.
Integrating out the spins then yields an effective action for the overlap field $Q^{\alpha\beta}(v)$ and the Lagrange multiplier field $\Lambda^{\alpha\beta}(v)$:
\begin{align}\label{eq:partition_function_Ieff}
    \overline{Z^n} = \int \prod_{\substack{v \\\alpha<\beta}} &dQ^{\alpha\beta}(v) \prod_{\substack{v \\\alpha\leq\beta}}d\Lambda^{\alpha\beta}(v)  
    e^{-NI_n[Q,\Lambda]}, 
\end{align}
where
\begin{align}
\begin{split}
    I_n[Q,\Lambda] = &-\frac{\beta^{2}J^{2}_{P}}{2}\sum_{\alpha ,\beta}\sum_{P}\prod_{v\in P}Q^{\alpha\beta}(v) \\
    &-\frac{\beta^{2}J^{2}_{l}p^{2}}{2}\sum_{\alpha ,\beta}\sum_{l= \langle vv' \rangle}Q^{\alpha\beta}(v)Q^{\alpha\beta}(v')  \\
    &- \frac{1}{2}\sum_{\substack{v \\\alpha, \beta}} Q^{\alpha\beta}(v)\Lambda^{\alpha\beta}(v) + \frac{1}{2}\Tr \log \mathcal{M}. \label{eq:Ieff} 
    \end{split}
\end{align}
Here the trace runs over both lattice sites and replica indices, and the spin kernel $\mathcal{M}$ is given by 
\begin{equation}
\begin{split}
    \mathcal{M}^{\alpha\beta}(v,v') = &-\beta(1-p)K(v-v')\delta^{\alpha\beta}  \\
    &+ \Lambda^{\alpha\beta}(v)\delta(v-v').
\end{split}
\end{equation}
Note that the first term is diagonal in replica space but acts off-diagonally in position space whereas the second term acts diagonally in position space but is off-diagonal in replica space.

\subsection{Replica-Resolved Correlation Functions}
\label{sec:replica_correlators}
The Gaussian spin integral also directly yields the replica-resolved spin correlation function,
\begin{align}
\label{eq:ss_corr}
    G^{\alpha\beta}(v,v') &\equiv \frac{1}{N} \big\langle S^\alpha(v)^T S^\beta(v') \big\rangle \nonumber \\
    &= \big\langle(\mathcal M^{-1})^{\alpha\beta}(v,v')\big\rangle,
\end{align}
where the final expectation value is taken over the fields $Q$ and $\Lambda$.
Its diagonal and off-diagonal replica components distinguish connected thermal correlations within a pure state from correlations between the frozen magnetization patterns of different pure states.

To make this distinction precise, fix a disorder realization $\mathcal J=(\mathcal J_l,\mathcal J_P)$ and decompose the corresponding Gibbs state into pure states in the thermodynamic limit. To factor out the trivial global spin-flip degeneracy, we choose one representative from each spin-flip pair and assign it the combined Gibbs weight of the pair. We use the same choice of representatives for every replica. Let $a$ label these states and $w_a(\mathcal J)$ denote their normalized weights. Averages in this reduced Gibbs measure are then $\langle \mathcal O\rangle_{\mathcal J} =\sum_a w_a(\mathcal J)\langle O\rangle_{a,\mathcal J}$; see~\appref{app:pure_states}. We suppress the explicit disorder label $\mathcal J$ below when no confusion can arise. 

For each pure state $a$, define the normalized local magnetization $\mu_a(v)=N^{-1/2}\langle S(v)\rangle_a$. The spin--spin correlator within that state separates into a connected thermal contribution and a frozen contribution,
\begin{align}\label{eq:pure_state_correlator_decomposition}
    \frac1N \langle S(v)^T S(v')\rangle_a = C_a(v,v') + \mu_a(v)^T\mu_a(v').
\end{align}
where $C_a(v,v')$ is the normalized connected thermal correlator.

The replica-diagonal and replica-off-diagonal components of~\eqnref{eq:ss_corr} probe different pieces of this decomposition. The diagonal correlator retains the full spin--spin correlation function and hence both thermal and frozen pieces. After averaging over pure states and disorder, we obtain
\begin{equation}
\label{eq:diag_replica_corr}
    G^{\alpha\alpha}(v,v') = \overline{ \sum_a w_a \left[ C_a(v,v') + \mu_a(v)^T\mu_a(v') \right] },
\end{equation}
For distinct replicas, by contrast, thermal fluctuations are independent once the pure states they occupy are fixed. If replica $\alpha$ occupies state $a$ and replica $\beta\neq\alpha$ occupies state $b$, then
\begin{equation}
    \frac1N\bigl\langle
    S^\alpha(v)^T S^\beta(v')
    \bigr\rangle_{a,b}
    =\mu_a(v)^T\mu_b(v').
\end{equation} 
Thus the off-diagonal correlator contains only frozen contributions. Sampling the two pure states independently from the reduced Gibbs measure and then averaging over disorder gives
\begin{equation}
    G^{\alpha\ne\beta}(v,v')
    =\overline{\sum_{a,b}w_aw_b\,
    \mu_a(v)^T\mu_b(v')}.
    \label{eq:offdiag_replica_corr}
\end{equation}
\eqnref{eq:offdiag_replica_corr} averages over all pairs of pure states, regardless of their mutual overlap. The Parisi hierarchy refines this average by organizing state pairs according to their overlap. The correlator associated with a given level of the hierarchy is therefore obtained by conditioning the state-pair average on the corresponding overlap.

At coincident points,~\eqnref{eq:ss_corr} gives
\begin{equation}\label{eq:Q_Lambda_avg_identity}
    \big\langle Q^{\alpha\beta}(v)\big\rangle = \big\langle(\mathcal M^{-1})^{\alpha\beta}(v,v)\big\rangle .
\end{equation}
Thus, the expectation value of the local overlap field is simply the coincident-point limit of the replica-resolved spin correlator.

The zero-momentum component of $\mathcal M^{-1}$ also directly diagnoses ferromagnetic order. We define the local magnetization in replica $\alpha$ as
\begin{equation}
    M^\alpha \equiv \frac{1}{\sqrt N}\langle S^\alpha(v)\rangle .
\end{equation}
It is independent of $v$ due to translation invariance of the disorder-averaged, replicated partition function and is normalized by $\sqrt N$ so that $(M^\alpha)^T M^\beta\leq1$. Introducing the replica-resolved connected correlator
\begin{equation}
    C^{\alpha\beta}(v-v') \equiv \frac1N\left[ \big\langle S^\alpha(v)^T S^\beta(v')\big\rangle -  \big\langle S^\alpha(v)\big\rangle^T \big\langle S^\beta(v')\big\rangle \right],
\end{equation}
we find that
\begin{align}
    \frac1V\big\langle(\mathcal M^{-1})^{\alpha\beta}(k=0)\big\rangle &= (M^\alpha)^T M^\beta +\frac{\chi^{\alpha\beta}}{V},
\end{align}
where $\chi^{\alpha\beta}\equiv\sum_r C^{\alpha\beta}(r)$ is the associated susceptibility. Provided $\chi^{\alpha\beta}$ grows subextensively (which is true in both gapped and gapless systems), the connected contribution vanishes in the thermodynamic limit. Thus, ferromagnetic order appears as an extensive contribution to the uniform mode of the replica-resolved spin correlation function,
\begin{equation}
\label{eq:ferro_order}
    ( M^\alpha)^T M^\beta =  \lim_{V\rightarrow\infty} \frac1V \big\langle(\mathcal M^{-1})^{\alpha\beta}(k=0)\big\rangle.
\end{equation}

\subsection{$N\to \infty$ Limit}
The overall factor of $N$ in~\eqnref{eq:partition_function_Ieff} suppresses fluctuations of $Q$ and $\Lambda$ about stationary configurations. Consequently, in the $N\to\infty$ limit the functional integral is dominated by saddle points of $I_n$. For a dominant saddle $(Q_*,\Lambda_*)$,
\begin{equation}
    \overline{Z^n}\simeq
    e^{-NI_n[Q_*,\Lambda_*]},\qquad N\to\infty,
\end{equation}
where $Q_{*}$ and $\Lambda_{*}$ solve
\begin{align}
    0 = \frac{\delta I_n[Q_{*},\Lambda_{*}]}{\delta Q^{\alpha\beta}} =  \frac{\delta I_n[Q_{*},\Lambda_{*}]}{\delta \Lambda ^{\alpha\beta}}.
\end{align}

We restrict our attention to translation-invariant saddle points,
\begin{align}
    Q_{*}^{\alpha\beta}(v) = Q_{*}^{\alpha\beta}, \qquad \Lambda_{*}^{\alpha\beta}(v) = \Lambda_{*}^{\alpha\beta}.
\end{align}
This restriction is consistent with the fluctuation analysis below: within the replicon sector, the softest overlap fluctuations occur at $k=0$, so no finite-wavevector replicon instability preempts the homogeneous one.

Translation invariance makes the spin kernel diagonal in momentum space:
\begin{align}
    \mathcal{M}_*^{\alpha\beta}(k) = [\Lambda_{*} - \beta(1-p) K(k)]^{\alpha\beta}.
\end{align}
The two saddle-point equations therefore reduce to
\begin{align}
    & Q_{*}^{\alpha\beta} 
    =\frac{1}{V}\sum_{k} \bigg[ \frac{1}{\Lambda_{*} - \beta(1-p) K(k)\mathbb{I}} \bigg]^{\alpha\beta}, \label{eq:invrse_eqn}\\
    & \beta^2 \left( A_P (Q_{*}^{\alpha\beta})^{3} 
    + A_l Q_{*}^{\alpha\beta} \right)
    +\frac{1}{2} \Lambda_{*}^{\alpha\beta} = 0. 
    \label{eq:local_eqn}\\
    \text{where}  \quad&A_P = J_P^2 D (D-1), \quad A_l = J_l^2 p^2 D \label{eq:ApAl_def}
\end{align}
Note that~\eqnref{eq:local_eqn} only holds for $\alpha\neq \beta$. The spatial dimensionality $D$ enters explicitly in~\eqnref{eq:ApAl_def}, enhancing the effective link- and plaquette-disorder contributions by factors of $D$ and $D(D-1)$, respectively.
The momentum sum in~\eqnref{eq:invrse_eqn} runs over the Brillouin zone, $k_{\mu}=2\pi n_{\mu}/L$ with $n_{\mu}=0,\dots,L-1$, and $V = L^{D}$.~\eqnref{eq:invrse_eqn} is a matrix equation in replica space, whereas~\eqnref{eq:local_eqn} acts componentwise on the replica indices.

The maximally disordered limit $p=1$ is particularly simple. In this case, the ferromagnetic coupling $\beta(1-p)$ vanishes, and $\mathcal M$ becomes momentum independent. Consequently, the first saddle-point equation reduces to $Q=\Lambda^{-1}$ while the second saddle-point equation coincides with that of the mixed $2+4$ spherical glass, up to the aforementioned dimension-dependent enhancement of the effective couplings. Further details on the solutions of the mixed spherical glass can be found in~\cite{CrisantiLeuzzi2004,CrisantiLeuzzi2006}.

In what follows, we restrict to $p<1$, where the spin kernel $\mathcal M(k)$ retains nontrivial momentum dependence and distinguishes the saddle-point equations from those of the mixed spherical spin-glass.
We solve the coupled saddle-point equations using a sequence of ansatze for $Q_{*}$ and $\Lambda_{*}$, analyze the stability of these solutions as a function of the parameters $p, J_{l}, J_{P}$ and $\beta$, and assess which ones are thermodynamically dominant.

Finally note that, in the large-$N$ limit, the replica-resolved spin correlation function is simply given by
\begin{align}
    G^{\alpha\beta}(v,v')
    =
    \bigl(\mathcal M_*^{-1}\bigr)^{\alpha\beta}(v-v'),
    \label{eq:ss_corr_saddle}
\end{align}
where the expectation value appearing in~\eqnref{eq:ss_corr} is replaced by its saddle-point value.
The operator $\mathcal M_*^{-1}$ can therefore be interpreted as the replica-resolved spin propagator.

\section{Replica Symmetric Ansatz}
\label{sec:RS}
We begin with the replica-symmetric ansatz, and use it to identify paramagnetic, ferromagnetic, and glassy solutions to the saddle-point equations.
Replica symmetry requires invariance under arbitrary permutations of the replicas, which constrains the overlap and Lagrange-multiplier matrices to the form
\begin{align}
    Q&= (1-q)\mathbb{I} + q\mathbf{1}\mathbf{1}^{T}, \\
    \Lambda &= (\lambda_{d}-\lambda)\mathbb{I} + \lambda\mathbf{1}\mathbf{1}^{T},
\end{align}
where $\mathbf{1}=(1,\ldots,1)$. The diagonal condition $Q^{\alpha\alpha}=1$ is automatically satisfied, and the single off-diagonal overlap $q$ is the Edwards--Anderson order parameter. 

Physically, replica symmetry corresponds to a phase where for each disorder realization, the Gibbs measure is supported on a single pure state, up to the global $\mathbb{Z}_2$ symmetry. All replicas therefore equilibrate within the same state, and no nontrivial hierarchy of overlaps arises.

The spin kernel $\mathcal{M}(k)$ inherits the same replica structure and is given by $\mathcal{M}(k)=\epsilon(k)\mathbb{I}+\lambda\textbf{1}\textbf{1}^{T}$ with
\begin{equation}
  \epsilon(k)\equiv\lambda_{d}-\lambda-\beta(1-p)K(k).
\end{equation}
In the replica limit $n \to 0$, its inverse is $\mathcal{M}^{-1}(k)=\frac{1}{\epsilon(k)}\mathbb{I}-\frac{\lambda}{\epsilon(k)^{2}}\textbf{1}\textbf{1}^{T}$. 
The first saddle point equation, \eqnref{eq:invrse_eqn}, therefore reduces to two equations, for the diagonal and off-diagonal components,
\begin{align}
    1 &= \frac{1}{V}\sum_{k}\,\frac{\epsilon(k)-\lambda}{\epsilon(k)^{2}}  \qquad \text{(diagonal)}  \label{eq:saddle_pt1}\\
    q &= -\frac{1}{V}\sum_{k}\,\frac{\lambda}{\epsilon(k)^{2}}  \qquad \text{(off-diagonal)}.\label{eq:saddle_pt2}
\end{align}
Taking their difference yields the useful sum rule
\begin{equation}\label{eq:sum_rule}
  1-q=\frac{1}{V}\sum_{k}\frac{1}{\epsilon(k)} .
\end{equation}
The remaining saddle point equation~\eqnref{eq:local_eqn} becomes the local algebraic relation
\begin{align}\label{eq:saddle_pt3}
    \beta^{2} \left( A_P q^{3} + A_l q \right)
    +\tfrac{1}{2}\lambda = 0.
\end{align}
Hence $\lambda<0$ whenever $q>0$, while $\lambda=0$ for $q=0$.

In the thermodynamic limit, momentum sums may generally be replaced by integrals over the Brillouin zone, $\frac{1}{V}\sum_{k}\to\int\frac{d^{D}k}{(2\pi)^{D}}$.
The infrared behaviour is then governed by the small-momentum expansion of the kernel, $K(k)\simeq 2D-k^{2}$. Special care is required when the uniform $k=0$ mode condenses.
Although it is only a single momentum point, its contribution can remain finite if the corresponding propagator grows extensively. We therefore separate the uniform mode before taking the thermodynamic limit,
\begin{align}
\frac{1}{V}\sum_{k}f(k)
  \;=\;\underbrace{\frac{1}{V}f(0)}_{\text{condensate}}
  \;+\;\int_{-\pi}^{\pi}\frac{d^{D}k}{(2\pi)^{D}}\,f(k),
\end{align}
where the isolated contribution survives as $V\to\infty$ only when $f(0)\sim V$.

\subsection{Saddle-Point Solutions}
\label{subsec:RS_sol}
We now classify the replica-symmetric saddle points into ferromagnetic, paramagnetic, and spin-glass branches.

\subsubsection{Ferromagnetic Branch}
Ferromagnetic order is signaled by the condensation of the uniform $k=0$ mode.
Since
\begin{align}
  ({M}^{\alpha})^{T}{M}^{\beta}&=\lim_{V\to\infty}\frac{1}{V}\big(\mathcal M_{*}^{-1}\big)^{\alpha\beta}(0) \nonumber \\
  &=-\lim_{V\to\infty}\frac{1}{V}\,
  \frac{\lambda}{\epsilon(0)^{2}},
\end{align}
a non-zero condensate requires that the \emph{gap} $\epsilon(0)$ shrink to 0 as $\epsilon(0)\sim V^{-1/2}$.
We denote the resulting replica-independent condensate by,
\begin{equation}
    M^{2}\equiv ({M}^{\alpha})^{T}{M}^{\beta}=-\lambda/[V\epsilon(0)^{2}] \,.
\end{equation}
The gap closing condition implies that $\lambda_{d}-\lambda=2D\beta(1-p)$, so the noncondensed modes $(k\neq 0)$ have dispersion
\begin{equation}
  \epsilon(k)=\beta(1-p)\big[2D-K(k)\big].
\end{equation}

It is now convenient to introduce the Watson integrals
\begin{align}
  w_{r}(D)&=\int_{-\pi}^{\pi}\frac{d^{D}k}{(2\pi)^{D}}\,
  \frac{1}{[2D-K(k)]^{r}},
\end{align}
which are convergent for $D>2r$. 
The saddle point equations Eqs.~\eqref{eq:saddle_pt1} and \eqref{eq:saddle_pt2} can be conveniently expressed in terms of the Watson integrals as,
\begin{align}
    1&=M^{2}+\frac{T\,w_{1}(D)}{1-p}
    -\frac{\lambda\,T^{2}w_{2}(D)}{(1-p)^{2}}, \\
    q&=M^{2}-\frac{\lambda\,T^{2}w_{2}(D)}{(1-p)^{2}} .
\end{align}
Their difference eliminates the condensate and determines $q$ in closed form
\begin{equation}\label{eq:RS_magnetizedq}
  1-q=\frac{T\,w_{1}(D)}{1-p}.
\end{equation}
Since $q\geq0$, a magnetized solution can exist only for $T\leq T^{*}=(1-p)/w_{1}(D)$. Eliminating $\lambda$ with \eqnref{eq:saddle_pt3} then gives the condensate in closed form,
\begin{equation}\label{eq:RS_condensate}
  M^{2}=q\left[1-\frac{2 (A_P q^{2}+ A_l)
  w_{2}(D)}{(1-p)^{2}}\right].
\end{equation}

Note that all explicit factors of temperature have canceled in \eqnref{eq:RS_condensate}. 
The existence of ferromagnetic order is therefore controlled by $w_{2}(D)$. For $D>4$, this integral is finite, and the magnetized branch exists whenever the right-hand side of~\eqnref{eq:RS_condensate} is positive. Since~\eqnref{eq:RS_magnetizedq} implies that $q$ increases upon cooling, the magnetized branch terminates when the condensate vanishes,
\begin{equation}
  1 = \frac{2\bigl(A_P q(T)^{2} + A_l \bigr) w_{2}(D)} {(1-p)^{2}}.
\end{equation}

For $D\leq4$, by contrast, $w_2(D)$ diverges. Finiteness of the saddle-point equations then requires $\lambda\to0$, and~\eqnref{eq:saddle_pt3} consequently forces $q\to0$. Since $M^2\leq q$, the condensate must vanish as well. Thus, in the presence of any nonzero anisotropic disorder, replica-symmetric ferromagnetic order is absent for $D\leq4$. We give the corresponding Imry–Ma interpretation in Sec.~\ref{subsec:ImryMa}.

\subsubsection{Paramagnetic Branch}
On the paramagnetic branch, $M^{2}=q=0$, and \eqnref{eq:saddle_pt3} therefore implies $\lambda=0$. The remaining variable $\lambda_{d}$ is fixed by the sum rule,
\begin{equation}\label{eq:paramagnet_sumrule}
  1=\frac{1}{\beta(1-p)}\int_{-\pi}^{\pi}\frac{d^{D}k}{(2\pi)^{D}}
  \,\frac{1}{\big[\Delta +2D- K(k)\big]} ,
\end{equation}
where $\Delta\equiv \frac{\epsilon(0)}{\beta(1-p)}$ is the dimensionless gap of the uniform mode. Solving for $\lambda_d$ is equivalent to solving for $\Delta$, since the two are related by $\Delta = \frac{\lambda_d}{\beta(1-p)}-2D$.

As the temperature is lowered, $\Delta$ decreases in order to maintain the sum rule. For $D>2$, the integral remains finite as $\Delta\to0$, and the paramagnetic gap closes when
\begin{align}
  1=\frac{1}{\beta(1-p)}\int_{-\pi}^{\pi}&\frac{d^{D}k}{(2\pi)^{D}}
  \,\frac{1}{2D- K(k)}=\frac{T\,w_{1}(D)}{1-p} \nonumber \\
  \Longrightarrow\qquad
  T^{*}&=\frac{1-p}{w_{1}(D)} .
\end{align}
This is precisely the same gap-closing scale obtained from the ferromagnetic branch. Below $T^*$, no gapped paramagnetic saddle exists, and the sum rule must instead be satisfied by an ordered solution, either through condensation of the uniform mode or through a
nonzero Edwards--Anderson order parameter.

For $D\leq2$, $w_1(D)$ diverges. The sum rule can then be satisfied with $\Delta>0$ at every finite temperature, so the gapped paramagnetic solution persists down to $T\to0$. Its existence alone does not establish its stability or thermodynamic dominance.

\subsubsection{Spin-Glass Branch}

On the spin-glass branch, the magnetization is zero $M^{2}=0$, but the overlap is nonzero  $q\neq0$. With no condensate, the sums in Eqs.~\eqref{eq:saddle_pt2} and \eqref{eq:sum_rule} become momentum integrals, with
\begin{equation}
    \epsilon(k)=\beta(1-p)[\Delta+2D-K(k)].
\end{equation}
Eliminating $\lambda$ using \eqnref{eq:saddle_pt3} and dividing by $q$ gives
\begin{align}\label{eq:sg_gain}
    1 &=\int_{-\pi}^{\pi}\frac{d^{D}k}{(2\pi)^{D}}
    \frac{2\beta^{2} A_P q^{2}+2\beta^{2} A_l} {\epsilon(k)^{2}} .
\end{align}
Together with the continuum form of \eqnref{eq:sum_rule}, this fixes $q$ and $\Delta$.

The plaquette contribution in \eqnref{eq:sg_gain} is proportional to $q^2$. It therefore drops out at a continuous onset from the paramagnet, where $q\to0^+$. Such an onset is controlled by the link disorder and satisfies
\begin{equation}
    1=2\beta^2A_l
    \int_{-\pi}^{\pi}\frac{d^Dk}{(2\pi)^D}
    \frac1{\epsilon(k)^2}.
    \label{eq:sg_cont}
\end{equation}
For $D>4$, its intersection with the paramagnetic gap-closing line is obtained by setting $\Delta=0$. The temperature factors cancel, leaving $2A_lw_2(D)=(1-p)^2$. Thus
\begin{align}
    p^*&=\frac1{1+J_l\sqrt{2D w_2(D)}},
    \label{eq:mcp}\\
    T^*(p^*)&=
    \frac{J_l\sqrt{2D w_2(D)}}
    {w_1(D)[1+J_l\sqrt{2D w_2(D)}]}.
\end{align}
When not preempted by a first-order transition, this intersection is the multicritical point where the continuous PM--FM, FM--glass, and PM--glass boundaries meet. Its location is independent of $J_P$ because the plaquette term does not enter the linear instability
condition.

\begin{figure*}[!t]
\centering
\includegraphics[width=\textwidth]{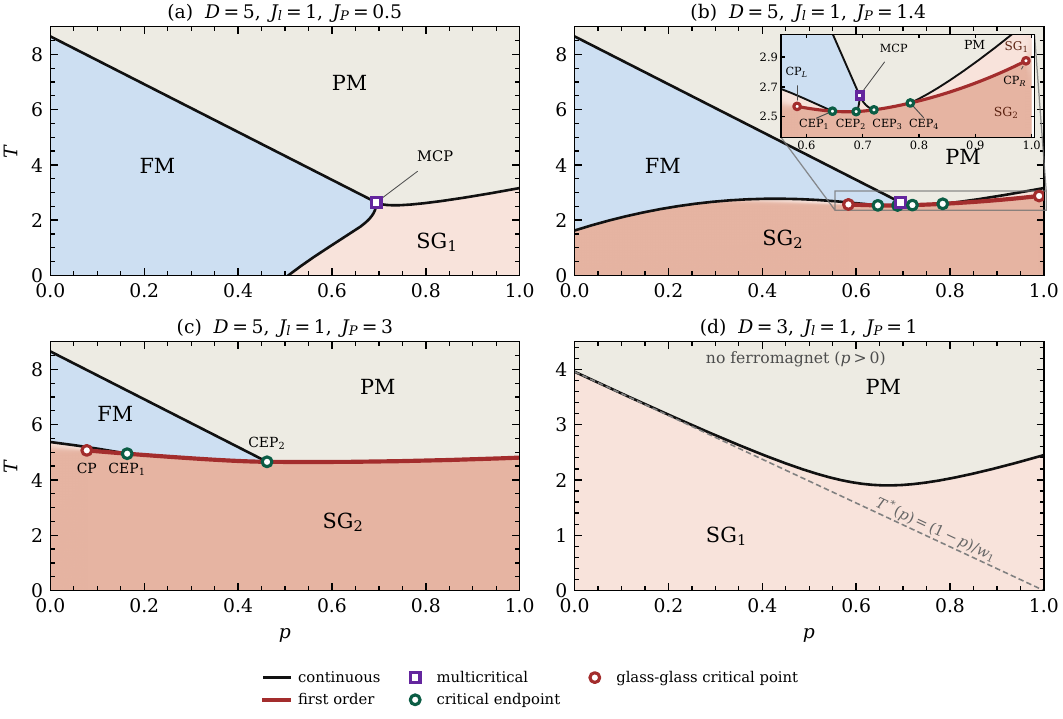}
\caption{\textbf{Replica-symmetric phase diagrams at $J_l=1$.}
At each point in the $(p,T)$ plane, we show the admissible RS saddle with the largest replica free energy~\eqref{eq:FreeEnergy_RS}. Black and red curves denote continuous and first-order transitions, respectively. Momentum integrals are evaluated using the exact lattice kernel. SG$_1$ and SG$_2$ denote the lower- and higher-overlap glass branches.
\textbf{(a)} Weak plaquette disorder, $D=5$ and $J_P=0.5$. Only the SG$_1$ glass branch appears, and all phase boundaries are continuous. The PM--FM, FM--SG$_1$, and PM--SG$_1$ boundaries meet at the multicritical point (square) of~\eqnref{eq:mcp}, $p^*=0.6945$ and $T^*=2.6420$. The FM phase is bounded above by the gap-closing line $T=(1-p)/w_1(D)$ and below by the locus $M^2\to0$ at finite $q$.
\textbf{(b)} Intermediate plaquette disorder, $D=5$ and $J_P=1.4$. The first-order coexistence curve connects two glass--glass critical points (open circles), CP$_L$ at $p\simeq0.5835$ and CP$_R$ at $p\simeq0.9899$. Four critical endpoints (filled circles), CEP$_1$, CEP$_2$, CEP$_3$, and CEP$_4$, occur at $p\simeq0.6463$, $0.6883$, $0.7199$, and $0.7845$, respectively. Between CEP$_1$ and CEP$_2$, the coexistence curve separates FM from SG$_2$; between CEP$_3$ and CEP$_4$, it separates PM from SG$_2$. The remaining three segments separate SG$_1$ from SG$_2$. At each critical endpoint, the two phases connected by the incident continuous boundary merge.
\textbf{(c)} Strong plaquette disorder, $D=5$ and $J_P=3$. A short first-order SG$_1$--SG$_2$ segment extends from a glass--glass critical point at $p\simeq0.0776$, $T\simeq5.0693$, to CEP$_1$ at $p\simeq0.1637$. Beyond CEP$_1$, the same coexistence curve separates FM from SG$_2$ and reaches CEP$_2$ at $p\simeq0.4617$, $T\simeq4.6554$, where it meets the continuous PM--FM boundary. For larger $p$, the PM--SG$_2$ transition remains first order up to $p=1$.
\textbf{(d)} $D=3$ and $J_P=1$. Imry-Ma excludes ferromagnetic order for $p>0$, so no finite-$p$ multicritical point remains. A single continuous PM--SG$_1$ boundary approaches the dashed gap-closing line $T=(1-p)/w_1(D)$ as $p\to0$. The transition temperature is nonmonotonic, reaching a minimum $T_{\rm SG}\simeq1.9007$ at $p\simeq0.6684$.}
\label{fig:phasediagram}
\end{figure*}

A continuous transition is not guaranteed. For sufficiently strong plaquette disorder, a solution with finite $q$ can become thermodynamically dominant before the paramagnet reaches \eqnref{eq:sg_cont}. The resulting transition is first order, as illustrated in \figref{fig:phasediagram}.

\subsection{Replica-Resolved Correlators}
\label{sec:RS_correlators}
We now examine the replica-resolved spin correlation functions introduced in \eqnref{eq:ss_corr_saddle} in the replica-symmetric phase, where all replicas occupy the same pure state once the global spin-flip choice is fixed. 
For a given disorder realization, denote this state by $a$ and its normalized local magnetization by $\mu_a(v)$. 
Distinct replicas fluctuate independently about the same frozen magnetization pattern. The off-diagonal correlator therefore contains only the frozen contribution,
\begin{equation}
    G^{\alpha\ne\beta}(r)
    =\overline{\frac1V\sum_v
    \mu_a(v)^T\mu_a(v+r)}.
    \label{eq:RS_offdiagonal}
\end{equation}
The diagonal correlator contains this same frozen contribution together with the connected thermal correlations within the pure state,
\begin{equation}
    G^{\alpha\alpha}(r)
    =\overline{\frac1V\sum_v
    \left[C_a(v,v+r)+\mu_a(v)^T\mu_a(v+r)\right]}.
    \label{eq:RS_diagonal}
\end{equation}
The spatial average over $v$ removes dependence on the absolute position and leaves a correlator that depends only on the separation $r$, as required by the translation invariant large-$N$ saddle. Importantly, this does not require the frozen magnetization pattern $\mu_a(v)$ itself to be spatially uniform.

Subtracting the off-diagonal correlator from the diagonal one cancels the frozen contribution and isolates the connected thermal correlator,
\begin{equation}
    G^{\alpha\alpha}(r)-G^{\alpha\ne\beta}(r)
    =\overline{\frac1V\sum_v C_a(v,v+r)}.
\end{equation}

\vspace{5pt} \noindent \textbf{Paramagnet.}  In the paramagnetic phase, $\mathcal M^{-1}_{*}(k)$ is diagonal in replica space, so $G^{\alpha\neq\beta}(r)=0$, consistent with the absence of frozen local magnetization.
The replica-diagonal correlator is therefore purely thermal and decays exponentially at large separation,
\begin{align}
    G^{\alpha\alpha}(r)
    \sim
    \frac{e^{-r/\xi}}{r^{(D-1)/2}},
\end{align}
For $r$ along a lattice axis, the correlation length is fixed exactly by the gap,
\begin{align}
    \xi^{-1} = \operatorname{arcosh}\!\left(1+\frac{\Delta}{2}\right) .
    \label{eq:xi_lattice}
\end{align}
In the small-gap limit $\Delta\ll1$, the decay becomes isotropic, with $\xi\simeq\Delta^{-1/2}$~\footnote{The gap remains small in all cases studied numerically, so $\xi\simeq\Delta^{-1/2}$ reproduces the exact correlation length up to negligible corrections. Along a lattice axis, the leading relative correction is $\Delta/24$.}.
The closing of the gap is therefore accompanied by a divergence of the thermal correlation length.

\vspace{5pt} \noindent \textbf{Replica-Symmetric Spin Glass:} 
In the replica-symmetric spin-glass phase, both diagonal and off-diagonal correlators are nonzero:
\begin{align}
   G^{\alpha\alpha}(r)&=  \frac{1}{V}\sum_k  e^{ik\cdot r} \bigg[\frac{1}{\epsilon(k)} - \frac{\lambda}{\epsilon(k)^2}\bigg],
    \\
    G^{\alpha\neq\beta}(r) &= \frac{1}{V}\sum_k e^{ik \cdot r} \bigg[-\frac{\lambda}{\epsilon(k)^2}\bigg].
\end{align}
Their difference isolates the connected thermal correlator,
\begin{align}
    G^{\alpha\alpha}(r) - G^{\alpha\neq \beta}(r) 
    &= \frac{1}{V}\sum_k
    \frac{e^{ik.r}}{\epsilon(k)} \sim\frac{e^{-r/\xi}}{r^{(D-1)/2}}.
\end{align}
The frozen contribution, by contrast, is governed by the double pole $\epsilon(k)^{-2}$:
\begin{equation}
    G^{\alpha\neq \beta}(r)
    \sim
    \frac{e^{-r/\xi}}{r^{(D-3)/2}}.
\end{equation}
The thermal and frozen correlations therefore share the same exponential correlation length $\xi$, but have different algebraic prefactors. The double pole makes the frozen correlations decay more slowly at long distances.

\vspace{5pt} \noindent \textbf{Ferromagnet:} 
In the ferromagnetic phase, the uniform mode condenses and the gap closes, $\Delta=0$. The $k=0$ contribution must therefore be separated from the continuum before taking the thermodynamic limit. The remaining noncondensed frozen fluctuations are governed by the double pole $\epsilon(k)^{-2}\sim k^{-4}$, and hence decay algebraically as
\begin{align}
  G^{\alpha\neq\beta}(r)-M^2 \sim \frac{1}{r^{D-4}},
  \qquad D>4.
  \label{eq:fm_frozen_corr}
\end{align}
The condensate contributes equally to the diagonal and off-diagonal correlators and therefore cancels in their difference. The connected thermal correlator is then controlled by the simple pole $\epsilon(k)^{-1}\sim k^{-2}$,
\begin{align}
  G^{\alpha\alpha}(r)-G^{\alpha\neq\beta}(r)
  \sim
  \frac{1}{r^{D-2}}.
  \label{eq:fm_thermal_corr}
\end{align}
Thus the ferromagnetic phase has true long-range order,
$\lim_{r\to\infty}G^{\alpha\neq\beta}(r)=M^2$,
with algebraically decaying fluctuations about the ordered state. As in the gapped spin-glass phase, the frozen correlations are longer ranged than the connected thermal correlations.

\subsection{Imry--Ma Instability of Ferromagnetic Order}
\label{subsec:ImryMa}
The absence of ferromagnetic order in $D\le4$ can be understood as an Imry--Ma instability~\cite{ImryMa,PelcovitsPytte1978,AizenmanWehr1989,AizenmanWehr1990}.
The essential competition is between ferromagnetic stiffness, which favors a uniform orientation, and the random anisotropic couplings, which favor different orientations in different spatial regions. For a region of size $L$, the elastic cost of a smooth rotation grows as $L^{D-2}$, whereas the energy gained by adapting to the local disorder grows as $L^{D/2}$. The latter therefore dominates on sufficiently large scales for $D<4$. The marginal case $D=4$ requires a logarithmic refinement of this balance~\cite{Fisher1985}.

To make this argument precise, fix a disorder realization $\mathcal{J}=(\mathcal{J}_l,\mathcal{J}_P)$ and a pure state $a$, with normalized frozen magnetization $\mu_a(v)=N^{-1/2}\langle S(v)\rangle_a$. 
Ferromagnetic order requires a nonzero spatial average, $M_a=\lim_{V\to\infty}\frac1V\sum_v\mu_a(v)\neq0$.
In what follows, we drop the subscript $a$ to avoid confusion and measure all lengths in lattice units.

\vspace{5pt} \noindent {\textbf{Random anisotropy}.}
Although the disorder ensemble is $O(N)$ invariant, an individual realization is not. Under a global rotation $R$ the couplings transform as
\begin{equation}
\begin{split}
    \tilde{\mathcal J}_l&\to R^T\tilde{\mathcal J}_lR,\\
    (\mathcal J_P)_{j_1j_2j_3j_4}&\to
    (\mathcal J_P)_{i_1i_2i_3i_4}R_{i_1j_1}R_{i_2j_2}R_{i_3j_3}R_{i_4j_4},
\end{split}
\end{equation}
which in general produces a different disorder realization. 
Rotating the spins in a spatial region therefore changes the quenched energy throughout the bulk of that region, not merely along its boundary. 
This is the crucial distinction from an $O(N)$-invariant random-bond model, $H_{\rm rb}=-\sum_{\langle vv'\rangle}J_{vv'}S(v)^TS(v')$ with $J_{vv'}=\pm1$, where a common rotation of all spins in a region $\Omega$ leaves every bond entirely inside $\Omega$ unchanged and affects only bonds crossing $\partial\Omega$. 
In the present model, by contrast, the matrix and tensor disorder defines random preferred directions in spin space throughout the volume, which drives the Imry--Ma instability.

This Imry--Ma scaling can be understood from a simple energy estimate.
To this end, we temporarily neglect the plaquette disorder, and consider a region $\Omega$ with a uniform magnetization profile, $\mu(v) = \hat e$.
Let $\mathcal L_\Omega$ denote the set of links in $\Omega$ of linear size $L$, and define $X_\Omega=\sum_{l\in\mathcal L_\Omega}\widetilde J_l$.
Since $|\mathcal L_\Omega|\sim L^D$, each entry of $X_\Omega$ has variance of order $J_l^2L^D/N$. Since the energy of the spins in $\Omega$ is proportional to $\hat e^T X_\Omega\hat e$, only the symmetric part, $Y_\Omega=(X_\Omega+X_\Omega^T)/2$, contributes.
At large $N$, the upper edge of the spectrum of $Y_\Omega$ is therefore of order $J_lL^{D/2}$.
Optimizing the orientation of the magnetization within $\Omega$ consequently yields a disorder-energy gain $\Delta E_{\rm rand}(L)\sim NpJ_lL^{D/2}$, while the elastic cost of a smooth rotation over a distance $L$ from the ferromagnetic exchange is $\Delta E_{\rm el}(L)\sim N(1-p)L^{D-2}$. In what follows, we show this more precisely for finite $N$ with plaquette disorders.

\vspace{5pt} \noindent {\textbf{Random field on the frozen moment}.} 
For a prescribed frozen profile $S(v)=\sqrt N\,\mu(v)$, the random interactions generate the local field
\begin{equation}
\begin{split}
    h_i(v) &=
    p\sqrt N\sum_{v'\sim v}
    (\tilde{\mathcal J}_{vv'})_{ij}\,\mu_j(v')
    \\
    &\quad+
    N^{3/2}\sum_{P\ni v}
    (\mathcal J_P)_{ijkl}\,
    \mu_j(v_2)\mu_k(v_3)\mu_l(v_4).
    \label{eq:imryma_field}
\end{split}
\end{equation}
Here $\tilde{\mathcal J}_{vv'}$ is the random matrix on the link $\langle vv'\rangle$, transposed when $v$ is its second vertex. The sites $v_2,v_3,v_4$ are the remaining corners of $P$, with tensor indices ordered accordingly.

To estimate the cost of long-wavelength orientational distortions, we restrict to a fixed-amplitude profile~\footnote{If amplitude fluctuations are massive, they only renormalize the coefficients and do not change the Imry--Ma scaling. An energy-lowering distortion within this restricted family is already sufficient to establish instability of the uniform state.},
\begin{equation}
    \mu(v)=\sqrt q\,n(v),
    \qquad |n(v)|=1.
\end{equation}
where $q=\frac1V\sum_v|\mu(v)|^2$ is the Edwards--Anderson self-overlap. Thus, the frozen moment $\langle S(v)\rangle=\sqrt{Nq}\,n(v)$ has fixed magnitude $\sqrt{Nq}$, while its orientation $n(v)$ is allowed to vary spatially.

For a fixed frozen profile, $h(v)$ is a zero-mean Gaussian random vector at each site $v$.
Its covariance follows by summing the independent link and plaquette contributions. Each of the $2D$ links incident on $v$ contributes a variance $p^2J_l^2q$ per component, while each of the $2D(D-1)$ incident plaquettes contributes $J_P^2q^3$. Hence
\begin{equation}
    \overline{h_i(v)h_j(v)}=2q\,(A_l+A_Pq^2)\,\delta_{ij}.
    \label{eq:imryma_cov}
\end{equation}
Remarkably, the same combination appears in the local saddle-point equation~\eqref{eq:saddle_pt3},
\begin{equation}
    -\frac{\lambda}{\beta^2}=2q\,(A_l+A_Pq^2).
    \label{eq:fieldvar}
\end{equation}
Thus $-\lambda/\beta^2$ admits a direct physical interpretation as the variance per spin component of the quenched random field acting on the prescribed frozen profile.

\vspace{5pt} \noindent {\textbf{Energy balance}.} 
Consider a uniformly magnetized state in which all frozen moments point along the same direction, $n(v)=\hat e_1$.
Let $\Omega$ be a region of linear size $L$, and suppose that the moments inside $\Omega$ are rotated through a small angle $\theta$ in a transverse direction $\pi$. To leading order,
\begin{equation}
    n=\hat e_1+\theta\pi+O(\theta^2),
\end{equation}
where $\pi$ is a unit vector, $|\pi|=1$, orthogonal to $\hat e_1$, $\pi\,\cdot\,\hat e_1=0$. 
To linear order in $\theta$, the corresponding change in the disordered part of the Hamiltonian is
\begin{equation}
    \delta E_{\rm rand}
    =
    -\theta\sqrt{Nq}\,
    \pi\cdot H_\Omega,
    \qquad
    H_\Omega
    \equiv
    \sum_{v\in\Omega} h(v).
\end{equation}
\eqnref{eq:imryma_cov} implies that each component of $H_\Omega$ has variance $2q(A_l+A_Pq^2)L^D$. Since only the $N-1$ components transverse to $\hat e_1$ contribute to the rotation, the transverse field typically has magnitude
\begin{equation}
    |H_{\Omega,\perp}|
    \simeq
    \sqrt{2Nq\,(A_l+A_Pq^2)}\,L^{D/2}.
\end{equation}
The region can choose its rotation direction after seeing the disorder, and therefore aligns $\pi$ with $H_{\Omega,\perp}$. The resulting energy gain is
\begin{equation}
    \Delta E_{\rm rand}(L,\theta)
    \simeq
    -\theta\,Nq
    \sqrt{2(A_l+A_Pq^2)}\,
    L^{D/2}.
    \label{eq:imryma_random}
\end{equation}

The competing elastic cost comes from the ferromagnetic exchange. If the rotation through angle $\theta$ is spread smoothly over a distance $L$, neighboring orientations differ by an amount of order $\theta/L$. Hence
\begin{equation}
    \Delta E_{\rm el}(L,\theta)
    \simeq
    Nq(1-p)\,\theta^2L^{D-2}.
    \label{eq:imryma_elastic}
\end{equation}
Both the optimized disorder gain and the elastic cost are therefore proportional to $N$. Their competition is set by their different scaling with $L$.

Minimizing $\Delta E_{\rm el}+\Delta E_{\rm rand}$ with respect to $\theta$ gives
\begin{equation}
    \theta_{\rm opt}(L)
    \sim
    \frac{\sqrt{A_l+A_Pq^2}}{1-p}\,
    L^{(4-D)/2}.
    \label{eq:imryma_thetaopt}
\end{equation}
For $D<4$, the preferred distortion grows with the region size. The small-angle description breaks down when it becomes order one, defining the Larkin scale
\begin{equation}
    L_c
    \sim
    \left[
        \frac{1-p}{\sqrt{A_l+A_Pq^2}}
    \right]^{\frac{2}{4-D}}.
    \label{eq:larkin}
\end{equation}
Below $L_c$, ferromagnetic stiffness maintains approximate orientational coherence. Above $L_c$, adapting the orientation to the disorder is energetically favorable, destabilizing ferromagnetic order.

The distinction from the random-field Ising model is important. A continuous vector order parameter can accommodate a change of orientation through a smooth twist, with cost $L^{D-2}$, rather than a sharp domain wall, with cost $L^{D-1}$.
At $D=4$, both terms in the elementary balance scale as $L^2$. The large-$N$ saddle resolves this marginal case. On a putative magnetized branch, the noncondensed frozen contribution to the self-overlap diverges logarithmically with system size, ruling out ferromagnetic order.

\vspace{5pt} \noindent {\textbf{Glass correlation length and the Larkin scale}.} 
Remarkably, the Larkin scale has the same scaling as the spin-glass correlation length $\xi$ derived in~\secref{sec:RS_correlators}. Recall that the latter characterizes spatial correlations of the frozen magnetization pattern. In the small-gap regime, $\xi\simeq\Delta^{-1/2}$, and the gapped Watson integrals, for $D<4$, behave as 
\begin{equation}
    w_2(D,\Delta)\sim \Delta^{(D-4)/2},
    \qquad D<4.
\end{equation}
Using this asymptotic form in~\eqnref{eq:sg_gain}, the saddle equation gives, up to dimension-dependent numerical constants,
\begin{equation}
    1
    \sim
    \frac{A_l+A_Pq^2}{(1-p)^2}\,
    \Delta^{-(4-D)/2}.
\end{equation}
The corresponding correlation length therefore scales as
\begin{equation}
    \xi
    \sim
    \left[
        \frac{1-p}{\sqrt{A_l+A_Pq^2}}
    \right]^{\frac{2}{4-D}} \sim L_{c},
    \qquad D<4.
    \label{eq:larkin_xi}
\end{equation}
The two lengths thus describe the same physical crossover from complementary perspectives. The Imry--Ma argument identifies $L_c$ as the scale beyond which uniform ferromagnetic alignment is lost, while the spin-glass saddle identifies $\xi$ as the distance over which the resulting frozen magnetization pattern remains spatially correlated.

\subsection{Phase Diagram}
\label{subsec:RS_phase_diagram}

We now construct the replica-symmetric phase diagram by solving the saddle-point equations for all coexisting paramagnetic, ferromagnetic, and spin-glass branches at fixed $T$, $p$, $J_P$, and $J_l$, and then selecting the thermodynamically dominant solution. Substituting a generic RS ansatz into~\eqnref{eq:Ieff} and taking the replica limit $n\to0$ gives
\begin{equation}
\begin{split}
 \frac{\beta}{NV} F
  &= \lim_{n\to 0}\frac{1}{nV}\,I_{n}[q,\lambda_{d},\lambda]
  \\
  &=\frac{\beta^2 A_{P}}{4}\left(q^{4}-1\right)
   +\frac{\beta^2 A_{l}}{2}\left(q^{2}-1\right) -\frac{\lambda_{d}}{2}
  \\
  &+\frac{\lambda q}{2}
   +\frac{1}{2V}\sum_{k\neq0}\log\epsilon(k)
   +\frac{1}{2V}\sum_{k\neq0}\frac{\lambda}{\epsilon(k)} .
  \label{eq:FreeEnergy_RS}
\end{split}
\end{equation}
Stationarity of~\eqnref{eq:FreeEnergy_RS} with respect to $q$, $\lambda$, and $\lambda_d$ reproduces the saddle-point equations derived above.
On the ferromagnetic branch, the uniform $k=0$ mode must be retained when taking these variations because it carries the condensate.
However, its direct contribution to the free-energy density vanishes as $V\to\infty$, so the on-shell ferromagnetic free energy may be evaluated using only the $k\neq0$ modes, with $\lambda_d-\lambda=2D\beta(1-p)$ and the corresponding saddle-point values of $q$ and $\lambda$.

As is standard in replica calculations, the variational principle requires us to maximize, rather than minimize, with respect to the overlap $q$. This reversed variational character is a consequence of the limit $n\to0$. The off-diagonal sector contains $n(n-1)/2$ components, so its per-replica multiplicity $(n-1)/2$ becomes negative as $n\to0$. Consequently, the minimum obtained at positive integer $n$ is analytically continued to a maximum in the replica limit.

Representative phase diagrams for $J_l=1$ are shown in~\figref{fig:phasediagram}. Black and red curves denote continuous and first-order transitions, respectively. Depending on the plaquette-disorder strength, the saddle-point equations can admit two competing glass branches. We label the lower-overlap branch SG$_1$, which connects continuously to the paramagnet at $q=0$, and the higher-overlap branch SG$_2$, which appears at finite $q$. Since both have $M^2=0$ and $q>0$, they share the same symmetry and can merge continuously at a glass--glass critical point.

For weak plaquette disorder, $J_P=0.5$, all phase boundaries in~\figref{fig:phasediagram}(a) are continuous, and SG$_1$ is the only glass branch. Its overlap $q$ grows from zero at the PM--SG$_1$ transition. The ferromagnetic phase is bounded above by the gap-closing line $T=(1-p)/w_1(D)$ and below by the boundary where the condensate $M^2$ vanishes while $q$ remains nonzero. These two boundaries meet the PM--SG$_1$ line at the multicritical point~\eqnref{eq:mcp},
\begin{equation}
    p^*\simeq0.6945,
    \qquad
    T^*\simeq2.6420.
\end{equation}
Its location is independent of $J_P$ because the plaquette interaction does not enter the linear instability condition for the onset of glass order. The multicritical point persists at intermediate plaquette disorder but is preempted by a first-order transition at stronger disorder.

At intermediate plaquette disorder, $J_P=1.4$, a higher-overlap glass branch, SG$_2$, appears and competes with SG$_1$ [Fig.~\ref{fig:phasediagram}(b)]. The first-order coexistence curve connects two glass--glass critical points, CP$_L$ at $p\simeq0.5835$ and CP$_R$ at $p\simeq0.9899$. At each critical point, the discontinuity in $q$ vanishes and the two glass branches merge. The coexistence curve also crosses four critical endpoints. Away from these points, most of the curve separates SG$_1$ from SG$_2$. Between CEP$_1$ and CEP$_2$, however, SG$_1$ is replaced by the ferromagnet, so the first-order boundary separates FM from SG$_2$. Similarly, between CEP$_3$ and CEP$_4$, SG$_1$ is replaced by the paramagnet, giving a PM--SG$_2$ boundary. Each critical endpoint occurs where a continuous phase boundary terminates on the first-order curve. At CEP$_1$ and CEP$_2$, the FM and SG$_1$ branches merge continuously as $M^2\to0$. At CEP$_3$ and CEP$_4$, the PM and SG$_1$ branches merge as $q\to0$. In each case, the merged phase remains in first-order coexistence with SG$_2$. These are therefore critical endpoints, not triple points at which three distinct phases coexist.

For strong plaquette disorder, $J_P=3$, SG$_2$ occupies most of the ordered region, although a small SG$_1$ region survives at low $p$ [Fig.~\ref{fig:phasediagram}(c)]. A short first-order SG$_1$--SG$_2$ segment connects the glass--glass critical point at $p\simeq0.0776$ to CEP$_1$ at $p\simeq0.1637$. Beyond CEP$_1$, the same first-order coexistence curve separates FM from SG$_2$. It reaches CEP$_2$ at $p\simeq0.4617$, where it meets the continuous PM--FM boundary. For larger $p$, the paramagnet transitions directly into SG$_2$ with a finite jump in $q$, and this first-order PM--SG$_2$ boundary persists to $p=1$.


Spatial dimension affects both the effective disorder strengths and the momentum integrals entering the saddle equations. 
Figure~\ref{fig:phasediagram}(d) shows the resulting phase diagram in $D=3$ for $J_P=1$. Here the divergence of $w_2(D)$ eliminates ferromagnetic order, and the finite-$p$ multicritical point of the $D=5$ diagrams is absent. Instead, a single continuous boundary separates PM from SG$_1$. As $p\to0$, this boundary approaches the dashed gap-closing line $T=(1-p)/w_1(D)$. Its temperature is nonmonotonic, reaching a minimum $T_{\mathrm{SG}}\simeq1.9007$ at $p\simeq0.6684$. At a fixed temperature just above this minimum, increasing $p$ therefore gives the re-entrant sequence
\begin{equation}
    \mathrm{SG}_1\longrightarrow\mathrm{PM}\longrightarrow\mathrm{SG}_1.
\end{equation}
These phase diagrams are restricted to the replica-symmetric ansatz. To assess their validity, we now analyze the stability of the replica-symmetric solutions against replica-symmetry-breaking fluctuations.

\subsection{Replicon Instability}
\label{subsec:RS_stability}
We now examine the stability of the replica-symmetric saddle against leading $1/N$ fluctuations.
We expand the effective action to quadratic order about the RS solution, writing
\begin{align}
    Q^{\alpha\beta}(v) &= Q_{\rm RS}^{\alpha\beta} + \frac{1}{\sqrt{N}}\delta Q^{\alpha\beta}(v),
     \\
    \Lambda^{\alpha\beta}(v) &= \Lambda_{\rm RS}^{\alpha\beta}
    + \frac{i}{\sqrt{N}}\delta\Lambda^{\alpha\beta}(v) .
\end{align}
The factors of $1/\sqrt{N}$ ensure that cubic and higher-order fluctuations are parametrically suppressed as $N\to\infty$.
Furthermore, the factor of $i$ multiplying $\delta\Lambda$ reflects the fact that $\Lambda$ was introduced in \eqnref{eq:delta_insertion_offdiag} and \eqnref{eq:delta_insertion_diag} through a Fourier representation of the delta function. Accordingly, its integration contour runs parallel to the imaginary axis, even after it is deformed through the real saddle.

The spherical constraint fixes the diagonal overlap, so diagonal fluctuations are absent, $\delta Q^{\alpha\alpha}=0$. Thus, $\delta Q$ belongs to the space of symmetric matrices with vanishing diagonal, whose dimension is $n(n-1)/2$. Because we expand about a saddle point, the linear terms vanish and the leading correction is quadratic,
\begin{align}
    I_{n}[Q,\Lambda] = I_{n}[Q_{\rm RS},\Lambda_{\rm RS}] + \frac{1}{N}\delta^{2}I_{n} + \cdots,
\end{align}

The RS saddle is invariant under all permutations of the replica indices. The Hessian therefore commutes with the action of $S_n$ and decomposes into three irreducible sectors:
\begin{itemize}[leftmargin=13pt]
  \item the \emph{longitudinal} sector, spanned by the uniform mode
  $\delta Q^{\alpha\beta}=\text{const}$, with dimension $1$;
  \item the \emph{anomalous} sector, spanned by
  $\delta Q^{\alpha\beta}=z^{\alpha}+z^{\beta}$ with $\sum_{\alpha}z^{\alpha}=0$, with dimension $n-1$;
  \item the \emph{replicon} sector, consisting of matrices with vanishing row and
  column sums, $\sum_{\beta}\delta Q^{\alpha\beta}=0$, with dimension $n(n-3)/2$.
\end{itemize}
These subspaces are mutually orthogonal, manifestly invariant under $S_n$, and their dimensions sum to $n(n-1)/2$.

\begin{figure}[!t]
    \centering
    \includegraphics[width=\linewidth]{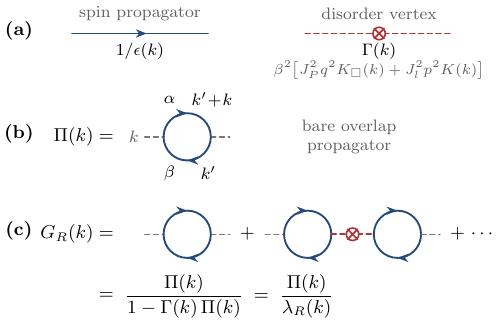}
    \caption{\textbf{Diagrammatic form of the replicon propagator.} (a) The two ingredients are the spin propagator $1/\epsilon(k)$ and the disorder vertex $\Gamma(k)$ of~\eqnref{eq:gamma_def}, which mediates interactions between overlap fluctuations. 
    (b) The bare overlap propagator $\Pi(k)$ is the one-loop bubble formed from two spin propagators carrying distinct replica indices $\alpha\neq\beta$.
    (c) The disorder vertex links successive bubbles, and the resulting geometric series resums into the full replicon propagator $G_{R}(k)=\Pi(k)/[1-\Gamma(k)\Pi(k)]=\Pi(k)/\lambda_{R}(k)$. Written as a Dyson equation, $G_{R}^{-1}=\Pi^{-1}-\Gamma$, we see that the disorder vertex plays the role of the self-energy of the replicon mode, and the instability condition $\Gamma\Pi=1$ is analogous to a Stoner criterion. 
    As $\lambda_R(0)\to0$, the zero-momentum replicon propagator diverges together with the spin-glass susceptibility, $\chi_{\rm SG}\propto G_{R}(0)$.}
\label{fig:replicon_diagrams}
\end{figure}

The three sectors have distinct physical meanings.
A longitudinal fluctuation changes all off-diagonal overlaps together and therefore moves along the RS manifold itself. It probes only whether the saddle is a proper extremum in $q$, not whether replica symmetry is stable. The anomalous direction corresponds to perturbations acting on a single replica at a time, and is often associated with replica-dependent fluctuations of the magnetization. In the limit $n\to 0$, the longitudinal and anomalous directions become degenerate.
The replicon sector is qualitatively different. Replicon fluctuations redistribute the overlaps among pairs of replicas without singling out any one replica. They are precisely the fluctuations from which a hierarchical Parisi structure develops. A negative replicon mode therefore signals an instability of the replica-symmetric saddle toward replica symmetry breaking. We consequently focus on this sector below.

\subsubsection{Quadratic Action}
We begin by expanding the effective action $I_n[Q,\Lambda]$ to quadratic order in $\delta Q$ and $\delta\Lambda$. The plaquette term contains $\prod_{v\in P}Q^{\alpha\beta}(v)$, whose quadratic part couples all six pairs of vertices on a plaquette. Four of these pairs are nearest neighbors, with each bond shared by $2(D-1)$ plaquettes, while the remaining two are plaquette diagonals. The plaquette contribution is therefore
\begin{align}
  -\frac{\beta^{2}J_{P}^{2}q^{2}}{4}\sum_{\substack{v,v'\\ \alpha,\beta}}
  \delta Q^{\alpha\beta}(v)\,K_{\square}(v-v')\,\delta Q^{\alpha\beta}(v'),
\end{align}
where the plaquette kernel $K_{\square}$ is defined by,
\begin{align}
  K_{\square}(v-v') = 2(D-1) K(v-v') + K_{2}(v-v') ,
\end{align}
and $K_2$ is the next-nearest-neighbour kernel, 
\begin{align}
    K_{2}(r) &= \begin{cases} 1, & r=\pm\hat{e}_{\mu}\pm\hat{e}_{\nu},\
  \mu<\nu,\\ 0, & \text{otherwise}.\end{cases} 
\end{align}

The link term contributes in the same way but only through the nearest-neighbour kernel,
\begin{align}
  -\frac{\beta^{2}J_{l}^{2}p^{2}}{4}\sum_{\substack{v,v'\\ \alpha,\beta}}
  \delta Q^{\alpha\beta}(v)\,K(v-v')\,\delta Q^{\alpha\beta}(v') .
\end{align}
Furthermore, the local $Q\Lambda$ term contributes
\begin{align}
  -\frac{i}{2}\sum_{\substack{v\\\alpha,\beta}}
  \delta Q^{\alpha\beta}(v)\,\delta\Lambda^{\alpha\beta}(v),
\end{align}
while expanding the logarithm to second order gives
\begin{align}
  \frac{1}{4}\Tr\big[\mathcal{M}_{\rm RS}^{-1}\delta\Lambda\,
  \mathcal{M}_{\rm RS}^{-1}\delta\Lambda\big] .
\end{align}
The positive sign of the latter term is due to the factor $i$ multiplying $\delta\Lambda$.

Collecting all the terms, we find that the quadratic action is
\begin{equation}
\begin{split}
  \delta^{2}I_{n} = &-\frac{1}{2}\sum_{\alpha<\beta}\sum_{v,v'}
  \delta Q^{\alpha\beta}(v)\,\Gamma(v-v')\,\delta Q^{\alpha\beta}(v')
  \\
  &- i\sum_{\alpha<\beta}\sum_{v}
  \delta Q^{\alpha\beta}(v)\,\delta\Lambda^{\alpha\beta}(v)  \\
  &+ \frac{1}{4}\Tr\big[\mathcal{M}_{\rm RS}^{-1}\,\delta\Lambda\,\mathcal{M}_{\rm RS}^{-1}\,\delta\Lambda\big],
\end{split}
\end{equation}
where we define the \emph{disorder vertex} $\Gamma$ as follows
\begin{equation}
\begin{split}
  \Gamma(v-v') = \beta^{2}J_{P}^{2}q^{2}\,&K_{\square}(v-v')  \\
  + \beta^{2}J_{l}^{2}p^{2}\,&K(v-v').
\end{split}
\end{equation}
\begin{figure}[!t]
    \centering
    \includegraphics[width=\linewidth]{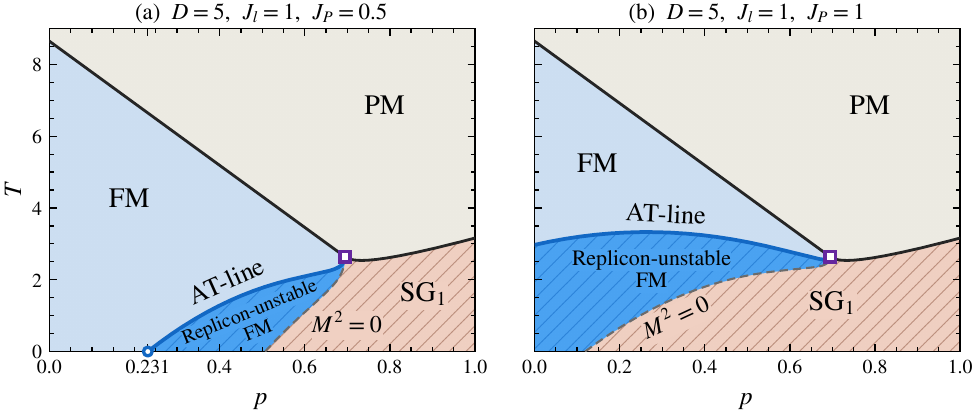}
    \caption{\textbf{Stability of the RS solution} at $D=5$, $J_{l}=1$, for (a) $J_{P}=0.5$ and (b) $J_{P}=1$. The replicon eigenvalue vanishes on the de Almeida--Thouless (AT) line (blue), which lies strictly above the locus $M^{2}=0$ at which the replica-symmetric condensate disappears (dashed). Between the two the saddle retains $M^{2}\neq0$ but breaks replica symmetry, giving a mixed phase (hatched, blue) in which ferromagnetic and glassy order coexist. Hatching marks every region where the replica-symmetric solution is unstable. Due to \eqnref{eq:replicon_sg}, this includes the whole glassy region for any $J_{P}\neq0$. In both panels, the AT line and $M^{2}=0$ lines merge at the multicritical point (square, $p^{*}=0.6945$), whose location is independent of $J_{P}$, and there the mixed phase closes. They differ at small $p$. For $J_{P}=0.5$ the AT line reaches zero temperature at $p=0.231$ (open circle), below which the ferromagnet remains replica symmetric down to $T=0$. For $J_{P}=1$ no such termination occurs: the AT line extends to $p=0$ and no replica-symmetric ferromagnet survives at zero temperature. The crossover between the two behaviours lies at $J_{P}\simeq0.66$.}
\label{fig:mixed}
\end{figure}
Because the replica-symmetric saddle is translation-invariant, it is natural to express the quadratic action in momentum space.
To this end, we write $\delta Q^{\alpha\beta}(v)=\frac{1}{V}\sum_{k}e^{ik\cdot v}\delta Q^{\alpha\beta}(k)$ and similarly for $\delta\Lambda$. The lattice kernels become
\begin{equation}
\begin{split}
  K_{\square}(k) &= 2(D-1)K(k)+K_{2}(k), \\
  \text{where} \quad K_{2}(k) &= 4\sum_{\mu<\nu}\!\cos k_{\mu}\cos k_{\nu}.
\end{split}
\end{equation}
Thus the quadratic action in momentum space assumes the form
\begin{align}
  \delta^{2}I_{n} = &-\frac{1}{2V}\sum_{k}\Gamma(k)
  \sum_{\alpha<\beta}\big|\delta Q^{\alpha\beta}(k)\big|^{2} \nonumber \\
  &-\frac{i}{V}\sum_{k}\sum_{\alpha<\beta}
  \delta Q^{\alpha\beta}(-k)\,\delta\Lambda^{\alpha\beta}(k)
  \nonumber\\
  +\frac{1}{4V^{2}}\sum_{k,k'}
  &\Tr\big[\mathcal{M}_{\rm RS}^{-1}(k')\delta\Lambda(k)\,
  \mathcal{M}_{\rm RS}^{-1}(k'-k)\delta\Lambda(-k)\big],
  \label{eq:delta2In_k}
\end{align}
where $\Gamma(k)$ is given by
\begin{align}\label{eq:gamma_def}
  \Gamma(k) = \beta^{2}J_{P}^{2}q^{2}\,K_{\square}(k)
  + \beta^{2}J_{l}^{2}p^{2}\,K(k) .
\end{align}
At zero momentum, $K_{\square}(0)=6D(D-1)$ and $K(0)=2D$, giving 
\begin{align}
    \Gamma(0) &= 6\beta^2 A_P q^2 + 2\beta^2 A_l .
\end{align}
Note that the plaquette and link disorders now enter the fluctuation spectrum through different lattice kernels.
This is a genuinely short-range effect with no analogue in the fully connected model.

\subsubsection{Replicon Eigenvalues}
We now restrict ourselves solely to fluctuations in the replicon sector, in which both $\delta Q$ and $\delta\Lambda$ are symmetric matrices with vanishing row and column sums. 

Recall that in the replica limit, $(\mathcal{M}_{\rm RS}^{-1})^{\alpha\beta}(k)=\epsilon(k)^{-1}\delta^{\alpha\beta}
-\lambda\,\epsilon(k)^{-2}$.
The second term is proportional to $\mathbf{1}\mathbf{1}^{T}$ and therefore annihilates any replicon vector,
\begin{align}
  \big[\mathcal{M}_{\rm RS}^{-1}(k)\,\delta\Lambda\big]^{\alpha\gamma}
  &= \frac{\delta\Lambda^{\alpha\gamma}}{\epsilon(k)}
  -\frac{\lambda}{\epsilon(k)^{2}}\sum_{\beta}\delta\Lambda^{\beta\gamma} \nonumber \\
  &= \frac{\delta\Lambda^{\alpha\gamma}}{\epsilon(k)} .
\end{align}
Accordingly, in $\delta^{2}I_{n}$, the term quadratic in $\delta \Lambda$ simplifies to,
\begin{align}
  \frac{1}{4V^{2}}\sum_{k,k'}
  \frac{\Tr\big[\delta\Lambda(k)\delta\Lambda(-k)\big]}
       {\epsilon(k')\epsilon(k'-k)}
  = \frac{1}{2V}\sum_{k}\Pi(k)\!\!\sum_{\alpha<\beta}\!\!
    \big|\delta\Lambda^{\alpha\beta}(k)\big|^{2},
\end{align}
where we have defined $\Pi(k) = \frac{1}{V}\sum_{k'} \epsilon(k')^{-1} \epsilon(k'-k)^{-1}$ which is the \emph{bare propagator} for the replicon; see~\figref{fig:replicon_diagrams}.

The quadratic action is now diagonal in both the momentum and the replica label. For each mode, it reads
\begin{align}
  \delta^{2}I_{n} = \frac{1}{2V}\sum_{k,\alpha<\beta}
  \begin{pmatrix}\delta Q^{*} & \delta\Lambda^{*}\end{pmatrix}
  \begin{pmatrix} -\Gamma(k) & -i \\ -i & \Pi(k)\end{pmatrix}
  \begin{pmatrix}\delta Q \\ \delta\Lambda\end{pmatrix} .
\end{align}
Since $\Pi(k)>0$, the $\delta\Lambda$ integral is Gaussian and convergent, and may be performed exactly.
Integrating out $\delta\Lambda$ amounts to taking the Schur complement of the $\delta \Lambda \delta \Lambda$ entry.
The resulting diagonal Hessian for the $\delta Q$ fluctuations is
\begin{align}
  H(k) = -\Gamma(k) - \frac{(-i)^{2}}{\Pi(k)}
  = \frac{1}{\Pi(k)} - \Gamma(k) ,
\end{align}
Stability requires $H(k)\geq0$ for every momentum. Since $\Pi(k)>0$, it is convenient to introduce the dimensionless replicon eigenvalue
\begin{align}
  \lambda_{R}(k) = \Pi(k)\,H(k) = 1 - \Gamma(k)\,\Pi(k) .
\end{align}
The stability condition is equivalently $\lambda_{R}(k)\geq0$ for all $k$.
Since both $\Gamma(k)$ and $\Pi(k)$ are maximal at $k=0$, the first instability always occurs in the uniform channel. It is therefore sufficient to examine $\lambda_{R}\equiv\lambda_{R}(0)$. 

Note that the form $\lambda_{R}=1-\Gamma\Pi$ invites a transparent reading. The Gaussian integral over $\delta\Lambda$ endows the overlap fluctuation
$\delta Q^{\alpha\beta}$ with the bare propagator $\Pi(k)$, while disorder generates an effective interaction $\Gamma(k)$. Resumming the resulting bubble series (Fig.~\ref{fig:replicon_diagrams}) gives the full propagator
\begin{align}
    G_R(k)
    = \frac{1}{H(k)}
    = \frac{\Pi(k)}{1-\Gamma(k)\Pi(k)}
    = \frac{\Pi(k)}{\lambda_R(k)} .
\end{align}
Thus, $\lambda_R$ plays the role of an RPA denominator, and replica symmetry becomes unstable when $\Gamma\Pi=1$. This is the analogue of a Stoner criterion for replicon fluctuations: $\Gamma$ measures their coupling through disorder, while $\Pi$ measures the strength of the underlying spin fluctuations and grows as the spin spectrum softens. The same denominator controls the divergence of the spin-glass susceptibility, since $G_R(0)\propto\chi_{\rm SG}$.

\vspace{5pt} \noindent \textbf{Paramagnet:} Here $q=0$, so the plaquette kernel $K_{\square}$ drops out of $\Gamma$ entirely, leaving $\Gamma(0) = 2\beta^{2} A_l$ and
\begin{align}
  \lambda_{R} = 1 - 2\beta^{2} A_l
  \int_{-\pi}^{\pi}\frac{d^{D}k}{(2\pi)^{D}}\frac{1}{\epsilon(k)^{2}}.
\end{align}
The condition $\lambda_R=0$ is exactly the continuous glass-onset condition in~\eqnref{eq:sg_cont}, obtained earlier from the bifurcation of the glass branch. This coincidence follows from the fact that $\lambda=0$ at $q=0$, so $\mathcal{M}_{\rm RS}^{-1}$ has no $\mathbf{1}\mathbf{1}^T$ component. Thus, the Hessian is proportional to the identity, and all three replica sectors are degenerate. The stability boundary of the paramagnet is therefore also the point at which a glass solution detaches from it, as must be the case for a continuous transition.

For sufficiently strong plaquette disorder, this continuous instability is preempted by a first-order transition to the higher-overlap glass SG$_2$. This transition occurs above the instability line while the paramagnetic saddle is still locally stable. Between the thermodynamic transition and the replicon instability, the paramagnet survives only as a metastable branch.

\vspace{5pt} \noindent \textbf{Replica-symmetric glass:} Using the saddle-point equation~\eqref{eq:sg_gain}, $2\beta^{2}(A_P q^{2} + A_l) \Pi(0)=1$, so the replicon eigenvalue collapses to
\begin{align}\label{eq:replicon_sg}
  \lambda_{R} = -4\beta^{2} A_P q^{2}\,\Pi(0) .
\end{align}
This is negative whenever $J_{P}\neq0$ and $q>0$. Any nonzero plaquette disorder thus destabilizes the replica-symmetric glass throughout its ordered phase, making replica symmetry breaking unavoidable. For $J_P=0$, by contrast, $\lambda_R$ vanishes identically. The model then reduces to a two-spin spherical glass, whose RS solution is marginally stable and exact. In this sense, the plaquette interaction is the source of replica symmetry breaking in the present model, paralleling the distinction between two-spin and higher-$p$ spherical glasses. The RS glass phases found above should therefore be interpreted as unstable continuations of the saddle-point equations. Their continuous transition temperatures may remain meaningful because the replicon instability begins precisely at the transition, but their low-temperature structure must be determined using a replica-symmetry-breaking ansatz.

\vspace{5pt} \noindent \textbf{Ferromagnet:} 
On the magnetized branch, the gap closes and the uniform mode develops a condensate. To determine the bulk stability of this phase, however, the thermodynamic limit must be taken before the zero-momentum limit. The relevant bare replicon propagator is therefore
\begin{align}
    \Pi(0^{+}) &\equiv
    \lim_{k\to 0}\lim_{V\to\infty}\Pi(k)
    =
    \frac{T^{2}w_{2}(D)}{(1-p)^{2}} .
\end{align}
Crucially, in this order of limits, the isolated condensate contribution to the bare propagator drops out, leaving only the continuum contribution. The explicit factors of $\beta$ then cancel from the replicon eigenvalue.
Combining the resulting expression with~\eqnref{eq:RS_condensate} gives
\begin{align}\label{eq:lambdaR_fm}
  \lambda_{R} = \frac{M^{2}}{q}
  - \frac{4 A_P q^{2}\,w_{2}(D)}{(1-p)^{2}} .
\end{align}
Near the onset of ferromagnetism $q\to0$ and $\lambda_{R}>0$. Thus, in this region the magnetized solution is stable. However, upon cooling, $q$ grows monotonically since $q=1-T\,w_{1}(D)/(1-p)$, and the second term in \eqref{eq:lambdaR_fm} grows as $q^{2}$ until it dominates. Replica symmetry is then lost at a temperature $T_{\rm AT}$, which we name by analogy with the de Almeida--Thouless (AT) line of the Sherrington--Kirkpatrick model.
It is determined by the condition
\begin{align}
  \big(6 A_P q_{\rm AT}^{2}+2 A_l \big)w_{2}(D)=(1-p)^{2} .
\end{align}
This differs from the condition $M^{2}=0$ only by the replacement $2\to6$ in the plaquette coefficient, which traces directly to $K_{\square}(0)=6D(D-1)$ in the fluctuation vertex against the $2D(D-1)$ appearing in the saddle-point equation.
Therefore, the threshold of the onset of magnetization and that at which replica symmetry is lost differ by a factor $\sqrt3$ in the overlap.
The latter is given by,
\begin{equation}
\begin{split}
  q_{\rm AT}&=\left[\frac{(1-p)^{2}-2 A_l \,w_{2}(D)}
  {6 A_P w_{2}(D)}\right]^{1/2},
   \\
  T_{\rm AT}&=\big(1-q_{\rm AT}\big)\frac{1-p}{w_{1}(D)} .
\end{split}
\end{equation}
Below $T_{\rm AT}$, but within the magnetized phase, the correct saddle therefore breaks replica symmetry while retaining $M\neq0$. This corresponds to a mixed phase in which ferromagnetic and glassy order coexist, see Fig.~\ref{fig:mixed}.

It should be remarked that just as with the replica symmetric glass phase, the instability of the replica symmetric ferromagnetic phase is entirely driven by the plaquette disorder.
For $J_{P}=0$ one has $\lambda_{R}=M^{2}/q\geq0$, so the magnetized solution is stable wherever it exists and no mixed phase appears.

\section{Replica Symmetry Breaking}
\label{sec:RSB}
The stability analysis above shows that the replica-symmetric saddle becomes unstable in part of the phase diagram. We therefore turn to saddle points that break replica symmetry, albeit without breaking translation invariance. A natural ansatz is to consider $k$-step replica-symmetry-breaking ($k$RSB) matrices, whose entries are organized according to a nested hierarchy of $k$ blocks. The resulting matrices are no longer invariant under the full permutation group $S_n$, but only under permutations that preserve this nested block structure---replicas may be permuted within the smallest blocks, and blocks may be permuted within larger blocks containing them, at each level of the hierarchy. 
$k$-step RSB matrices form a finite-dimensional commutative, associative, unital algebra, and therefore provide a natural invariant space in which to seek saddle-point solutions.

We begin with the simplest symmetry-broken ansatz, one-step replica symmetry breaking (1RSB), and determine where its solutions are locally stable and thermodynamically favored. We then turn to full replica symmetry breaking (FRSB), in which the discrete hierarchy becomes the continuous Parisi function $q(x)$ and the pure states organize into an ultrametric tree. A central result of this work is that, in the present short-range model, this thermodynamic hierarchy acquires a direct spatial interpretation through a corresponding hierarchy of correlation lengths. Pure states that separate near the outer levels of the Parisi tree differ through long-wavelength variations of their frozen magnetization patterns, whereas states that remain together to deeper levels differ only on progressively shorter spatial scales. We also study one-step full replica symmetry breaking (1FRSB), which combines a continuous sector with a single discrete jump. Finally, we compare these competing saddle points to construct the full phase diagram and argue that RS, 1RSB, FRSB, and 1FRSB exhaust the replica structures realized by the model. We further show that a continuous Parisi hierarchy leads to Goldstone modes, whose long-wavelength fluctuations destabilize continuous replica symmetry breaking for $D\leq2$ through a Mermin--Wagner-type mechanism.

\subsection{1RSB Solution}
\label{subsec:1RSB_sol}
We begin with the 1RSB ansatz, in which the replicas are partitioned into blocks of size $m$. $Q$ and $\Lambda$ then take the form
\begin{align}
    Q^{\alpha\beta} &= \begin{cases}
        1 & \alpha = \beta \\
        q_1 & \alpha \neq \beta,\ \alpha,\beta \text{ in the same block}  \\
        q_0 & \alpha,\beta \text{ in different blocks}
    \end{cases} \\
    \Lambda^{\alpha\beta} &= \begin{cases}
        \lambda_{d} & \alpha = \beta \\
        \lambda_1 & \alpha \neq \beta,\ \alpha,\beta \text{ in the same block}  \\
        \lambda_0 & \alpha,\beta \text{ in different blocks}.
    \end{cases}
\end{align}
The two off-diagonal overlaps have a simple physical interpretation. Replicas in the same block occupy the same thermodynamic pure state and have overlap $q_1$, the Edwards--Anderson self-overlap. Replicas in different blocks occupy distinct pure states and have the smaller overlap $q_0$. The difference $q_1-q_0$ therefore measures the separation between distinct pure states.
For integer replica number $n$, the block size $m$ is an integer divisor of $n$. After analytic continuation to $n\to0$, it becomes the Parisi breaking parameter, with $0\leq m\leq1$. At either endpoint, the 1RSB ansatz reduces to RS form: as $m\to 1$, all off-diagonal overlaps reduce to $q_0$, whereas as $m\to 0$, they reduce to $q_1$.


Substituting the one-step ansatz into the effective action and taking the replica limit gives

\begin{equation}
\begin{split}
   \frac{\beta}{NV} F
   =-&\frac{\beta^{2}A_{P}}{4} \big[1+(m-1)q_{1}^{4}-mq_{0}^{4}\big] \\
   -&\frac{\beta^{2}A_{l}}{2} \big[1+(m-1)q_{1}^{2}-mq_{0}^{2}\big]  \\
   -&\frac{1}{2}\big[\lambda_{d}+(m-1)q_{1}\lambda_{1}-mq_{0}\lambda_{0}\big]
    \\
   +&\frac{1}{2V}\sum_{k}\Bigg[\bigg(1 - \frac{1}{m}\bigg)\log\epsilon_{1}(k)  \\
    +&\frac{1}{m}\log\epsilon_{0}(k)
    +\frac{\lambda_{0}}{\epsilon_{0}(k)}\Bigg].
\end{split}
\end{equation}
Here $\epsilon_{1}(k)$ and $\epsilon_{0}(k)$ are the two dispersions associated with the \emph{intra-block} and \emph{inter-block eigenchannels} of $\mathcal{M}(k)$, respectively:
\begin{align}
    \epsilon_{1}(k) &= \lambda_{d}-\beta(1-p)K(k)-\lambda_{1}, \\
    \epsilon_{0}(k) &= \lambda_{d}-\beta(1-p)K(k)+(m-1)\lambda_{1}-m\lambda_{0}.
\end{align}

Stationarity of the 1RSB free energy with respect to $q_0$, $q_1$, $\lambda_0$, $\lambda_1$, $\lambda_d$, and $m$ yields the complete set of saddle-point equations.

\begin{figure*}[!t]
\centering
\includegraphics[width=\textwidth]{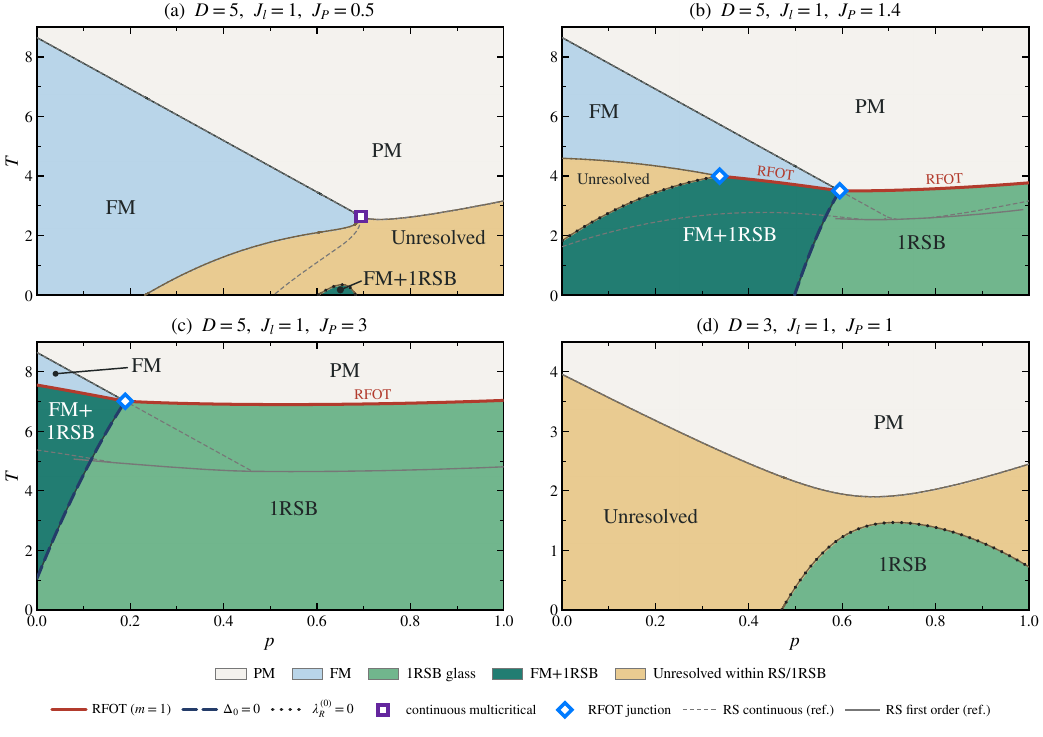}
\caption{\textbf{1RSB phase diagrams in the $(p,T)$ plane for $J_l=1$.}
At each point, we compare the admissible RS and 1RSB saddles and show the replicon-stable solution with the largest on-shell free energy. Off-white, blue, green, and teal denote the RS paramagnet (PM), RS ferromagnet (FM), unmagnetized 1RSB glass, and magnetized FM+1RSB phase, respectively. Amber marks regions unresolved within the RS/1RSB comparison. Solid red, dashed dark-blue, and dotted black curves denote the static $m=1$ (random-first order transition) boundary, closure of the outer spin gap $\Delta_0=0$, and outer-replicon marginality $\lambda_R^{(0)}=0$ with $\lambda_R^{(1)}>0$, respectively.
Gray curves show the RS phase boundaries for reference; additional
markers indicate the junctions labeled in the legend.
\textbf{(a)} $D=5$, $J_P=0.5$: no stable unmagnetized 1RSB glass is found, although a small low-temperature FM+1RSB pocket appears. The square marks the multicritical point of~\eqnref{eq:mcp}, which now borders the unresolved region.
\textbf{(b)} $D=5$, $J_P=1.4$: FM+1RSB occurs at lower frustration, while the unmagnetized 1RSB glass occupies the higher-frustration region, preempting the RS glass--glass structure.
\textbf{(c)} $D=5$, $J_P=3$: the unmagnetized 1RSB glass occupies most of the ordered region, together with a low-$p$ FM+1RSB wedge and a small RS ferromagnetic region.
\textbf{(d)} $D=3$, $J_P=1$: ferromagnetic order is absent, and an unresolved region separates the paramagnet from the low-temperature 1RSB dome.}
\label{fig:stable_1rsb_phase_diagram}
\end{figure*}

Variation with respect to the overlaps $q_{1}$ and $q_{0}$ gives, 
\begin{align}
\lambda_{i} &= -2\beta^2 \left(A_P q_{i}^{3} + A_l q_{i}\right), \qquad i=0,1,
\label{eq:sp_l1RSB}
\end{align}
while variation with respect to $\lambda_{d}$, $\lambda_{1}$ and $\lambda_{0}$ returns the diagonal, intra-block and inter-block components of the saddle-point equation $Q=(\mathcal{M}^{-1})(0)$,
\begin{align}
    1 &= \frac{1}{V}\sum_{k}\Bigg[
    \left(1-\frac{1}{m}\right)\frac{1}{\epsilon_{1}(k)}
    +\frac{1}{m}\frac{1}{\epsilon_{0}(k)}
    -\frac{\lambda_{0}}{\epsilon_{0}(k)^{2}}
    \Bigg], \label{eq:sp_diag} \\
    q_{1} &= \frac{1}{V}\sum_{k}\Bigg[
    \frac{1}{m}\left(\frac{1}{\epsilon_{0}(k)}-\frac{1}{\epsilon_{1}(k)}\right)
    -\frac{\lambda_{0}}{\epsilon_{0}(k)^{2}}
    \Bigg], \label{eq:sp_q1}  \\
    q_{0} &= -\frac{\lambda_{0}}{V}\sum_{k}\frac{1}{\epsilon_{0}(k)^{2}} . 
\label{eq:sp_q0}
\end{align}
Suitable combinations of these are more transparent, and generalize the replica-symmetric sum rule in \eqnref{eq:sum_rule}:
\begin{equation}\label{eq:1RSB_sumrule}
\begin{split}
  1-q_{1} &= \frac{1}{V}\sum_{k}\frac{1}{\epsilon_{1}(k)},
  \\
  q_{1}-q_{0} &= \frac{1}{mV}\sum_{k}
  \left(\frac{1}{\epsilon_{0}(k)}-\frac{1}{\epsilon_{1}(k)}\right) .
\end{split}
\end{equation}
Finally, stationarity with respect to the breaking parameter $m$ gives
\begin{align}
0 = \;&-\frac{\beta^2 A_P}{4}\left(q_{1}^{4}-q_{0}^{4}\right)
-\frac{\beta^2 A_l}{2}\left(q_{1}^{2}-q_{0}^{2}\right)
-\frac{1}{2}\left(q_{1}\lambda_{1}-q_{0}\lambda_{0}\right) \nonumber\\
+\frac{1}{2V}&\sum_{k}\Bigg[
\frac{1}{m^{2}}\log\frac{\epsilon_{1}(k)}{\epsilon_{0}(k)}
+\frac{\lambda_{1}-\lambda_{0}}{m\,\epsilon_{0}(k)}
-\frac{\lambda_{0}\left(\lambda_{1}-\lambda_{0}\right)}{\epsilon_{0}(k)^{2}}
\Bigg]. \label{eq:sp_m}
\end{align}

The condition on $m$ has a somewhat different status from the other saddle-point equations. For integer $n$, $m$ is a discrete block size rather than an independent matrix element. Different values of $m$ therefore correspond to different one-step block structures. Only after the analytic continuation $n\to0$ does $m$ become a continuous parameter. The condition $\partial F/\partial m=0$ then selects the thermodynamically dominant member of this family of 1RSB solutions.

This interpretation is particularly clear in terms of the configurational entropy, or complexity density~\cite{Monasson1995},
\begin{align}
    \Sigma(m)
    =
    \frac{\beta m^2}{NV}
    \frac{\mathrm d F(m)}{\mathrm dm}.
\end{align}
Schematically, the number of metastable states contributing at a given free-energy density scales as $\exp[NV\Sigma(m)]$. An interior solution $m_\star\in(0,1)$ of $\partial F/\partial m=0$ therefore has $\Sigma(m_\star)=0$ and describes a phase in which the equilibrium Gibbs measure is supported on only a subexponential number of pure states. If the on-shell free energy has no stationary point for $0<m<1$, the zero-complexity condition cannot be satisfied in the interior of the allowed interval. The one-step solution then terminates at the endpoint $m=1$, where the glassy states have positive complexity and occur in exponentially large number.

If instead the equilibrium branch reaches the endpoint $m=1$ before an interior stationary point appears, the system remains in the positive-complexity regime associated with an exponential number of metastable states. This is the regime between the \emph{dynamical} and \emph{Kauzmann} transitions~\cite{Kauzmann1948, Monasson1995}. The dynamical line does not correspond to a  thermodynamic singularity; rather, in the large-$N$ limit it marks the onset of dynamical arrest, where relaxation times diverge even though the equilibrium free energy remains analytic. The Kauzmann line marks the lower-temperature point at which the equilibrium complexity vanishes and the static condensed glass phase sets in.

Our aim here is the static thermodynamic phase diagram, so in what follows we restrict to the zero-complexity sector $\Sigma=0$. We leave the positive-complexity branches and their dynamical transition to future work.

\subsubsection{Ferromagnetic Branch}
The magnetized 1RSB branch arises from condensation of the uniform $k=0$ mode. As in the replica-symmetric case, such condensation requires a vanishing gap. Here it is the inter-block gap that closes, $\epsilon_{0}(0)\sim V^{-1/2}\to0^{+}$, fixing
\begin{align}
  \lambda_{d}+(m-1)\lambda_{1}-m\lambda_{0}=2D\beta(1-p) .
\end{align}
The resulting ferromagnetic condensate is $M^{2}=-\lambda_{0}/[V\epsilon_{0}(0)^{2}]$. Only the inter-block channel can become gapless. Indeed, the splitting between the two dispersions is momentum independent, $\epsilon_{1}(k)-\epsilon_{0}(k)=m(\lambda_{0}-\lambda_{1}) = 2m \beta^2 \big[A_P(q_{1}^{3}-q_{0}^{3}) + A_l(q_{1}-q_{0})\big]$, and is strictly positive whenever $q_{1}>q_{0}$. Thus, as $\epsilon_{0}(0)\to0^{+}$, the intra-block channel remains gapped, $\epsilon_{1}(0)=m(\lambda_{0}-\lambda_{1})>0$. The splitting vanishes only in the replica-symmetric limit $q_{1}=q_{0}$, where the two channels merge.

For $k\neq0$, the two dispersions may therefore be written as
\begin{equation}
\begin{split}
  \epsilon_{0}(k)&=\beta(1-p)\big[2D-K(k)\big], \\
  \epsilon_{1}(k)&=\beta(1-p)\big[\Delta_{1}+2D-K(k)\big],  \\
  &\text{where}\quad \Delta_{1}\equiv\frac{\epsilon_{1}(0)}{\beta(1-p)} .
\end{split}
\end{equation}
where $\Delta_{1}$ is the dimensionless intra-block gap. It is now convenient to extend the Watson integrals to finite gap as follows
\begin{align}
  w_{r}(D,\Delta)=\int_{-\pi}^{\pi}\frac{d^{D}k}{(2\pi)^{D}}
  \frac{1}{[\Delta+2D-K(k)]^{r}} ,
\end{align}
with $w_{r}(D,0)=w_{r}(D)$.

Separating the condensed $k=0$ mode from the continuum in the saddle-point equations~\eqref{eq:sp_diag}--\eqref{eq:sp_q0}, we find that the condensate contributes equally to the diagonal, intra-block, and inter-block components:
\begin{align}\label{eq:magnetized_1RSB_lambda}
    1 &=M^{2} + \int_{k}\Bigg[
    \left(1-\frac{1}{m}\right)\frac{1}{\epsilon_{1}(k)}
    +\frac{1}{m}\frac{1}{\epsilon_{0}(k)}
    -\frac{\lambda_{0}}{\epsilon_{0}(k)^{2}}\Bigg], \nonumber\\
    q_{1} &= M^{2}+\int_{k}\Bigg[
    \frac{1}{m}\left(\frac{1}{\epsilon_{0}(k)}-\frac{1}{\epsilon_{1}(k)}\right)
    -\frac{\lambda_{0}}{\epsilon_{0}(k)^{2}}\Bigg], \nonumber \\
    q_{0} &= M^{2}-\lambda_{0}\int_{k}\frac{1}{\epsilon_{0}(k)^{2}} .
\end{align}
Because the condensate appears identically in all three, it cancels from their differences, and the two sum rules remain the same,
\begin{align}
  1-q_{1} &= \frac{T}{1-p}\,w_{1}(D,\Delta_{1}), \label{eq:fm1rsb_a}\\
  q_{1}-q_{0} &= \frac{T}{m(1-p)}\big[w_{1}(D)-w_{1}(D,\Delta_{1})\big].
  \label{eq:fm1rsb_b}
\end{align}
Eliminating $w_{1}(D,\Delta_{1})$ then  yields a closed relation involving only the overlaps,
\begin{align}\label{eq:fm1rsb_sumrule}
  m\,q_{0}+(1-m)\,q_{1} = 1-\frac{T\,w_{1}(D)}{1-p} .
\end{align}
This is the one-step analogue of the RS relation
$q=1-Tw_1(D)/(1-p)$, with the single RS overlap replaced by the weighted average $m q_{0}+(1-m)q_{1}$. Consequently, the upper temperature bound for the existence of a magnetized solution remains $T^{*}=(1-p)/w_{1}(D)$, unchanged by replica symmetry breaking.

The full magnetized 1RSB solution is determined by solving \eqnref{eq:sp_m}, \eqnref{eq:fm1rsb_a}, and \eqnref{eq:fm1rsb_sumrule}, together with $\Delta_{1}=2m \beta [A_P(q_{1}^{3}-q_{0}^{3})+A_l(q_{1}-q_{0})]/(1-p)$. The remaining inter-block saddle-point equation~\eqref{eq:magnetized_1RSB_lambda} then fixes the condensate. Using \eqnref{eq:sp_l1RSB} and $\int_{k}\epsilon_{0}^{-2}=T^{2}w_{2}(D)/(1-p)^{2}$, the factors of $\beta$ cancel, yielding
\begin{align}
  M^{2}=q_{0}\left[1-\frac{\big(2J_{P}^{2}D(D-1)q_{0}^{2}
  +2J_{l}^{2}p^{2}D\big)\,w_{2}(D)}{(1-p)^{2}}\right] .
\end{align}
This has exactly the same form as the RS result, with the RS overlap $q$ replaced by the inter-state overlap $q_0$. Since a finite condensate again requires convergence of  $w_{2}(D)$, the magnetized branch still exists only for $D>4$. Replica-symmetry breaking therefore does not restore ferromagnetic order in $D\leq4$.


\subsubsection{Spin-Glass Branch}
On this branch, the magnetization vanishes, $M=0$, while the intra-block overlap remains nonzero, $q_{1}\neq0$.
Both eigenchannels, $\epsilon_{1}(k)$ and $\epsilon_{0}(k)$, remain gapped.
The global $\mathbb{Z}_{2}$ symmetry naturally selects the branch with $q_{0}=0$ since pure states occur in symmetry-related pairs $a$ and $-a$.
As shown in~\appref{app:q0}, any gapped unmagnetized solution with $q_{0}\neq0$ is unstable.
We therefore restrict throughout to $q_{0}=\lambda_{0}=0$.

The two eigenchannel dispersions may then be written as
\begin{align}\label{eq:sg1rsb_eps}
  \epsilon_i(k)=\beta(1-p)[\Delta_i+2D-K(k)], \qquad i = 0,1
\end{align}
where $\Delta_i$ denotes the corresponding dimensionless gap. Their splitting is fixed by $\Delta_{1}-\Delta_{0}= -\frac{m\lambda_1}{\beta(1-p)}$. Since $\lambda_{1}=-2\beta^2 \bigl(A_P q_{1}^{3} + A_l q_{1}\bigr)<0$, it follows that $\Delta_{1}>\Delta_{0}$: the inter-block eigenchannel is the softer of the two.

The momentum integrals appearing in the sum rules can now be expressed in terms of the massive Watson integrals. The sum rules in~\eqnref{eq:1RSB_sumrule} reduce to
\begin{align}
  1-q_{1}
  &=
  \frac{T}{1-p}\,
  w_{1}(D,\Delta_{1}),
  \label{eq:sg1rsb_in}\\
  1-(1-m)q_{1}
  &=
  \frac{T}{1-p}\,
  w_{1}(D,\Delta_{0}).
  \label{eq:sg1rsb_out}
\end{align}

Stationarity with respect to the Parisi breaking parameter $m$ gives
\begin{equation}
\begin{split}
  0={}&
  \frac{3A_P}{4}q_{1}^{4}
  +\frac{A_l}{2}q_{1}^{2}
  +\frac{T^2}{2m^{2}}\,
  \mathcal L(\Delta_{1},\Delta_{0})
  \\
  &-
  \frac{1}{m}
  \bigl(A_P q_{1}^{3} + A_l q_{1}\bigr)
  \bigl[1-(1-m)q_{1}\bigr],
  \label{eq:sg1rsb_m}
\end{split}
\end{equation}
where we have used \eqnref{eq:sg1rsb_out} to eliminate
$\int_k \epsilon_{0}(k)^{-1}$.
We have also defined
\begin{align}
  \mathcal L(\Delta_{1},\Delta_{0})
  \equiv
  \int_{-\pi}^{\pi}
  \frac{d^{D}k}{(2\pi)^{D}}
  \log\!\left[
  \frac{\Delta_{1}+2D-K(k)}
       {\Delta_{0}+2D-K(k)}
  \right] > 0 .
  \label{eq:sg1rsb_L}
\end{align}
The common factor $\beta(1-p)$ cancels inside the logarithm, so the temperature dependence of $\mathcal L$ enters only implicitly through the two gaps $\Delta_{1}$ and $\Delta_{0}$. For fixed $(p,T)$ with $p<1$, ~\eqnref{eq:sg1rsb_in}, \eqnref{eq:sg1rsb_out}, and \eqnref{eq:sg1rsb_m} form three coupled equations for $(q_{1},\Delta_{1},m)$. Admissible gapped solutions must satisfy $0<q_{1}<1$, $\Delta_{1}>0$ and $0<m\leq1$. 


\subsubsection{Stability of the 1RSB Solution}\label{sec:1RSB_stability}
We now analyze fluctuations about the 1RSB saddle. As in the replica-symmetric case, we write
\begin{align}
    Q^{\alpha\beta}(v) &= Q_{\rm 1RSB}^{\alpha\beta} + \frac{1}{\sqrt{N}}\delta Q^{\alpha\beta}(v),
     \\
    \Lambda^{\alpha\beta}(v) &= \Lambda_{\rm 1RSB}^{\alpha\beta}
    + \frac{i}{\sqrt{N}}\,\delta\Lambda^{\alpha\beta}(v) ,
\end{align}
with $\delta Q^{\alpha\alpha}=0$ as before. Two features distinguish this case from the replica-symmetric one.

First, the residual symmetry is smaller. The one-step saddle is invariant not under all of $S_{n}$ but only under the subgroup preserving the block structure, namely permutations within each block together with permutations of the blocks themselves. Consequently the Hessian decomposes into more sectors than before, and in particular there are now \emph{two} replicon channels rather than one:
\begin{itemize}
  \item The \emph{inner replicon}, spanned by $\delta Q^{\alpha\beta}$ supported on pairs within a single block $I$ and obeying $\sum_{\beta\in I}\delta Q^{\alpha\beta}=0$ for every $\alpha\in I$.
  \item The \emph{outer replicon}, spanned by fluctuations constant on
  block-pairs, $\delta Q^{\alpha\beta}=\delta Q_{IJ}$ for $\alpha\in I$,
  $\beta\in J$ with $I\neq J$, obeying $\sum_{J}\delta Q_{IJ}=0$.
\end{itemize}
The remaining longitudinal and anomalous sectors, as in the RS case, do not govern the onset of additional hierarchical breaking.

Second, the disorder vertex now depends on the level of the hierarchy. Expanding the plaquette interaction to quadratic order produces a factor $(Q^{\alpha\beta})^2$ multiplying the kernel $K_{\square}$. This takes values $q_1^2$ and $q_0^2$ for intra-block and inter-block pairs, respectively. The corresponding vertices are
\begin{align}
  \Gamma_{1}(k) &= \beta^{2}J_{P}^{2}q_{1}^{2}\,K_{\square}(k)
  + \beta^{2}J_{l}^{2}p^{2}\,K(k) ,\nonumber\\
  \Gamma_{0}(k) &= \beta^{2}J_{P}^{2}q_{0}^{2}\,K_{\square}(k)
  + \beta^{2}J_{l}^{2}p^{2}\,K(k) .
\end{align}
Thus, only the four-body coupling distinguishes the two levels of the hierarchy. Since $q_1>q_0$, one has $\Gamma_1(0)\geq\Gamma_0(0)$. 

The essential simplification is again that the replicon projections annihilate all but one component of $\mathcal{M}^{-1}_{\rm 1RSB}$. One finds that for an inner-replicon vector with $\sum_{\beta\in
I}\delta\Lambda^{\beta\gamma}=0$,
\begin{align}
  \big[\mathcal{M}_{\rm 1RSB}(k)^{-1}\delta\Lambda\big]^{\alpha\gamma}
  = \frac{\delta\Lambda^{\alpha\gamma}}{\epsilon_{1}(k)} .
\end{align}
On the other hand, for an outer-replicon vector with $\sum_{J}\delta\Lambda_{JK}=0$, 
\begin{align}
  \big[\mathcal{M}_{\rm 1RSB}(k)^{-1}\delta\Lambda\big]_{IK}
  = \frac{\delta\Lambda_{IK}}{\epsilon_{0}(k)} .
\end{align}
Thus, the inner replicon propagates through the intra-block eigenchannel $\epsilon_1(k)$, while the outer replicon propagates through the inter-block eigenchannel $\epsilon_0(k)$.

Each channel therefore reduces to the same two-by-two structure as before, and performing the Gaussian $\delta\Lambda$ integral, the resulting Schur complement gives two dimensionless replicon eigenvalues,
\begin{align}\label{eq:replicon_1rsb}
  \lambda_{R}^{(1)}(k) &= 1-\Gamma_{1}(k)\,\Pi_{1}(k) ,\nonumber\\
  \lambda_{R}^{(0)}(k) &= 1-\Gamma_{0}(k)\,\Pi_{0}(k) ,
\end{align}
with the bare propagators
\begin{align}
  \Pi_{i}(k) = \frac{1}{V}\sum_{k'}
  \frac{1}{\epsilon_{i}(k')\,\epsilon_{i}(k'+k)} .
\end{align}
Both are maximal at $k=0$, so it again suffices to examine the stability of the $k=0$ uniform modes. Setting $q_{1}=q_{0}$ collapses $\epsilon_{1}=\epsilon_{0}$ and $\Gamma_{1}=\Gamma_{0}$, and the two expressions reduce to the single RS replicon eigenvalue.

The two replicons diagnose instabilities at different levels of the hierarchy. If $\lambda_R^{(1)}<0$, the internal structure of a one-step block is unstable and must be further resolved. If $\lambda_R^{(0)}<0$, the organization of the blocks themselves is unstable and further breaking is required at the outer level. A 1RSB solution is locally stable only when both replicons are nonnegative.

\vspace{5pt} \noindent \textbf{Spin Glass:} On the spin-glass branch, $q_{0}=0$, and hence $\Gamma_{0}$ loses its plaquette contribution entirely. The two eigenvalues at $k=0$ read
\begin{align}
  \lambda_{R}^{(1)} &= 1-\beta^2 \big(6A_P q_{1}^{2} + 2A_l\big)
  \int_{k}\frac{1}{\epsilon_{1}(k)^{2}} , \nonumber\\
  \lambda_{R}^{(0)} &= 1-2 \beta^2 A_l
  \int_{k}\frac{1}{\epsilon_{0}(k)^{2}} .
\end{align}
The outer-replicon condition has the same form as the RS paramagnetic stability condition, but with the inter-block dispersion $\epsilon_0(k)$. The inner-replicon condition generalizes~\eqnref{eq:replicon_sg}. Unlike in the RS glass, the 1RSB saddle equations do not force either replicon to be negative. We therefore evaluate both on every numerical solution and retain only saddles for which both are nonnegative.

\vspace{5pt} \noindent \textbf{Ferromagnet:} 
On the magnetized 1RSB branch, the inter-block gap closes, whereas the intra-block gap remains open. As in the replica-symmetric case, the bulk replicon stability is obtained by taking the thermodynamic limit before the zero-momentum limit. The corresponding bare propagators are therefore
\begin{align} 
    \int_{k}\frac{1}{\epsilon_{0}(k)^{2}} &= \frac{T^{2}}{(1-p)^{2}}\,w_{2}(D), \\ \int_{k}\frac{1}{\epsilon_{1}(k)^{2}} &= \frac{T^{2}}{(1-p)^{2}}\,w_{2}(D,\Delta_{1}). 
\end{align} 
The factors of $\beta$ therefore cancel in both replicon eigenvalues, which become 
\begin{align} 
    \lambda_{R}^{(0)} &= 1- \frac{ ( 6 A_P q_{0}^{2} +2 A_l ) w_{2}(D) }{(1-p)^{2}}, \label{eq:fm1rsb_outer_replicon} \\ 
    \lambda_{R}^{(1)} &= 1- \frac{ ( 6 A_P q_{1}^{2} +2 A_l ) w_{2}(D,\Delta_{1}) }{(1-p)^{2}}. \label{eq:fm1rsb_inner_replicon} 
\end{align} 
The outer-replicon condition has the same algebraic form as the RS de Almeida--Thouless condition, with $q$ replaced by $q_0$. Its location in the phase diagram nevertheless changes, since $q_0$ is now collectively determined with $q_1$, $\Delta_1$, and $m$ by the 1RSB saddle equations. Combining it with the condensate expression above gives
\begin{align}
    \lambda_R^{(0)}
    =\frac{M^2}{q_0}
    -\frac{4A_Pq_0^2w_2(D)}{(1-p)^2}.
\end{align}
At a demagnetization point where $M^2\to0$ while $q_0$ remains nonzero, this becomes
\begin{align}
    \left.\lambda_R^{(0)}\right|_{M^2=0}
    =-\frac{4A_Pq_0^2w_2(D)}{(1-p)^2}<0
\end{align}
for $J_P\neq0$. A magnetized branch approaching such a point must therefore lose outer-replicon stability before its condensate vanishes. This argument does not exclude stable magnetized 1RSB order. Nor does it exclude a continuous transition to the unmagnetized 1RSB branch, provided $q_0\to0$ together with $M^2\to0$. The distinction matters when interpreting the gap-closing boundaries in the phase diagram below. There is also no general ordering between the two replicon eigenvalues. The softer outer spin channel gives a larger bare propagator, $\Pi_0(0)>\Pi_1(0)$, but the inner channel has the stronger disorder vertex, $\Gamma_1(0)>\Gamma_0(0)$. These effects compete, so the gapless outer spin channel does not by itself determine which replicon becomes unstable first. We check both eigenvalues when selecting magnetized 1RSB solutions.

\subsubsection{Phase Diagram}
\label{sec:1rsb-phase-diagram}

We construct the 1RSB phase diagram by solving the magnetized and unmagnetized 1RSB saddle-point equations together with the competing RS paramagnetic and ferromagnetic solutions. At each point in the $(p,T)$ plane, we retain only replicon-stable saddles and select the thermodynamically dominant one by comparing their on-shell free energies. For a 1RSB saddle, both the inner and outer replicons must be nonnegative. The resulting phase diagrams for $J_l=1$ are shown in~\figref{fig:stable_1rsb_phase_diagram}. Off-white and blue denote the RS paramagnet and ferromagnet, while green and teal denote the unmagnetized 1RSB glass and the magnetized FM+1RSB phase, respectively. Regions not resolved by any stable RS or 1RSB saddle are shown in amber and are analyzed below using more general RSB ansatzes. The RS phase boundaries are shown in gray for reference and are preempted wherever another saddle is selected.

In the interior of the unmagnetized 1RSB glass, $M^2=q_0=0$ and $0<\Delta_0<\Delta_1$. The magnetized FM+1RSB phase instead has $M^2>0$, $q_0>0$, and $\Delta_0=0<\Delta_1$. In both phases, $0<m<1$, $q_1>q_0$, and $\lambda_R^{(0,1)}(0)\ge0$.
The relevant 1RSB boundaries are identified by
\begin{align}
    m&=1,
    &&\text{(solid red)},\nonumber\\
    \Delta_0&=0,
    &&\text{(dashed dark blue)},\nonumber\\
    \lambda_R^{(0)}(0)&=0,\qquad
    \lambda_R^{(1)}(0)>0,
    &&\text{(dotted black)}.
\end{align}
The $\Delta_0=0$ boundary connects the unmagnetized and magnetized one-step phases where $M^2\to0$ and does not by itself signal further replica symmetry breaking. By contrast, outer-replicon marginality marks an instability of the one-step organization toward a more refined Parisi hierarchy.

\vspace{5pt} \noindent {\textbf{The $m=1$ boundary}.}
The $m=1$ boundary marks the static PM--1RSB and FM--FM+1RSB transitions. In both cases, the transition has the characteristic random-first-order structure: the Edwards--Anderson self-overlap remains at finite $q_1$ on the glassy side, while its statistical weight vanishes as $m\to1$.

For example, on the unmagnetized branch $q_0=0$, and the equilibrium overlap distribution is
\begin{align}
    P(q) = m\,\delta(q) +(1-m)\,\delta(q-q_1).
\end{align}
As $m\to1$, $P(q)\longrightarrow\delta(q)$, recovering the paramagnet even though $q_1$ remains finite on the 1RSB side. The same argument applies on the magnetized branch, where the surviving peak is instead at the finite RS overlap $q_0$.

The thermodynamics has the same structure. At $m=1$, the 1RSB free energy reduces to that of the corresponding RS phase. Along the equilibrium branches, the first thermodynamic derivatives also match at the transition. The entropy is therefore continuous and no latent heat is released despite the discontinuous onset of $q_{1}$.

At the opposite endpoint, $m=0$, the $q_0$ peak has zero weight and $P(q)=\delta(q-q_1)$. This corresponds to a replica symmetric glass, so this endpoint is not counted as a distinct 1RSB phase.

\vspace{5pt} \noindent {\textbf{Structure of the phase diagrams}.}
Allowing one-step replica symmetry breaking substantially changes the RS phase diagram and preempts the glass--glass coexistence structure.

For $D=5$ and weak plaquette disorder,
$J_P=0.5$ [Fig.~\ref{fig:stable_1rsb_phase_diagram}(a)],
the RS paramagnet occupies the high-temperature region, while the RS ferromagnet survives at sufficiently low frustration. No stable unmagnetized 1RSB glass is found. A small low-temperature FM+1RSB pocket does appear, however, showing that one-step glass order can coexist with ferromagnetism even in this weak-plaquette regime. The remaining amber region is unresolved within the RS/1RSB comparison.

At intermediate plaquette disorder,
$J_P=1.4$ [Fig.~\ref{fig:stable_1rsb_phase_diagram}(b)], both magnetized and unmagnetized 1RSB phases are present. FM+1RSB occupies the lower frustration region, while the unmagnetized glass occupies the higher-frustration region. The static $m=1$ boundary separates FM+1RSB from the RS ferromagnet at lower frustration and 1RSB from the paramagnet at higher frustration.
At lower temperatures, the $\Delta_0=0$ boundary connects the unmagnetized glass to its magnetized counterpart. Together, these 1RSB phases preempt the RS glass--glass coexistence structure.

For strong plaquette disorder,
$J_P=3$ [Fig.~\ref{fig:stable_1rsb_phase_diagram}(c)], the unmagnetized 1RSB glass occupies most of the ordered region. 
A narrow FM+1RSB wedge and a small RS ferromagnetic region remain at low $p$, with the $m=1$ boundary again marking the static FM--FM+1RSB and PM--1RSB transitions where they occur.
Increasing $J_P$ therefore progressively favors one-step glass order over a broader part of the phase diagram.

Finally, for $D=3$ and $J_P=1$
[\figref{fig:stable_1rsb_phase_diagram}(d)], ferromagnetic order is absent, so the selected 1RSB phase is entirely unmagnetized. It forms a low-temperature dome, separated from the paramagnet by an amber region that remains unresolved in the present comparison.

\subsubsection{Replica-Resolved Spin Correlators} \label{sec:1rsb-two-correlation-lengths} 

The 1RSB kernel $\mathcal{M}_{\rm 1RSB}(k)$ contains two distinct hierarchical eigenchannels, with dispersions $\epsilon_1(k)$ and $\epsilon_0(k)$. We now examine how these two channels appear in the replica-resolved spin correlator $G^{\alpha\beta}(r)$ and interpret their physical meaning in terms of the underlying pure-state structure. For convenience, we define
\begin{align*}
    G^{\mathrm d}(r)
    &\equiv G^{\alpha\alpha}(r), \\
    G^{\mathrm{in}}(r)
    &\equiv G^{\alpha\beta}(r),
    \qquad
    \alpha\neq\beta
    \ \text{in the same Parisi block}, \\
    G^{\mathrm{out}}(r)
    &\equiv G^{\alpha\beta}(r),
    \qquad
    \alpha,\beta
    \ \text{in different Parisi blocks}.
\end{align*}
Evaluating~\eqnref{eq:ss_corr_saddle} for the one-step kernel and taking the replica limit $n\to0$ gives
\begin{align}
    G^{\mathrm d}(r)-G^{\mathrm{in}}(r)
    &=
    \frac{1}{V}\sum_k
    \frac{e^{ik\cdot r}}{\epsilon_1(k)},
    \label{eq:1rsb-inner-correlator}\\
    G^{\mathrm{in}}(r)-G^{\mathrm{out}}(r)
    &=
    \frac{1}{mV}\sum_k
    \left[
        \frac{e^{ik\cdot r}}{\epsilon_0(k)}
        -
        \frac{e^{ik\cdot r}}{\epsilon_1(k)}
    \right].
    \label{eq:1rsb-outer-correlator}
\end{align}
Note that at $r=0$, these relations reduce to the two 1RSB sum rules, $G^{\mathrm d}(0)-G^{\mathrm{in}}(0)=1-q_1$ and $G^{\mathrm{in}}(0)-G^{\mathrm{out}}(0)=q_1-q_0$. Eqs.~\eqref{eq:1rsb-inner-correlator} and \eqref{eq:1rsb-outer-correlator} may therefore be viewed as their spatially resolved counterparts.

The pure-state decomposition gives the two projections distinct physical interpretations. Let $\mu_a(r) \equiv \frac{1}{\sqrt N} \langle S(r)\rangle_a$ again denote the normalized frozen magnetization in pure state $a$, and let $C_a(v,v')$ denote the corresponding connected thermal correlator, as in~\eqnref{eq:pure_state_correlator_decomposition}.
The replica-diagonal correlator $G^{\mathrm d}(r)$ contains the full spin--spin correlation function and therefore includes both connected thermal fluctuations and correlations of the frozen magnetization.
By contrast, two distinct replicas in the same Parisi block occupy the same pure state but fluctuate independently within it, so $G^{\mathrm{in}}(r)$ contains only the frozen contribution.
Their difference therefore isolates the connected thermal correlations within a pure state,
\begin{align}
    G^{\mathrm d}(r)-G^{\mathrm{in}}(r)
    =
    \overline{
    \sum_a w_a
    \frac{1}{V}\sum_v
    C_a(v,v+r)
    }.
    \label{eq:1rsb-inner-pure-state}
\end{align}
By contrast, $G^{\mathrm{out}}$ correlates frozen magnetization patterns belonging to distinct pure states. Defining the difference field
\begin{align}
    \delta\mu_{ab}(v)
    \equiv
    \mu_a(v)-\mu_b(v),
\end{align}
we may write
\begin{align}
    G^{\mathrm{in}}(r)&-G^{\mathrm{out}}(r) \nonumber \\
    &=
    \frac{1}{2}
    \overline{
    \left\langle
    \frac{1}{V}\sum_v
    \delta\mu_{ab}(v)^T
    \delta\mu_{ab}(v+r)
    \right\rangle}_{q_{ab}=q_0}
    .
    \label{eq:1rsb-state-difference-correlator}
\end{align}
Here the conditional average runs over pairs of distinct pure states with mutual overlap $q_{ab}=q_0$. At coincident points ($r=0$), the difference-field correlator reduces to the overlap deficit $q_{\rm EA}-q_0 \equiv q_{1}-q_0$, the mean-square separation between their frozen magnetization patterns. At finite $r$, it resolves how this difference is organized in real space, with its decay defining the characteristic length scale over which differences between the frozen magnetization patterns of the two states remain spatially correlated.
Further details on the difference-field correlator and its relation to thermodynamic overlaps are given in~\appref{app:pure_states}.

On a gapped branch, $\Delta_i>0$, and the corresponding eigenchannel propagators decay at large distance as
\begin{align}
    \frac{1}{V}\sum_k
    \frac{e^{ik\cdot r}}{\epsilon_i(k)}
    \sim
    \frac{e^{-r/\xi_i}}{r^{(D-1)/2}},
    \qquad
    r\gg1,
\end{align}
where $\xi^{-1}_i = \operatorname{arcosh}\!\left(1+\frac{\Delta_{i}}{2}\right)$ for $r$ along a lattice axis, which reduces to $\xi_i\simeq\Delta_i^{-1/2}$ for $\Delta_i\ll1$. Since $\Delta_0<\Delta_1$, the two correlation lengths satisfy $\xi_0>\xi_1$.
Eqs.~\eqref{eq:1rsb-inner-correlator} and
\eqref{eq:1rsb-outer-correlator} therefore imply
\begin{align}
    G^{\mathrm d}(r)-G^{\mathrm{in}}(r)
    &\sim
    \frac{e^{-r/\xi_1}}{r^{(D-1)/2}},
    \\
    G^{\mathrm{in}}(r)-G^{\mathrm{out}}(r)
    &\sim
    \frac{1}{m}
    \frac{e^{-r/\xi_0}}{r^{(D-1)/2}}.
\end{align}
We therefore identify $\xi_1$ as the thermal correlation length within a selected pure state, while $\xi_0$ characterizes the longer spatial scale associated with the frozen difference between distinct pure states. 
Thus $q_{\rm EA}-q_0$ measures the mean-square difference between the frozen magnetization patterns of the two pure states, while $\xi_0$ sets the spatial scale over which these differences remain organized. One-step replica symmetry breaking therefore introduces a second, longer correlation length, $\xi_0>\xi_1$, associated with spatial correlations between distinct pure-state configurations.


In the interior of the unmagnetized 1RSB branch $0<\Delta_0<\Delta_1$, so both correlation lengths are finite, with $\xi_0>\xi_1$. On the magnetized branch, by contrast, the outer gap closes, $\Delta_0=0$, while the inner channel remains gapped, $\Delta_1>0$. The replica-uniform condensate contributes equally to $G^{\mathrm d}$, $G^{\mathrm{in}}$, and $G^{\mathrm{out}}$, and therefore cancels from the projected differences. The outer projection is then governed by a massless simple pole and decays algebraically,
\begin{align}
    G^{\mathrm{in}}(r)-G^{\mathrm{out}}(r)
    \sim
    r^{2-D},
\end{align}
whereas the inner projection, $G^{\mathrm d}(r)-G^{\mathrm{in}}(r)$, remains exponentially short ranged.


\subsection{FRSB Solution}
\label{subsec:FRSB}

The instability of the 1RSB saddle toward additional replica-symmetry breaking motivates us to consider the full replica-symmetry-breaking (FRSB) limit. In this limit, the discrete hierarchy of Parisi blocks is replaced by a continuous coordinate $x\in[0,1]$. A hierarchical matrix is then specified by its diagonal element together with a function of $x$. We therefore write
\begin{align}
  Q\longleftrightarrow \bigl(1,q(x)\bigr),
  \qquad
  \Lambda\longleftrightarrow \bigl(\lambda_{d},\lambda(x)\bigr).
\end{align}
The Parisi function $q(x)$ is nondecreasing. Small $x$ labels the outer levels of the hierarchy, corresponding to more distant pure states, while larger $x$ probes progressively deeper levels associated with more closely related states. The 1RSB ansatz is recovered when $q(x)$ and $\lambda(x)$ reduce to step functions with a single breakpoint.

For the FRSB ansatz, the replica-space spectrum of $\mathcal M(k)$ decomposes into a continuous hierarchical band and an isolated replica-uniform mode. The band is parametrized by $x$, with level-dependent eigenvalue,
\begin{align}
    \epsilon(k,x)
    \equiv
    \lambda_d-\beta(1-p)K(k)
    -x\lambda(x)
    -\int_x^1dy\,\lambda(y).
    \label{eq:frsb_epsilon}
\end{align}
At finite replica number, the isolated replica-uniform mode has eigenvalue
\begin{align}
    \epsilon_{\mathrm{iso}}(k) =
    \epsilon(k,0)+n\lambda(0)+O(n^2).
    \label{eq:frsb_iso}
\end{align}
Although this eigenvalue merges with the outer edge $\epsilon(k,0)$ of the hierarchical band as $n\to0$, its $O(n)$ splitting leaves a finite contribution in $\lim_{n\to 0}\frac{1}{n}\Tr\log\mathcal M(k)$.
In particular, we find that
\begin{align}
    &\mathcal T(k)
    \equiv
    \lim_{n\to0}\frac{1}{n}\Tr\log\mathcal M(k) \nonumber \\
    &\quad =
    \log\epsilon(k,0)
    -\int_0^1\frac{dx}{x^2}
    \log\left[
        \frac{\epsilon(k,x)}{\epsilon(k,0)}
    \right]
    +\frac{\lambda(0)}{\epsilon(k,0)}.
    \label{eq:frsb_Tk}
\end{align}
A derivation of the hierarchical spectrum and Eq.~\eqref{eq:frsb_Tk} is given in~\appref{app:frsb_spectrum}.

Consequently, the FRSB free-energy functional in the replica limit becomes
\begin{align}
  \frac{\beta F}{NV} = &-\frac{\beta^2 A_P}{4}\left[ 1-\int_{0}^{1}dx\,q(x)^{4} \right] \nonumber \\
  &- \frac{\beta^2 A_l}{2}\left[ 1-\int_{0}^{1}dx\,q(x)^{2}\right] \nonumber \\
  - &\frac{1}{2}\left[\lambda_{d}-\int_{0}^{1}dx\,q(x)\lambda(x)\right] +\frac{1}{2V}\sum_{k}\mathcal T(k).
  \label{eq:frsb_free_energy}
\end{align}

Variation with respect to $q(x)$ is local in the Parisi coordinate and gives
\begin{align}
    \lambda(x) = -2\beta^2 \bigl[A_P q(x)^3 + A_lq(x)\bigr].
    \label{eq:frsb_lambda}
\end{align}
The remaining variations impose the continuum analogue of the matrix saddle equation $Q=\frac{1}{V}\sum_k\mathcal M^{-1}(k)$. This relation is most transparent in the hierarchical eigenbasis. To this end, we define the Parisi transform of the overlap as
\begin{align}
    \widehat q(x)
    \equiv
    1-xq(x)-\int_x^1dy\,q(y).
    \label{eq:qhat}
\end{align}
The quantity $\widehat q(x)$ is the hierarchical eigenvalue of $Q$ at level $x$. Since the corresponding eigenvalue of $\mathcal M(k)$ is $\epsilon(k,x)$, the matrix saddle equation reduces to
\begin{align}
    \widehat q(x)
    =
    \frac{1}{V}\sum_k
    \frac{1}{\epsilon(k,x)},
    \qquad x\in[0,1].
    \label{eq:frsb_sumrule}
\end{align}
This is the level-resolved FRSB sum rule.

The hierarchical band does not, however, contain all of the saddle-point information. The isolated replica-uniform mode supplies an additional outer-edge condition. Its leading $O(n^0)$ contribution reproduces the $x\to0$ limit of~\eqnref{eq:frsb_sumrule}, whereas its $O(n)$ splitting gives
\begin{align}
  q(0) = -\lambda(0)\,\frac{1}{V}\sum_k
  \frac{1}{\epsilon(k,0)^2}.
  \label{eq:frsb_isolated_sumrule}
\end{align}
A derivation of this condition from the finite-$n$ spectrum is given in~\appref{app:frsb_spectrum}.

It is again convenient to parametrize each hierarchical eigenchannel by a dimensionless gap $\Delta(x)$,
\begin{align}
    \epsilon(k,x)
    =
    \beta(1-p)
    \bigl[
        \Delta(x)+2D-K(k)
    \bigr].
    \label{eq:frsb_epsilon_gap}
\end{align}
Using Eqs.~\eqref{eq:frsb_epsilon} and \eqref{eq:frsb_lambda}, its derivative along the hierarchy is
\begin{align}
    \Delta'(x)
    &=
    -\frac{x\lambda'(x)}{\beta(1-p)}
    \nonumber\\
    &=
    \frac{
        2\beta x\bigl[3A_P q(x)^2 + A_l\bigr]
    }{
        (1-p)
    }
    q'(x)
    \geq 0.
    \label{eq:frsb_DeltA_Prime}
\end{align}
Thus $\Delta(x)$ is constant on plateaus of $q(x)$ and increases along continuously breaking intervals. Outer levels of the Parisi hierarchy therefore correspond to softer spin eigenchannels, while
deeper levels are progressively more strongly gapped.

We now restrict to a continuously breaking interval, on which $q'(x)\neq0$. In terms of the massive Watson integral, Eq.~\eqref{eq:frsb_sumrule} becomes
\begin{align}
    \widehat q(x)
    =
    \frac{T}{1-p}\,
    w_1\!\left(D,\Delta(x)\right).
    \label{eq:frsb_sumrule_watson}
\end{align}
Differentiating with respect to $x$ and using
\begin{align}
    \widehat q\,'(x)
    &=
    -xq'(x),
    \\
    \partial_{\Delta}w_1(D,\Delta)
    &=
    -w_2(D,\Delta),
\end{align}
gives
\begin{align}
    -xq'(x) = -\frac{T}{1-p}\,
    \Delta'(x)\,
    w_2\!\left(D,\Delta(x)\right).
\end{align}
Substituting Eq.~\eqref{eq:frsb_DeltA_Prime} and cancelling the common factor $xq'(x)$ gives
\begin{align}
    1 = \frac{2}{(1-p)^2} \bigl[3 A_P q(x)^2 + A_l\bigr]\, w_2\!\left(D,\Delta(x)\right).
    \label{eq:frsb_marginality}
\end{align}
Equivalently,
\begin{align}
    w_2\!\left(D,\Delta(x)\right) = \frac{(1-p)^2}{2 \bigl[3A_P q(x)^2 + A_l\bigr]} .
    \label{eq:frsb_gap_equation}
\end{align}
This relation has two useful consequences. First, it contains no explicit dependence on $x$ since the Parisi coordinate enters only through the overlap $q(x)$ and the gap $\Delta(x)$.
Second, the explicit temperature dependence has canceled out.
The dimensionless gap is therefore determined by a temperature-independent relation, $\Delta(x)=\Delta\!\left(q(x)\right)$.

Differentiating \eqnref{eq:frsb_gap_equation} again determines the location of each overlap value along the Parisi hierarchy,
\begin{align}
    &x(q)
    =
    \frac{
        3T A_P q(1-p)^3
    }{
        4 \bigl[3A_P q^2 + A_l\bigr]^3
        w_3\!\left(D,\Delta(q)\right)
    }.
    \label{eq:frsb_x_of_q}
\end{align}
Thus $x(q)$ scales linearly with temperature, $x(q,T)=T f(q)$, where $f(q)$ is temperature independent. Equivalently, on every continuously breaking interval,
\begin{align}
    q(x,T) = \mathcal Q\!\left(\frac{x}{T}\right).
    \label{eq:frsb_scaling}
\end{align}
Temperature therefore rescales the Parisi coordinate without altering the relation between the overlap $q$ and the gap $\Delta$. This is the short-range, spatially dispersive analogue of the Parisi--Toulouse scaling~\cite{ParisiToulouse1980}. Since the real-space correlation length is fixed by the gap, $\xi(\Delta)$, the same overlap $q$ is associated with the same correlation length at all temperatures. Cooling only changes where that overlap appears along the Parisi hierarchy.

For a pure FRSB solution, the continuously breaking interval terminates at $x=x_c$, where $q(x)$ reaches the Edwards--Anderson value $q_{\rm EA}$.
For $x_c<x\leq1$, the Parisi function remains on the upper plateau, $q(x)=q_{\mathrm{EA}}$. Evaluating the undifferentiated sum  rule~\eqref{eq:frsb_sumrule} at $x=1$ gives
\begin{align}
    1-q_{\mathrm{EA}}
    =
    \frac{T}{1-p}
    w_1\!\left(D,\Delta(q_{\mathrm{EA}})\right).
    \label{eq:frsb_EA_condition}
\end{align}
Since the local equations already determine $\Delta(q)$, this condition fixes $q_{\rm EA}$. The endpoint of the continuous sector then follows from $x(q_{\mathrm{EA}})=x_c$. The undifferentiated condition is essential because Eq.~\eqref{eq:frsb_marginality} was obtained by differentiating the full sum rule and therefore loses one integration condition. For a pure FRSB profile, Eq.~\eqref{eq:frsb_EA_condition} supplies the required boundary condition at the upper endpoint.

\subsubsection{Ferromagnetic Branch} \label{sec:frsb_magnetized} 
The ferromagnetic condensate contributes the replica-uniform term
$M^2\mathbf{1}\mathbf{1}^{T}$ to the overlap matrix. This rank-one contribution has vanishing eigenvalue in every nonuniform hierarchical channel, $M^2\left[1-x-\int_x^1dy\right]=0$, so it does not enter the continuum sum rule~\eqref{eq:frsb_sumrule}. Instead, the condensate resides entirely in the isolated replica-uniform mode. A nonzero condensate therefore requires the outermost spin-kernel gap to close, $\Delta(0)=0$. If the Parisi function contains a lower plateau, then $q'(x)=0$ on that interval and Eq.~\eqref{eq:frsb_DeltA_Prime} implies $\Delta'(x)=0$. The gap consequently remains zero throughout the lower plateau and becomes positive only upon entering a continuously breaking sector.

Evaluating the continuum sum rule at the outer edge gives
\begin{align} 
    1-\int_0^1dx\,q(x) = \frac{T}{1-p}\,w_1(D,0). \label{eq:frsb_FM_sumrule} 
\end{align} 
This fixes the average overlap along the Parisi hierarchy, and generalizes the corresponding RS (\eqnref{eq:RS_magnetizedq}) and 1RSB (\eqnref{eq:fm1rsb_sumrule}) relations.
Since $q(x)\geq0$, it also implies
\begin{align} 
    T \leq T^{*} \equiv \frac{1-p}{w_1(D,0)} 
\end{align} 
for any magnetized solution. Thus, whenever the PM--FM transition is continuous and is not preempted by a glassy phase, it occurs at the same gap-closing temperature as in the RS theory.

Separating the condensed $k=0$ contribution in the isolated replica-uniform saddle condition of \eqnref{eq:frsb_isolated_sumrule} gives
\begin{align} 
    q(0) = M^2 - \lambda(0)\, \frac{T^2}{(1-p)^2}\, w_2(D,0). \label{eq:frsb_FM_q0} 
\end{align}
Using \eqnref{eq:frsb_lambda}, we obtain
\begin{align} 
    M^2 = q(0) \left[ 1- \frac{ 2\bigl[A_P q(0)^2+ A_l\bigr] w_2(D,0)}{ (1-p)^2 }  \right]. \label{eq:frsb_FM_condensate} 
\end{align} 
A finite condensate therefore requires $w_2(D,0)$ to converge, which in the presence of disorder requires $D>4$. Full replica-symmetry breaking thus does not restore ferromagnetic order for $D\leq4$.


Let $x_\ell$ denote the lower endpoint of the continuously breaking interval. A pure FRSB profile has no finite jumps, so $q(x)=q(0)$ on the lower plateau and increases continuously from $q(0)$ at $x_\ell$. This plateau may have zero width. Since $\Delta(x)$ is constant on the plateau and $\Delta(0)=0$, the continuous sector necessarily begins at
$q=q(0)$ and $\Delta=0$.

Taking the limit toward this endpoint from within the continuous sector,~\eqnref{eq:frsb_marginality} gives
\begin{align}
    1
    =\frac{2\bigl[3A_Pq(0)^2+A_l\bigr]}{(1-p)^2}
    w_2(D,0).
    \label{eq:frsb_FM_outer_marginal}
\end{align}
Combining this with~\eqnref{eq:frsb_FM_condensate} yields
\begin{align}
    M^2
    =\frac{2A_Pq(0)^3}{3A_Pq(0)^2+A_l}.
    \label{eq:frsb_FM_M2_closed}
\end{align}
For $q(0)>0$ and nonzero plaquette disorder, this condensate is positive. 

\subsubsection{Spin-Glass Branch}
\label{sec:frsb_sg}

In the absence of magnetization, the spin-flip-symmetric branch $q(0)=\lambda(0)=0$ is an exact solution of the saddle-point equations. Moreover, as shown in~\appref{app:q0}, any gapped unmagnetized solution with $q(0)\neq0$ is unstable in the outer-replicon channel for nonzero plaquette disorder. We therefore restrict to $q(0)=0$ when constructing stable FRSB spin-glass solutions.

If the FRSB phase emerges continuously from the paramagnet, the continuously breaking sector begins at the outer edge, where $q(0)=0$. Taking the limit $x\to0^+$ in Eq.~\eqref{eq:frsb_gap_equation} gives
\begin{align}
    1 = \frac{
        2J_l^2p^2D
    }{
        (1-p)^2
    }
    w_2\!\left(D,\Delta(0)\right).
    \label{eq:frsb_PM_onset}
\end{align}
This is precisely the condition for marginal stability of the RS paramagnet in the replicon channel, derived in Sec.~\ref{subsec:RS_stability}. At a continuous PM--FRSB transition, the outer FRSB gap $\Delta(0)$ therefore matches continuously onto the paramagnetic gap.

\begin{figure*}[!t]
    \centering
    \includegraphics[width=\textwidth]{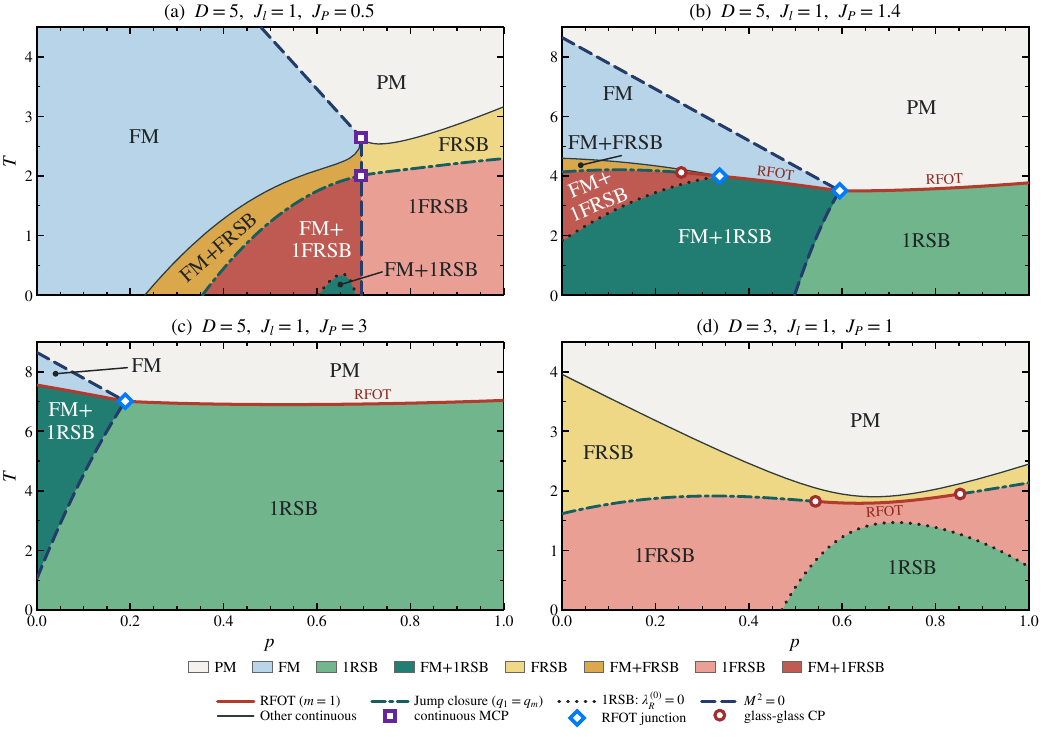}
    \caption{
    \textbf{Complete RSB phase diagrams in the $(p,T)$ plane at $J_l=1$.}
    At each point, we select the admissible, replicon-stable saddle with the largest on-shell replica free energy among the RS, 1RSB, FRSB, and 1FRSB solutions found numerically.
    Off-white and blue denote the RS paramagnet and ferromagnet.
    Green and teal denote unmagnetized 1RSB and FM+1RSB order.
    Yellow and gold denote FRSB and FM+FRSB, while salmon and rust denote 1FRSB and FM+1FRSB.
    Solid red curves denote static RFOT boundaries at $m=1$.
    Dash-dotted teal curves mark closure of the finite 1FRSB jump, $q_1=q_m$.
    Dotted black curves denote outer-replicon marginality of the 1RSB phase, $\lambda_R^{(0)}=0$, while dashed dark-blue curves mark $M^2=0$ and separate magnetized from unmagnetized phases.
    Purple squares mark continuous multicritical points, blue diamonds mark RFOT junctions, and red circles mark glass--glass critical points.
    \textbf{(a)} $D=5$, $J_P=0.5$:
    FRSB and 1FRSB occupy the unmagnetized glassy region, with FM+FRSB and FM+1FRSB at lower frustration. A small low-temperature FM+1RSB pocket also survives, while no unmagnetized 1RSB phase is selected.
    \textbf{(b)} $D=5$, $J_P=1.4$:
    FM+1RSB occupies a broad low-frustration region, while the unmagnetized 1RSB glass extends to larger $p$.
    Narrow FM+FRSB and FM+1FRSB regions remain near the upper boundary of the magnetized glass, and the PM--1RSB and FM--FM+1RSB transitions contain RFOT segments.
    \textbf{(c)} $D=5$, $J_P=3$:
    the unmagnetized 1RSB glass occupies most of the ordered region, together with a low-$p$ FM+1RSB region and a small RS ferromagnetic phase.
    No continuous-RSB region is resolved in this panel.
    \textbf{(d)} $D=3$, $J_P=1$:
    all magnetized phases are absent.
    The paramagnet gives way to FRSB, followed at lower temperature by 1FRSB, while a 1RSB dome appears at larger frustration.
    The FRSB--1FRSB boundary contains both continuous jump-closure and RFOT segments, separated by glass--glass critical points, while the 1FRSB--1RSB boundary is set by outer-replicon marginality.
    }
    \label{fig:full_phase_diagram}
\end{figure*}

\subsubsection{Stability of the FRSB Solution} \label{sec:frsb_stability} 
Local stability of the FRSB saddle against hierarchical refinements is controlled by the replicon sector.
As before, we expand about the FRSB saddle as
\begin{align} 
    Q &= Q_{\rm FRSB}+\frac{1}{\sqrt{N}}\delta Q,\,\,\, \Lambda = \Lambda_{\rm FRSB}+\frac{i}{\sqrt{N}}\,\delta\Lambda. 
\end{align} 
Expanding the effective action to quadratic order and integrating out $\delta\Lambda$ yields the effective Hessian governing the replicon fluctuations $\delta Q$.

In the FRSB phase, the replicon sector is continuously resolved by the Parisi coordinate $x$. Each level of the hierarchy therefore carries its own spatially dispersing replicon mode, with dimensionless replicon eigenvalue
\begin{align} 
    \lambda_R(k;x) = 1-\Gamma_x(k)\Pi_x(k),\label{eq:frsb_replicon_k} 
\end{align} 
where
\begin{align} 
    \Gamma_x(k) &= \beta^2J_P^2q(x)^2K_{\square}(k) + \beta^2J_l^2p^2K(k), \\ \Pi_x(k) &= \int_{k'} \frac{1}{ \epsilon(k',x)\epsilon(k'+k,x) }. \label{eq:frsb_GammaPi} 
\end{align}
Here $\Gamma_x(k)$ is the disorder-induced interaction vertex in the replicon channel at hierarchy level $x$, while $\Pi_x(k)$ is the corresponding bare replicon propagator constructed from a pair of level-$x$ spin propagators. 

As in the RS and 1RSB analyses, the softest mode occurs at $k=0$. Using
\begin{align}
    \Gamma_x(0)
    &=
    6 \beta^2 A_P q(x)^2+ 2\beta^2 A_l, \\
    \Pi_x(0)
    &=
    \frac{T^2}{(1-p)^2}
    w_2\!\left(D,\Delta(x)\right),
\end{align}
the zero-momentum replicon eigenvalue becomes
\begin{align}
    \lambda_R(0;x)
    =
    1
    -
    \frac{2}{(1-p)^2}
    \bigl[3A_P q(x)^2 + A_l\bigr]
    w_2\!\left(D,\Delta(x)\right).
    \label{eq:frsb_replicon}
\end{align}
Comparison with Eq.~\eqref{eq:frsb_marginality} immediately gives $\lambda_R(0;x)=0$ throughout every continuously breaking interval. Replicon marginality therefore follows directly from the saddle-point equations.

This marginality also extends to any plateau attached to a continuously breaking interval. On such a plateau, both $q(x)$ and $\Delta(x)$
remain fixed at their endpoint values, so Eq.~\eqref{eq:frsb_replicon} gives the same zero replicon eigenvalue throughout.

In particular, the terminal plateau of a pure FRSB solution remains marginal,
\begin{align}
    \lambda_R(0;x)=0,
    \qquad x_c\leq x\leq1.
\end{align}
The same reasoning applies to a lower plateau when present. An independent stability check is needed only when a plateau is separated from the continuous sector by a finite jump. This occurs for the upper plateau of a 1FRSB solution, which need not inherit the marginality of the continuous sector below it.

The vanishing of $\lambda_R(0;x)$ implies a gapless replicon mode at every level of the continuous Parisi hierarchy. Although this result has so far been established only at leading order in $1/N$, in~\secref{sec:mermin_wagner} we show that the gaplessness persists at finite $N$ and has a more fundamental origin: continuous Parisi symmetry breaking gives rise to a band of Goldstone modes. At leading order in $1/N$, these modes behave as non-interacting spin waves with level-dependent stiffness. Their long-wavelength fluctuations therefore diverge for $D\leq2$, producing a Mermin--Wagner instability of any saddle containing a continuously broken Parisi sector.

\subsection{1FRSB Solution} 
\label{sec:frsb_1frsb} 
A 1FRSB solution combines a continuously breaking FRSB sector with a single finite jump. It arises naturally when the outer replicon of a 1RSB saddle becomes unstable while the inner replicon remains stable. In this case, the outer part of the hierarchy develops continuous breaking, while the deeper one-step jump survives; see~\figref{fig:1FRSBvsFRSB}. The local FRSB equations determine the continuous trajectory itself, but not how it is matched to the remaining plateaus and discrete jump. We therefore consider the following general 1FRSB profile:
\begin{align}
    q(x)=
    \begin{cases}
        q_0,
        & 0\leq x<x_{\ell},
        \\
        q_{\mathrm c}(x),
        & x_{\ell}<x<x_c,
        \\
        q_m,
        & x_c<x<m,
        \\
        q_1,
        & m<x\leq1,
    \end{cases}
    \qquad
    q_1>q_m .
    \label{eq:one_frsb_general_form}
\end{align}
Here $q_0\equiv q(0)$ denotes the outermost overlap, while $q_1\equiv q_{\rm EA}$ is the Edwards--Anderson self-overlap. The continuous sector extends from $x_\ell$ to $x_c$ and terminates at $q(x_c)=q_m$. It may then be followed by a plateau at $q=q_m$ before the finite jump to $q_1$ at $x=m$. Thus $x_c\leq m$, with $x_c=m$ corresponding to a plateau of zero width.

We denote the gaps associated with $q_0$, $q_m$, and $q_1$ by $\Delta_0$, $\Delta_m$, and $\Delta_1$, respectively. On the unmagnetized branch, the stable outer boundary condition is $q_0=0$. For $A_l >0$, Eq.~\eqref{eq:frsb_x_of_q} implies $x(q)\propto q$ as $q\to0$, so the continuous sector begins at the outer edge, $x_\ell=x(0)=0$. On the magnetized branch, by contrast, $q_0>0$ and the replica-uniform mode is condensed. The lower plateau is therefore gapless, $\Delta_0=0 $, as discussed in Sec.~\ref{sec:frsb_magnetized}.

\vspace{5pt} \noindent \textbf{Matching and breakpoint conditions.} The local FRSB gap equation~\eqref{eq:frsb_gap_equation} applies only on smooth continuously breaking intervals and therefore cannot be continued through the finite jump at $x=m$. The two sides of the jump must instead be matched using the undifferentiated Parisi sum rule.

On the upper plateau, where $q=q_1$, evaluating the sum rule at $x=1$ gives
\begin{align}
    1-q_1 = \frac{T}{1-p}\,
    w_1(D,\Delta_1).
    \label{eq:frsb_1frsb_inner}
\end{align}
Immediately below the jump,
\begin{align}
    \widehat q(m^-) = 1-mq_m-(1-m)q_1,
\end{align}
and hence
\begin{align}
    1-mq_m-(1-m)q_1
    =
    \frac{T}{1-p}\,
    w_1(D,\Delta_m).
    \label{eq:frsb_1frsb_outer}
\end{align}
Subtracting these two conditions gives the equivalent jump relation
\begin{align}
    m(q_1-q_m) = 
    \frac{T}{1-p}
    \left[
        w_1(D,\Delta_m)
        -
        w_1(D,\Delta_1)
    \right].
    \label{eq:frsb_1frsb_jumprule}
\end{align}
For a given endpoint $q_m$, the local FRSB equations fix the end of the continuous sector through $\Delta_m=\Delta(q_m)$ and $x_c=x(q_m)$.
The gap $\Delta_1$ on the upper plateau is then related to $\Delta_m$ by the splitting across the finite jump, 
\begin{align}
    \Delta_1-\Delta_m
    &=
    \frac{m(\lambda_m-\lambda_1)}
         {\beta(1-p)}
    \nonumber\\
    &=
    \frac{2m \beta}{1-p}
    \left[
        A_P(q_1^3-q_m^3)
        +
        A_l(q_1-q_m)
    \right],
    \label{eq:frsb_1frsb_split}
\end{align}
where $\lambda_m\equiv\lambda(q_m)$ and $\lambda_1\equiv\lambda(q_1)$.

To determine the breakpoint $m$, we first solve the continuous sector and substitute the resulting profile back into the full free-energy functional. Let $F_{\rm red}$ denote the resulting reduced free-energy density. The remaining variables are then $(q_m,q_1,m)$.

If $m>x_c$, the plateau at $q=q_m$ has finite width and $m$ remains an independent variational parameter. In addition to the two matching conditions, one imposes
\begin{align}
    \left.
    \frac{\partial F_{\rm red}}{\partial m}
    \right|_{q_m,q_1}
    =
    0.
    \label{eq:frsb_1frsb_transversality}
\end{align}
The matching conditions, \eqnref{eq:frsb_1frsb_inner} and \eqnref{eq:frsb_1frsb_outer}, together with the breakpoint stationarity condition, \eqnref{eq:frsb_1frsb_transversality}, determine the interior 1FRSB saddle.

A boundary solution occurs when $m=x_c=x(q_m)$, so that the $q_m$ plateau has zero width. 
The two matching conditions determine a candidate $(q_m,q_1)$, but do not alone establish its stationarity.
In addition, one must require
\begin{align}
  & \frac{\partial F_{\rm red}}{\partial q_m}
  +x'(q_m)\frac{\partial F_{\rm red}}{\partial m}
  =0,
  \\
  & \left.\frac{\partial F_{\rm red}}{\partial m}
  \right|_{q_m,q_1}
  \hspace{-15pt} \leq0,
  \qquad m=x(q_m).
    \label{eq:frsb_1frsb_kkt}
\end{align}
The first condition tests motion along the boundary, including the induced variation of $m$ when $q_m$ changes. The second excludes a first-order increase in the free energy when the intermediate plateau opens at fixed $q_m,q_1$. Since $\partial F_{\rm red}/\partial q_m$ vanishes at $m=x_c$, the first condition further implies $\partial F_{\rm red}/\partial m=0$ whenever $x'(q_m)>0$. These are necessary first-order conditions, not a complete stability criterion.

On a magnetized branch, these jump and breakpoint equations must be supplemented by the gapless outer boundary $\Delta_0=0$ and the isolated-mode condensate condition of~\secref{sec:frsb_magnetized}. A finite magnetized 1FRSB solution again requires $D>4$ in the presence of disorder.

\vspace{5pt} \noindent \textbf{Stability.}
The continuously breaking sector is replicon-marginal by construction. Thus, in $D\leq 2$, any 1FRSB phase containing such a sector therefore inherits the same Mermin--Wagner infrared instability as the FRSB phase.

A finite plateau at $q=q_m$ has the same overlap and gap as its endpoint and therefore inherits this marginality.
The upper plateau at $q=q_1$, however, is separated from the continuous sector by a finite jump and need not be marginal. Its stability must therefore be checked independently:
\begin{align}
    \lambda_R^{(1)}= 1-
    \frac{2 \bigl[3A_P q_1^2 + A_l\bigr]}{(1-p)^2}
    w_2(D,\Delta_1)
    \geq0.
    \label{eq:frsb_1frsb_upper_replicon}
\end{align}


\subsection{Goldstone Modes and Mermin--Wagner Instability}\
\label{sec:mermin_wagner}

We now show that the gapless replicon modes associated with a continuous Parisi profile have a symmetry origin that persists beyond the large-$N$ limit. Although the replica-permutation group $S_n$ is discrete, a continuously broken Parisi hierarchy admits permutations whose action on the ordered state becomes arbitrarily small. These infinitesimal deformations generate Goldstone modes of the continuously broken replica symmetry~\cite{Bray_1979NonIsing, DominicisTemesvariKondor1998,TemesváriKondorDominicis2000,Brezin_2002, DeDominicisGiardina2006}.  
At leading order in $1/N$, these modes behave as free spin waves with level-dependent stiffness, producing a Mermin--Wagner-type infrared instability in $D\leq2$. 

We emphasize the absence of $\O(N)$ symmetry in our setup: related Mermin--Wagner-type arguments for short-range vector spin glasses show that conventional XY spin-glass order is absent in $D\leq2$~\cite{Schwartz_1991}. Here the relevant soft modes have a different origin: they are Goldstone modes associated with continuous breaking of replica permutation symmetry rather than physical spin-rotation symmetry.

 \vspace{5pt} \noindent \textbf{Symmetries of the effective action.}
Consider the disorder-averaged, replicated partition function in \eqnref{eq:partition_function_Ieff}, which is exact at finite $N$. To probe fluctuations of the overlap field, we introduce a source $h^{\alpha\beta}$ conjugate to $Q^{\alpha\beta}$ and define
\begin{equation}
    e^{NW_n[h]}=\int dQ d\Lambda 
    e^{-NI_n[Q,\Lambda]+N\sum_{v}h^{\alpha\beta}(v)Q^{\alpha\beta}(v)} .
\end{equation}
where repeated replica indices are summed. The corresponding Legendre transform is
$\Gamma_n[\bar Q]=\sum_{v}h^{\alpha\beta}(v)\bar Q^{\alpha\beta}(v)-W_n[h]$, where $\bar Q=\partial W_n/\partial h$.  $\Gamma_n[\bar Q]$ is the exact effective action for the overlap field. Its first derivative satisfies
\begin{equation}
    \frac{\partial\Gamma_n}
    {\partial\bar Q^{\alpha\beta}(v)}
    =
    h^{\alpha\beta}(v),
\end{equation}
so at zero source the physical overlap configuration $\bar Q_*$ is a stationary point of $\Gamma_n$. Its second derivative $\Gamma_n^{(2)}$ defines the Hessian, or equivalently the exact inverse overlap propagator. In the large-$N$ limit, $\Gamma_n$ reduces to the saddle-point effective action, and $\Gamma_n^{(2)}$ reproduces the fluctuation Hessian studied above.

The crucial point is that the replicated theory is exactly invariant under permutations of the
replica labels. Therefore, for every permutation $P\in S_n$,
\begin{equation}
\begin{split}
    \Gamma_n[P\bar Q]
    &=\Gamma_n[\bar Q],\\
    \Gamma_n^{(1)}[P\bar Q]
    &=P \Gamma_n^{(1)}[\bar Q],\\
    \Gamma_n^{(2)}[P\bar Q]
    &=P \Gamma_n^{(2)}[\bar Q] P^{-1}.
    \label{eq:Sn_covariance}
\end{split}
\end{equation}
In particular, if $\bar Q_*$ is a stationary point, so is $P\bar Q_*$. For a finite-step Parisi profile, these symmetry-related stationary points are separated by finite distances in overlap space. The situation changes when the Parisi hierarchy becomes continuous. In the limit of infinitely many breaking steps, there exist replica permutations whose action on the ordered state becomes arbitrarily small~\cite{DominicisTemesvariKondor1998}.

\begin{figure}[!t]
    \centering
    \includegraphics[width=\linewidth]{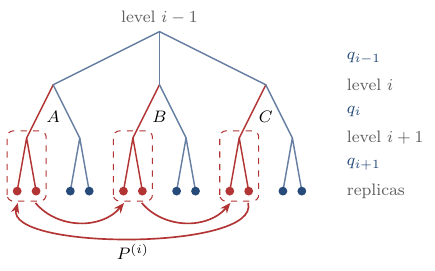}
    \caption{\textbf{Small permutation of the Parisi tree.}  
    $P^{(i)}$ moves a level-$(i+1)$ sub-block from one level-$i$ block into a neighboring level-$i$ block within their common level-$(i-1)$ block. Only overlap matrix elements involving the moved replicas change. These elements shift between $q_i$ and $q_{i-1}$, so the deformation is proportional to $q_i-q_{i-1}$. On a continuously broken sector this difference vanishes as the Parisi hierarchy is refined.}
\label{fig:small_permutation}
\end{figure}
 
\vspace{5pt} \noindent \textbf{Small permutations of the Parisi tree.}
Consider a $k$-step Parisi matrix with breakpoints $0=m_0<m_1<\cdots<m_k<1$ and overlap values $q_0<q_1<\cdots<q_k$. Let $P^{(i)}$ exchange a level-$(i+1)$ sub-block between two neighboring level-$i$ blocks belonging to the same level-$(i-1)$ block, see \figref{fig:small_permutation}. Only matrix elements involving the moved replicas are affected.
Their overlap with replicas in the original block changes from $q_i$ to $q_{i-1}$, while their overlap with replicas in the destination block changes from $q_{i-1}$ to $q_i$.
The resulting change in the Parisi matrix therefore has the form
\begin{equation}
    P^{(i)}\bar Q_*-\bar Q_*
    =(q_i-q_{i-1})\,\Xi_i ,
    \label{eq:small_perm}
\end{equation}
where $\Xi_i$ is the symmetric matrix that records which replica pairs change hierarchical level under the permutation: $(\Xi_i)^{\alpha\beta}=+1$ for pairs whose overlap changes from $q_{i-1}$ to $q_i$, $-1$ for pairs whose overlap changes from $q_i$ to $q_{i-1}$, and $0$ otherwise.

Along a continuously broken sector, let $x=m_i$ and
$\delta x=m_{i}-m_{i-1}$. Then
\begin{equation}
    q_i-q_{i-1}
    =q'(x)\,\delta x \longrightarrow 0,
\end{equation}
Thus, although $P^{(i)}$ is a discrete permutation, its action on the ordered state becomes infinitesimal as the hierarchy is refined. These are the ``small permutations'' of Ref.~\cite{DominicisTemesvariKondor1998}. This construction applies only where $q(x)$ varies continuously. Across a finite jump, as in 1RSB or at the jump of a 1FRSB profile, $q_i-q_{i-1}$ remains finite so the corresponding permutation does not generate an infinitesimal deformation. On a plateau, $q_i=q_{i-1}$, so the permutation leaves the state unchanged and is unbroken.

\vspace{5pt} \noindent \textbf{Goldstone modes.}
Before deriving the zero mode, it is useful to clarify the notation for the replicon Hessian. 
Since the overlap field carries two replica indices, $Q^{\alpha\beta}$, its Hessian carries four, $\Gamma^{(2)}_{\alpha\beta,\gamma\delta}$. In an ultrametric Parisi state, the matrix elements of the Hessian depend on the relative positions of these four replicas in the hierarchy.

In the replicon sector, the pairs $(\alpha,\beta)$ and $(\gamma,\delta)$ both lie at overlap level $x$, while their relative geometry is characterized by two cross-overlap levels $y$ and $z$. We write the corresponding component as $\Gamma^{(2)}_{n,R}(k;x,y,z)$~\cite{TemesváriKondorDominicis2000}. 

A small permutation at level $x$ probes the limit in which the cross-overlap levels approach $x$ from above. We denote this component by
\begin{equation}
    \Gamma^{(2)}_{R}(k;x)
    \equiv 
    \lim_{y,z\to x^{+}}
    \Gamma^{(2)}_{n,R}(k;x,y,z).
\end{equation}

We now show that this mode is massless at zero momentum. Let $\delta_P Q \equiv P\bar Q_*-\bar Q_*$. On a continuously broken interval, the small-permutation construction gives $\delta_P Q = q'(x)\,\delta x\,\Xi_x +O(\delta x^2)$. Replica-permutation invariance implies that both $\bar Q_*$ and $P\bar Q_*$ are stationary points of the exact effective action,
\begin{equation}
    \Gamma_n^{(1)}[\bar Q_*] = \Gamma_n^{(1)}[P\bar Q_*] =0.
\end{equation}
Expanding the second condition about $\bar Q_*$ gives
\begin{equation}
    \Gamma_n^{(2)}\,\delta_PQ = -\frac12\Gamma_n^{(3)}
    \,\delta_PQ\,\delta_PQ -\cdots .
    \label{eq:WT_expansion}
\end{equation}
Since $\delta_P Q=O(\delta x)$, the left-hand side is linear in $\delta x$, whereas the right-hand side begins at order $\delta x^2$. Dividing by $\delta x$ and taking $\delta x\to0$ therefore yields
\begin{equation}
    \sum_{\gamma<\delta}
    \Gamma^{(2)}_{n;\alpha\beta,\gamma\delta}(k=0)
    (\Xi_x)^{\gamma\delta} = 0,
    \qquad
    q'(x)\neq0.
\label{eq:goldstoneband}
\end{equation}
The mode has zero momentum because the same replica permutation is applied at every spatial point. Thus $\Xi_x$ is a zero eigenvector of the zero-momentum overlap Hessian. Projecting this relation onto the replicon sector gives the corresponding Ward--Takahashi identity,
\begin{equation}
    \Gamma^{(2)}_{R}(0;x)=0
\end{equation}
throughout every continuously broken interval. 

Each level of the continuous Parisi hierarchy therefore carries a massless replicon mode, producing a continuous band of Goldstone modes.
This conclusion is more general than the large-$N$ saddle-point calculation. The Ward identity follows from the exact replica-permutation symmetry of the effective action. Consequently, within a symmetry-preserving $1/N$ expansion, the mode therefore remains massless order by order, provided the stationary Parisi profile is corrected consistently at the same order.

\vspace{5pt} \noindent \textbf{Large-$N$ spin waves and Mermin--Wagner instability.}
The Ward identity fixes the zero-momentum mass of the diagonal replicon to vanish, but symmetry alone does not determine its spatial dispersion. At leading order in the $1/N$ expansion, the latter follows from the level-resolved disorder vertex and bare replicon propagator. Expanding at small momentum, we find
\begin{align}
    \Gamma_x(k) &= \Gamma_x(0)-\gamma_x k^2+O(k^4),
    \\
    \Pi_x(k) &= \Pi_x(0)-\pi_x k^2+O(k^4),
\end{align}
where 
\begin{align}
    \gamma_x &= \frac{\beta^2}{D}\left[4A_Pq(x)^2+A_l\right],  \\
    \pi_{x} &=\frac{
    [2D+\Delta(x)]w_3(D,\Delta(x))
    -w_2(D,\Delta(x))}{6D\beta^2(1-p)^2}.
\end{align}
On a continuously broken interval, the marginality gives $\Gamma_x(0)\Pi_x(0)=1$, and hence
\begin{align}
    \lambda_R(k;x) 
    &=
    \kappa_R(x)k^2+O(k^4),
    \\ 
    \text{where }\quad
    \kappa_R(x)
    &=
    \gamma_x\Pi_x(0)
    +\Gamma_x(0)\pi_x
    >0.
    \label{eq:frsb_replicon_dispersion}
\end{align}
Since the inverse replicon propagator is
\begin{align}
    \Gamma^{(2)}_{R}(k;x) \equiv \frac{1}{G_R(k;x)} =\frac{\lambda_R(k;x)}{\Pi_x(k)}
\end{align}
its long-wavelength form is 
\begin{align}
    \Gamma^{(2)}_{R}(k;x)\approx \rho_R(x)k^2+O(k^4),
\end{align}
where $\rho_R(x)=\kappa_R(x)/\Pi_x(0)>0$. Thus, at leading order in $1/N$, each level of a continuously broken sector supports a massless replicon with a quadratic inverse propagator and finite positive stiffness. Collectively, these modes form a continuum of free spin waves with level-dependent stiffness $\rho_R(x)$.

The corresponding connected overlap fluctuations behave as
\begin{align}
    \big\langle |\delta Q(k;x)|^2\big\rangle_{\rm c}
    &=\frac{1}{\Gamma^{(2)}_{R}(k;x)}
    \simeq
    \frac{1}{\rho_R(x)k^2},
    \label{eq:mw_fluct}
\end{align}
Their real-space variance therefore contains the infrared contribution
\begin{align}
    \big\langle \delta Q(v;x)^2\big\rangle_{\rm c}
    &\sim
    \frac{1}{\rho_R(x)} \int\frac{d^Dk}{(2\pi)^D}\,\frac{1} {k^2}
\end{align}
The integral diverges logarithmically in $D=2$ and as a power for $D<2$. Thus, for $D\leq 2$, the leading $1/N$ fluctuations about the continuous Parisi saddle are infrared divergent at every level of the broken hierarchy. This is the replica analogue of the Mermin--Wagner mechanism: the Ward identity enforces a massless band of Goldstone modes, while the short-range spatial stiffness produces the $k^{-2}$ propagator responsible for the infrared divergence. Higher-order corrections may renormalize the stiffness and other long-wavelength parameters, but the Ward identity prevents the Goldstone modes from acquiring a mass perturbatively. 
Within the present large-$N$ framework, $D=2$ therefore plays the role of the lower critical dimension for continuous replica symmetry breaking. Both FRSB and the continuously broken sector of a 1FRSB phase are infrared unstable for $D\leq2$. By contrast, a gapped 1RSB phase has no associated Goldstone band and is not subject to this instability.

\subsection{Allowed Parisi Structures}
\label{sec:no-other-saddles}

More complicated Parisi profiles could, in principle, contain several continuously breaking sectors or multiple finite jumps. In \appref{app:no_other_phases}, we show that within the translation-invariant, large-$N$ theory, the saddle-point equations together with replicon stability restrict the allowed structures to
\begin{align}
    \mathrm{RS},\qquad
    \mathrm{1RSB},\qquad
    \mathrm{FRSB},\qquad
    \mathrm{1FRSB},
\end{align}
together with their magnetized counterparts where allowed. This classification holds in every integer spatial dimension and does not rely on a numerical search over higher-step RSB ansatzes.

The proof has two main ingredients. First, continuous FRSB sectors are highly constrained. On any continuously breaking interval, the local FRSB equations uniquely determine
\begin{align}
    \Delta=\Delta(q),\qquad x=x(q),
\end{align}
so all continuous sectors must follow the same relation between overlap, gap, and Parisi coordinate. In particular, two continuous sectors cannot be separated by a finite plateau. A plateau at fixed overlap $q=q_*$ would have to leave and rejoin the continuous solution at the same value $x(q_*)$, forcing the plateau to have zero width.

To further constrain the continuous solution, it is useful to introduce the integrated overlap
\begin{align}
    \chi(q)=\int_q^1 du\,x(u),
\end{align}
which is concave because $x(q)$ is nondecreasing. Replicon stability bounds $\chi(q)$ by an envelope $h(q)$,
\begin{align}
    \chi(q)\leq h(q),
\end{align}
with equality on every continuously breaking interval. The key result of \appref{app:no_other_phases} is that $h(q)$ can change curvature at most once, from concave to convex. Once $h(q)$ enters its convex regime, a concave $\chi(q)$ lying below it cannot leave and later rejoin the envelope without violating this bound. This strongly restricts the possible arrangement of continuously breaking sectors.

Second, finite jumps are also strongly constrained. The jump-matching condition, together with replicon stability and the curvature property above, implies that any finite jump must be terminal. More specifically, after a jump the profile remains strictly below the envelope $h(q)$ at all larger overlaps, so it cannot subsequently enter another continuous sector, which would require $\chi(q)=h(q)$, or satisfy the matching condition for a second jump. Thus, once a finite jump occurs, no further hierarchical breaking is possible.

It follows that an admissible, replicon-stable profile can contain at most one continuously breaking sector, followed by at most one finite jump. The only possible structures are therefore RS, 1RSB, FRSB, and 1FRSB, together with their magnetized counterparts where allowed.

The only lattice-specific ingredient in the proof is a spectral inequality involving the Watson integrals $w_r(D,\Delta)$. We establish this inequality for every integer dimension using analytic bounds supplemented by a finite exact-arithmetic verification, as detailed in \appref{app:no_other_phases}. The classification is therefore computer-assisted, but does not rely on a numerical search over higher-step RSB ansatzes. 

\subsection{Full Phase Diagram}
\label{sec:full-rsb-phase-diagram}
We now compare the RS, 1RSB, FRSB, and 1FRSB saddles, including their magnetized counterparts where allowed. At each point in the $(p,T)$ plane, we retain the admissible, replicon-stable solutions found numerically and select the one with the largest on-shell replica free energy. The resulting phase diagrams are shown in~\figref{fig:full_phase_diagram}. For a continuous sector, admissibility requires $dx/dq>0$ throughout the occupied overlap interval; see~\appref{app:frsb_monotonicity}. A 1FRSB solution must also satisfy the jump-matching and breakpoint conditions, together with replicon stability of the upper plateau. 


For weak plaquette disorder, $D=5$ and $J_P=0.5$ [Fig.~\ref{fig:full_phase_diagram}(a)], the regions left unresolved by the RS/1RSB analysis are replaced predominantly by FRSB and 1FRSB order.
Their magnetized counterparts occupy the lower-frustration region, while the RS ferromagnet survives at small $p$. 
A small low-temperature FM+1RSB pocket also remains.
Thus continuous RSB dominates much of the glassy regime in this panel, although one-step order can still coexist with ferromagnetism.

At intermediate plaquette disorder, $D=5$ and $J_P=1.4$
[Fig.~\ref{fig:full_phase_diagram}(b)], one-step order expands substantially. 
FM+1RSB occupies a broad region at lower $p$, while the unmagnetized 1RSB glass extends to larger $p$.
Narrow FM+FRSB and FM+1FRSB regions survive near the upper boundary of the magnetized glass.
The magnetized and unmagnetized 1RSB phases join as the condensate vanishes, while at larger frustration the paramagnet transitions directly into the 1RSB glass.

For strong plaquette disorder, $D=5$ and $J_P=3$
[Fig.~\ref{fig:full_phase_diagram}(c)], one-step order dominates the ordered region. The unmagnetized 1RSB glass occupies most of the phase diagram, with an FM+1RSB region and a small RS ferromagnetic phase
remaining at low $p$. No thermodynamically selected FRSB or 1FRSB region is resolved in this panel.

Finally, for $D=3$ and $J_P=1$ [\figref{fig:full_phase_diagram}(d)], ferromagnetic order is absent, so all glassy phases are unmagnetized. Upon cooling, the paramagnet first enters an FRSB phase, followed at lower temperature by 1FRSB. At sufficiently large frustration, the low-temperature phase instead becomes 1RSB, producing the 1RSB dome seen in the figure.

Taken together, the $D=5$ panels show a clear evolution with increasing plaquette disorder: weak $J_P$ favors extended regions of continuous RSB, while stronger $J_P$ progressively favors one-step order. Within the large-$N$ theory, these results establish that a strictly short-range, finite-dimensional spin glass can support thermodynamically
stable continuously broken Parisi phases.

\subsection{Real-Space Ultrametricity}

\label{sec:frsb-correlation-hierarchy}
Consider a continuously breaking interval described by $q(x)$ and $\Delta(x)$. Along this interval, the spin kernel $\mathcal M(k)$ has a continuum of hierarchical eigenchannels, with each level $x$ characterized by the dispersion $\epsilon(k,x) = \beta(1-p)[ \Delta(x)+2D-K(k)]$. We now show that this hierarchy of eigenchannels directly gives rise to a corresponding hierarchy of real-space correlation lengths.

We define the diagonal and level-resolved off-diagonal correlators by
\begin{align}
    G^{\rm d}(r)
    &\equiv
    G^{\alpha\alpha}(r) \\
    G^x(r)
    &\equiv
    \left.
    G^{\alpha\beta}(r)
    \right|_{Q_*^{\alpha\beta}=q(x)}.
\end{align}
Thus $G^{x}(r)$ is the correlator between two replicas whose mutual overlap is $q(x)$. At $x=1$, the replicas occupy the same pure state, with $q(1)=q_{\rm EA}$, so $G^{1}(r)$ is the corresponding same-state frozen correlator. 

The replica-resolved spin correlator is given by $G^{\alpha\beta}(r) = (\mathcal M^{-1})^{\alpha\beta}(r)$. Its Parisi transform is 
\begin{align} 
    \widehat G(r,x) \equiv G^{\rm d}(r) -xG^x(r) -\int_x^1dy\,G^y(r).
\end{align}
Since the hierarchical eigenvalue of $\mathcal M^{-1}(k)$ at level $x$ is $1/\epsilon(k,x)$, we obtain
\begin{align} 
    \widehat G(r,x) = \frac{1}{V} \sum_k \frac{e^{ik\cdot r}}{\epsilon(k,x)}. \label{eq:frsb_Ghat} 
\end{align}
This is the spatially resolved counterpart of the level-resolved saddle-point sum rule. In particular, at $r=0$ it reduces to $\widehat G(0,x) = \widehat q(x)$.

Evaluating Eq.~\eqref{eq:frsb_Ghat} at the terminal level $x=1$ immediately gives
\begin{align} 
    G^{\rm d}(r)-G^{1}(r) = \frac{1}{V} \sum_k \frac{e^{ik\cdot r}}{\epsilon(k,1)}. \label{eq:frsb_thermal_projection} 
\end{align}
The frozen contribution cancels in this difference, leaving the connected thermal correlator averaged over pure states and quenched disorder. Its long-distance decay is therefore governed by the $x=1$ eigenchannel. For $\Delta(1)>0$, this projection asymptotically behaves as
\begin{align}
    G^{\rm d}(r)-G^{1}(r)
    \sim
    \frac{e^{-r/\xi(1)}}{r^{(D-1)/2}},
    \qquad
    r\gg1,
    \label{eq:frsb_level_asymptotic}
\end{align}
with $\xi(1)^{-1}=\operatorname{arcosh}\bigl(1+\Delta(1)/2\bigr)$ for $r$ along a lattice axis.

To characterize the spatial differences between distinct pure states, we again introduce the difference field $\delta\mu_{ab}(v) \equiv \mu_a(v)-\mu_b(v)$ for a pair of pure states $a$ and $b$, and consider its two-point correlator, conditioned on their mutual overlap,
\begin{align}
    D^x(r)
    \equiv
    \frac{1}{2}
    \overline{
    \left\langle
    \frac{1}{V}\sum_v
    \delta\mu_{ab}(v)^T
    \delta\mu_{ab}(v+r)
    \right\rangle_{q_{ab}=q(x)}
    }.
    \label{eq:frsb_difference_field}
\end{align}
As shown in~\appref{app:pure_states}, this quantity is equivalently 
\begin{align}
    D^x(r) = G^{1}(r)-G^x(r).
    \label{eq:frsb_difference_replica}
\end{align}
At zero separation, $D^x(0) = q(1)-q(x) \equiv q_{\rm EA} - q(x)$. Thus, the usual overlap deficit $q_{\rm EA} - q(x)$ measures the mean-square difference between the frozen magnetization patterns of two pure states, while $D^{x}(r)$ resolves how this difference is organized in real space. In this way, each level $x$ of the hierarchy is characterized both by a thermodynamic overlap $q(x)$ and by a corresponding spatial scale over which differences between the associated frozen configurations remain correlated.

Differentiating Eq.~\eqref{eq:frsb_Ghat} with respect to $x$, and using $\partial_x \widehat G(r,x) = -x\,\partial_x G^x(r)$, gives
\begin{align}
    D^x(r)
    =
    -\int_x^1
    \frac{dy}{y}\,
    \partial_y\Bigg[\frac{1}{V} \sum_k \frac{e^{ik\cdot r}}{\epsilon(k,y)}\Bigg].
    \label{eq:frsb_difference_hierarchy}
\end{align}
Thus, the frozen spatial difference between two pure states that branch at level $x$ receives contributions from all hierarchical spin eigenchannels between $x$ and the terminal level of the Parisi tree.

Along a continuously breaking interval, the saddle-point equations imply
\begin{align}
    \Delta'(x)\geq0 \quad
    \implies \quad
    \xi'(x)\leq0.
    \label{eq:frsb_xi_monotonicity}
\end{align}
Thus, the outer levels of the hierarchy correspond to softer spin eigenchannels and longer correlation lengths, while deeper levels are progressively more strongly gapped and shorter ranged.

The long-distance behavior of $D^x(r)$ follows directly from Eq.~\eqref{eq:frsb_difference_hierarchy}. Integrating by parts gives
\begin{align}
    D^x(r)
    &=
    \frac{1}{xV}
    \sum_k
    \frac{e^{ik\cdot r}}{\epsilon(k,x)}
    -
    \frac{1}{V}
    \sum_k
    \frac{e^{ik\cdot r}}{\epsilon(k,1)}
    \nonumber\\
    &\quad
    -
    \int_x^1\frac{dy}{y^2}
    \left[
        \frac{1}{V}
        \sum_k
        \frac{e^{ik\cdot r}}{\epsilon(k,y)}
    \right].
    \label{eq:frsb_difference_parts}
\end{align}
Because $\xi(y)$ decreases with increasing $y$, the level-$x$ eigenchannel is the longest-ranged contribution to Eq.~\eqref{eq:frsb_difference_parts}. It therefore determines the asymptotic decay of the difference-field correlator, 
\begin{align}
    D^x(r) \sim \frac{e^{-r/\xi(x)}}{r^{(D-1)/2}}, \qquad r\gg1, \label{eq:frsb_difference_asymptotic} 
\end{align} 
up to an $x$-dependent prefactor~\footnote{
At the outer endpoint $x=0$, the algebraic prefactor is modified. For a gapped unmagnetized continuous branch with $A_l>0$, $q(x)\propto x$ and $\Delta(x)-\Delta(0)\propto x^2$, so Eq.~\eqref{eq:frsb_difference_asymptotic} gives $D^0(r)\sim e^{-r/\xi(0)}/r^{(D-2)/2}$. The exponential correlation length $\xi(0)$ is unchanged.}.
Eqs.~\eqref{eq:frsb_thermal_projection} and \eqref{eq:frsb_difference_asymptotic} thus give a direct real-space interpretation of the Parisi hierarchy. 

The continuous saddle equations determine a temperature-independent relation $\Delta=\Delta(q)$ and hence a corresponding relation $\xi=\xi(q)$. Each overlap is thus associated with a characteristic real-space scale. Temperature only changes where a given overlap appears along the Parisi hierarchy, through $x(q,T)=Tf(q)$, but not the correlation length associated with that overlap. The terminal scale $\xi(1)$ governs connected thermal fluctuations within a pure state, while $\xi(x)$ characterizes the spatial correlations of the frozen differences between pure states that branch at level $x$. Since $\xi(x)$ decreases along a continuously breaking interval, states that separate near the outer levels of the Parisi tree differ on longer spatial scales, while states sharing a deeper common branch differ only on shorter scales.

This hierarchy provides a direct real-space counterpart of ultrametricity. For any three pure states $a$, $b$, and $c$, their thermodynamic overlaps satisfy
\begin{align} 
    q_{ab} \geq \min\!\left(q_{ac},q_{bc}\right). \label{eq:frsb_overlap_ultrametricity} 
\end{align}
Because $\xi(q)$ is nonincreasing with $q$, the corresponding correlation lengths $\xi_{ab} = \xi(q_{ab})$ obey
\begin{align} 
    \xi_{ab} \leq \max\!\left[ \xi_{ac}, \xi_{bc} \right]. \label{eq:frsb_length_ultrametricity} 
\end{align}
Equivalently, the two largest of the three pairwise correlation lengths are equal. Thus, the ultrametric organization of pure states is mirrored by an ultrametric organization of real-space scales.

When all hierarchical channels are gapped, these correlation lengths remain finite. If the outer channel becomes gapless, as on a magnetized branch, its correlation length diverges and the corresponding difference-field correlator becomes algebraically long ranged, while the deeper gapped channels remain exponentially short ranged.

\section{Conclusion and outlook}

In this work, we have constructed a local lattice model of vector spin glasses whose large-$N$ limit supports both magnetized and unmagnetized realizations of one-step, full, and one-full replica-symmetry breaking. Together with the replica-symmetric paramagnetic and ferromagnetic phases, these exhaust the Parisi structures allowed by the saddle-point equations and the replicon-stability conditions in every integer spatial dimension.

The proposed lattice model exposes two distinct infrared mechanisms constraining the structure of phase diagrams, which extend beyond the large-$N$ limit. First, continuous Parisi breaking gives rise to replica Goldstone modes whose masslessness is protected by a replica Ward identity. At leading order in $1/N$, these modes behave as free spin waves with level-dependent stiffness. Their infrared fluctuations therefore diverge in $D\leq2$, producing a Mermin--Wagner-type instability of the continuously broken phase. Second, quenched random anisotropy destabilizes ferromagnetic order in $D\leq4$ through an Imry-Ma mechanism, including on replica-symmetry-broken branches.

Our central result links the thermodynamic overlap between pure states to the spatial correlation length of their frozen magnetization differences.
For two pure states with mutual overlap $q$, the deficit $q_{\mathrm{EA}}-q$ measures the mean-square separation of their frozen spin configurations, while the difference-field correlator, conditioned on $q$, describes how these differences are organized in real space. Its characteristic decay length defines a correlation length $\xi(q)$, which is independent of temperature, consistent with Parisi--Toulouse scaling, and decreases monotonically with overlap.
Pure states that branch near the root of the Parisi tree therefore differ through long-wavelength frozen patterns, whereas states that share a deeper common branch differ only over progressively shorter spatial scales. The ultrametric hierarchy of overlaps is thus inherited by a hierarchy of real-space correlation lengths, giving replica-symmetry breaking a direct spatial interpretation.

Our analysis opens several directions for refining and extending these results. We highlight four in particular:
\begin{itemize}[leftmargin=13pt]
    \item {\bf Finite-$N$ corrections.} The framework above provides a starting point for a systematic large-$N$ program to study the stability of hierarchical glassy order in a strictly local spin-glass model. A natural next step is to compute higher-order $1/N$ corrections to the saddle-point equations and Hessian eigenvalues, thereby refining both the phase diagram and the hierarchy of spatial correlation lengths. It would also be valuable to determine whether finite-$N$ calculations can access the free-energy cost of imposing different overlaps in spatially separated regions, enabling a direct comparison with stiffness-based predictions of the droplet picture.
    Finally, it is also essential to supplement these analytical studies with finite-$N$ Monte Carlo simulations to directly test the predicted phase structure away from the large-$N$ limit.

    \item {\bf Spatially resolved dynamics}. The static phase diagram also motivates a study of dynamical arrest and hierarchical aging. For the 1RSB and 1FRSB ansatzes, candidate dynamical transition lines can be obtained by replacing equilibrium stationarity with respect to the breakpoint $m$ by the upper-plateau marginality condition $\lambda_R^{(1)}(0)=0$, following the approach used for spherical glasses~\cite{Monasson1995,CrisantiSommers1994}. Deriving the corresponding local Langevin dynamics would then determine whether these marginal lines indeed mark the onset of dynamical arrest and identify the associated relaxation processes.
    Furthermore, the aging dynamics of the SK model exhibits an infinite hierarchy of asymptotically separated timescales~\cite{CugliandoloKurchan1994}.
    Our results raise the interesting question of how this temporal hierarchy relates to the hierarchy of spatial correlation lengths $\xi(q)$. Deriving the large-$N$ equations for local Langevin dynamics would allow us to test for dynamical ultrametricity and determine whether distinct aging time scales can be associated with different equilibrium length scales.
    Such a correspondence would connect the spatial organization of pure-state differences to the hierarchy of relaxation times, a relation not implied by the static Parisi structure alone.
    

    \item {\bf Ultrametricity as an RG scale.} 
    Perhaps most strikingly, our results suggest that the Parisi hierarchy itself may provide an organizing principle for spatial renormalization, with the Parisi level $x$ acting as an emergent scale coordinate. The monotonic relation $x\mapsto\xi[q(x)]$ associates deeper levels of the hierarchy with progressively shorter-ranged differences between pure states. This raises a concrete question: as the observation scale is increased, can spatial coarse-graining be described by a progressive loss of resolution among the deeper branches of the Parisi tree, effectively inducing a flow of the ultrametric hierarchy? Related ideas have appeared in earlier work by Oppermann and Sherrington, who formulated an RG-like scaling directly in hierarchical space, with the number of breaking steps playing the role of a decimation parameter~\cite{OppermannSherrington2005}. Furthermore, similar results have been discovered in forthcoming work~\cite{Vijay2026}, where replica-symmetry breaking resolves a dimensionally transmuted RG scale into an ordered hierarchy of constituent gaps, with successive gap ratios fixed by the Parisi level. Together, these results suggest a deeper connection between ultrametric organization and the emergence of physical scales. 
    A natural starting point in this direction would be to define overlaps from block-averaged frozen magnetizations and follow their distribution as the block size is increased. This would test whether spatial coarse-graining can be represented directly as a flow of the Parisi tree, so that the hierarchy describes not only how pure states are organized, but also how that organization changes with spatial resolution.

    \item {\bf Quantum Spin Glasses.} 
    A natural extension is to ask whether the hierarchical organization of real-space length scales found here survives in short-range quantum spin glasses~\cite{YeSachdevRead1993,ReadSachdevYe1995,GeorgesParcolletSachdev2000,GeorgesParcolletSachdev2001,CugliandoloGrempelSilva2001}. A useful starting point is an $O(N)$ quantum-rotor generalization of the present model with random link and plaquette interactions. In an imaginary-time replica formulation, the overlap field becomes bilocal in time, $Q^{\alpha\beta}(\mathbf{x};\tau,\tau')$, allowing spatial and temporal correlations to be treated within the same framework. One can then ask whether the overlap-dependent hierarchy of spatial length scales, $\xi(q)$, survives quantum fluctuations and whether it is accompanied by a corresponding hierarchy of temporal scales, namely \emph{ultrametric temporal structure}. Coupling this short-range quantum spin-glass sector to itinerant fermions would also provide a setting in which to ask whether different levels of the Parisi hierarchy produce distinct channels of singular fermion scattering and, collectively, non-Fermi-liquid metallic behavior.

\end{itemize}

\section*{Acknowledgments}
We thank Matthew O'Brien, Eduardo Fradkin, Paul Goldbart, Subir Sachdev, Federico Ricci-Tersenghi, and Daniel Stein for discussions and feedback. AV and JYL are supported by the IIDAI-IBM Accelerator Institute and faculty startup grant at the University of Illinois, Urbana-Champaign. JK acknowledges support from the National Science Foundation under Grant No. DMR-2225920 (Eduardo Fradkin). JYL acknowledges the support from KIAS through the Quantum Universe Center scholar program. 

\bibliography{bibfile}

@article{Brezin_2002,
   title={Twist free energy in a spin glass},
   volume={30},
   ISSN={1434-6036},
   url={http://dx.doi.org/10.1140/epjb/e2002-00360-1},
   DOI={10.1140/epjb/e2002-00360-1},
   number={1},
   journal={The European Physical Journal B},
   publisher={Springer Science and Business Media LLC},
   author={Brézin, E. and De Dominicis, C.},
   year={2002},
   month=Nov, pages={71–76} }

@article{Bray_1979NonIsing,
doi = {10.1088/0022-3719/12/7/023},
url = {https://doi.org/10.1088/0022-3719/12/7/023},
year = {1979},
month = {apr},
publisher = {},
volume = {12},
number = {7},
pages = {1349},
author = {A J Bray and M A Moore},
title = {Replica symmetry and massless modes in spin glasses. II. Non-Ising spins},
journal = {Journal of Physics C: Solid State Physics}
}

@article{Schwartz_1991,
doi = {10.1209/0295-5075/15/2/017},
url = {https://doi.org/10.1209/0295-5075/15/2/017},
year = {1991},
month = {may},
publisher = {},
volume = {15},
number = {2},
pages = {209},
author = {M. Schwartz and A. P. Young},
title = {On the Lower Critical Dimension of XY Spin Glasses},
journal = {Europhysics Letters}
}

@article{EdwardsAnderson1975,
  author = {Edwards, S. F. and Anderson, P. W.},
  title = {Theory of Spin Glasses},
  journal = {Journal of Physics F: Metal Physics},
  volume = {5},
  number = {5},
  pages = {965--974},
  year = {1975},
  doi = {10.1088/0305-4608/5/5/017}
}

@article{SherringtonKirkpatrick1975,
  author = {Sherrington, David and Kirkpatrick, Scott},
  title = {Solvable Model of a Spin-Glass},
  journal = {Physical Review Letters},
  volume = {35},
  number = {26},
  pages = {1792--1796},
  year = {1975},
  doi = {10.1103/PhysRevLett.35.1792}
}

@article{SherringtonKirkpatrick1978,
  title = {Infinite-ranged models of spin-glasses},
  author = {Kirkpatrick, Scott and Sherrington, David},
  journal = {Phys. Rev. B},
  volume = {17},
  issue = {11},
  pages = {4384--4403},
  numpages = {0},
  year = {1978},
  month = {Jun},
  publisher = {American Physical Society},
  doi = {10.1103/PhysRevB.17.4384},
  url = {https://link.aps.org/doi/10.1103/PhysRevB.17.4384}
}

@article{ThoulessAndersonPalmer1977,
author = {D. J. Thouless and P. W. Anderson and R. G. Palmer},
title = {Solution of 'Solvable model of a spin glass'},
journal = {The Philosophical Magazine: A Journal of Theoretical Experimental and Applied Physics},
volume = {35},
number = {3},
pages = {593--601},
year = {1977},
publisher = {Taylor \& Francis},
doi = {10.1080/14786437708235992},
URL = {https://doi.org/10.1080/14786437708235992},
eprint = {https://doi.org/10.1080/14786437708235992}
}

@article{AlmeidaThouless1978,
doi = {10.1088/0305-4470/11/5/028},
url = {https://doi.org/10.1088/0305-4470/11/5/028},
year = {1978},
month = {may},
publisher = {},
volume = {11},
number = {5},
pages = {983},
author = {J R L de Almeida and D J Thouless},
title = {Stability of the Sherrington-Kirkpatrick solution of a spin glass model},
journal = {Journal of Physics A: Mathematical and General}
}

@article{BrayMoore1978,
  title = {Replica-Symmetry Breaking in Spin-Glass Theories},
  author = {Bray, A. J. and Moore, M. A.},
  journal = {Phys. Rev. Lett.},
  volume = {41},
  issue = {15},
  pages = {1068--1072},
  numpages = {0},
  year = {1978},
  month = {Oct},
  publisher = {American Physical Society},
  doi = {10.1103/PhysRevLett.41.1068},
  url = {https://link.aps.org/doi/10.1103/PhysRevLett.41.1068}
}

@article{BrayMoore1980,
doi = {10.1088/0022-3719/13/19/002},
url = {https://doi.org/10.1088/0022-3719/13/19/002},
year = {1980},
month = {jul},
publisher = {},
volume = {13},
number = {19},
pages = {L469},
author = {A J Bray and M A Moore},
title = {Metastable states in spin glasses},
journal = {Journal of Physics C: Solid State Physics}
}

@article{Parisi1979,
  author = {Parisi, Giorgio},
  title = {Infinite Number of Order Parameters for Spin-Glasses},
  journal = {Physical Review Letters},
  volume = {43},
  number = {23},
  pages = {1754--1756},
  year = {1979},
  doi = {10.1103/PhysRevLett.43.1754}
}

@article{Parisi1980a,
  author = {Parisi, Giorgio},
  title = {The Order Parameter for Spin Glasses: A Function on the Interval 0-1},
  journal = {Journal of Physics A: Mathematical and General},
  volume = {13},
  number = {3},
  pages = {1101--1112},
  year = {1980},
  doi = {10.1088/0305-4470/13/3/042}
}

@article{Parisi1980b,
doi = {10.1088/0305-4470/13/4/009},
url = {https://doi.org/10.1088/0305-4470/13/4/009},
year = {1980},
month = {apr},
publisher = {},
volume = {13},
number = {4},
pages = {L115},
author = {G Parisi},
title = {A sequence of approximated solutions to the S-K model for spin glasses},
journal = {Journal of Physics A: Mathematical and General}
}

@article{ParisiToulouse1980,
  title={A Simple hypothesis for the spin glass phase of the infinite-ranged SK model},
  author={Parisi, G and Toulouse, G},
  journal={Journal de Physique Lettres},
  volume={41},
  number={15},
  pages={361--364},
  year={1980},
  publisher={Les Editions de Physique}
}

@article{Parisi1983,
  title = {Order Parameter for Spin-Glasses},
  author = {Parisi, Giorgio},
  journal = {Phys. Rev. Lett.},
  volume = {50},
  issue = {24},
  pages = {1946--1948},
  numpages = {0},
  year = {1983},
  month = {Jun},
  publisher = {American Physical Society},
  doi = {10.1103/PhysRevLett.50.1946},
  url = {https://link.aps.org/doi/10.1103/PhysRevLett.50.1946}
}

@article{Plefka1982,
doi = {10.1088/0305-4470/15/6/035},
url = {https://doi.org/10.1088/0305-4470/15/6/035},
year = {1982},
month = {jun},
publisher = {},
volume = {15},
number = {6},
pages = {1971},
author = {T Plefka},
title = {Convergence condition of the TAP equation for the infinite-ranged Ising spin glass model},
journal = {Journal of Physics A: Mathematical and General}
}

@article{DominicsKonder1983,
  title = {Eigenvalues of the stability matrix for Parisi solution of the long-range spin-glass},
  author = {De Dominicis, C. and Kondor, I.},
  journal = {Phys. Rev. B},
  volume = {27},
  issue = {1},
  pages = {606(R)--608(R)},
  numpages = {0},
  year = {1983},
  month = {Jan},
  publisher = {American Physical Society},
  doi = {10.1103/PhysRevB.27.606},
  url = {https://link.aps.org/doi/10.1103/PhysRevB.27.606}
}

@article{Guerra2003,
   title={Broken Replica Symmetry Bounds in the Mean Field Spin Glass Model},
   volume={233},
   ISSN={1432-0916},
   url={http://dx.doi.org/10.1007/s00220-002-0773-5},
   DOI={10.1007/s00220-002-0773-5},
   number={1},
   journal={Communications in Mathematical Physics},
   publisher={Springer Science and Business Media LLC},
   author={Guerra, Francesco},
   year={2003},
   month=Feb, pages={1–12} }

@article{Talagrand2006,
  title={The parisi formula},
  author={Michel Talagrand},
  journal={Annals of Mathematics},
  year={2006},
  volume={163},
  pages={221-263},
  url={https://api.semanticscholar.org/CorpusID:56101828}
}

@article{Young1983,
  title = {Direct Determination of the Probability Distribution for the Spin-Glass Order Parameter},
  author = {Young, A. P.},
  journal = {Phys. Rev. Lett.},
  volume = {51},
  issue = {13},
  pages = {1206--1209},
  numpages = {0},
  year = {1983},
  month = {Sep},
  publisher = {American Physical Society},
  doi = {10.1103/PhysRevLett.51.1206},
  url = {https://link.aps.org/doi/10.1103/PhysRevLett.51.1206}
}

@article{MezardParisiSourlasToulouseVirasoro1984a,
  title = {Nature of the Spin-Glass Phase},
  author = {M\'ezard, M. and Parisi, G. and Sourlas, N. and Toulouse, G. and Virasoro, M.},
  journal = {Phys. Rev. Lett.},
  volume = {52},
  issue = {13},
  pages = {1156--1159},
  numpages = {0},
  year = {1984},
  month = {Mar},
  publisher = {American Physical Society},
  doi = {10.1103/PhysRevLett.52.1156},
  url = {https://link.aps.org/doi/10.1103/PhysRevLett.52.1156}
}

@article{MezardParisiSourlasToulouseVirasoro1984b,
	author = {{Mézard, M.} and {Parisi, G.} and {Sourlas, N.} and {Toulouse, G.} and {Virasoro, M.}},
	title = {Replica symmetry breaking and the nature of the spin glass phase},
	DOI= "10.1051/jphys:01984004505084300",
	url= "https://doi.org/10.1051/jphys:01984004505084300",
	journal = {J. Phys. France},
	year = 1984,
	volume = 45,
	number = 5,
	pages = "843-854",
}

@article{MezardVirasoro1985,
	author = {{Mézard, M.} and {Virasoro, M. A.}},
	title = {The microstructure of ultrametricity},
	DOI= "10.1051/jphys:019850046080129300",
	url= "https://doi.org/10.1051/jphys:019850046080129300",
	journal = {J. Phys. France},
	year = 1985,
	volume = 46,
	number = 8,
	pages = "1293-1307",
}

@article{MézardParisiVirasoro1986,
doi = {10.1209/0295-5075/1/2/006},
url = {https://doi.org/10.1209/0295-5075/1/2/006},
year = {1986},
month = {jan},
publisher = {},
volume = {1},
number = {2},
pages = {77},
author = {M. Mézard and G. Parisi and M. A. Virasoro},
title = {SK Model: The Replica Solution without Replicas},
journal = {Europhysics Letters}
}

@article{RammalToulouseVirasoro1986,
  title = {Ultrametricity for physicists},
  author = {Rammal, R. and Toulouse, G. and Virasoro, M. A.},
  journal = {Rev. Mod. Phys.},
  volume = {58},
  issue = {3},
  pages = {765--788},
  numpages = {0},
  year = {1986},
  month = {Jul},
  publisher = {American Physical Society},
  doi = {10.1103/RevModPhys.58.765},
  url = {https://link.aps.org/doi/10.1103/RevModPhys.58.765}
}

@book{MezardParisiVirasoro1986b,
author = {Mezard, M and Parisi, G and Virasoro, M},
title = {Spin Glass Theory and Beyond},
publisher = {WORLD SCIENTIFIC},
year = {1986},
doi = {10.1142/0271},
address = {},
edition   = {},
URL = {https://www.worldscientific.com/doi/abs/10.1142/0271},
eprint = {https://www.worldscientific.com/doi/pdf/10.1142/0271}
}

@article{BinderYoung1986,
  title = {Spin glasses: Experimental facts, theoretical concepts, and open questions},
  author = {Binder, K. and Young, A. P.},
  journal = {Rev. Mod. Phys.},
  volume = {58},
  issue = {4},
  pages = {801--976},
  numpages = {0},
  year = {1986},
  month = {Oct},
  publisher = {American Physical Society},
  doi = {10.1103/RevModPhys.58.801},
  url = {https://link.aps.org/doi/10.1103/RevModPhys.58.801}
}

@article{GhirlandaGuerra1998,
doi = {10.1088/0305-4470/31/46/006},
url = {https://doi.org/10.1088/0305-4470/31/46/006},
year = {1998},
month = {nov},
publisher = {},
volume = {31},
number = {46},
pages = {9149},
author = {Stefano Ghirlanda and Francesco Guerra},
title = {General properties of overlap probability distributions in disordered spin systems. Towards Parisi ultrametricity},
journal = {Journal of Physics A: Mathematical and General}
}

@article{ParisiTersenghi2000,
doi = {10.1088/0305-4470/33/1/307},
url = {https://doi.org/10.1088/0305-4470/33/1/307},
year = {2000},
month = {jan},
publisher = {},
volume = {33},
number = {1},
pages = {113},
author = {Giorgio Parisi and Federico Ricci-Tersenghi},
title = {On the 
origin of ultrametricity},
journal = {Journal of Physics A: Mathematical and General}
}

@misc{Panchenko2015,
      title={The Parisi ultrametricity conjecture}, 
      author={Dmitry Panchenko},
      year={2015},
      eprint={1112.1003},
      archivePrefix={arXiv},
      primaryClass={math.PR},
      url={https://arxiv.org/abs/1112.1003}, 
}

@article{Derrida1980,
  title = {Random-Energy Model: Limit of a Family of Disordered Models},
  author = {Derrida, B.},
  journal = {Phys. Rev. Lett.},
  volume = {45},
  issue = {2},
  pages = {79--82},
  numpages = {0},
  year = {1980},
  month = {Jul},
  publisher = {American Physical Society},
  doi = {10.1103/PhysRevLett.45.79},
  url = {https://link.aps.org/doi/10.1103/PhysRevLett.45.79}
}

@article{Derrida1981,
  title = {Random-energy model: An exactly solvable model of disordered systems},
  author = {Derrida, Bernard},
  journal = {Phys. Rev. B},
  volume = {24},
  issue = {5},
  pages = {2613--2626},
  numpages = {0},
  year = {1981},
  month = {Sep},
  publisher = {American Physical Society},
  doi = {10.1103/PhysRevB.24.2613},
  url = {https://link.aps.org/doi/10.1103/PhysRevB.24.2613}
}

@article{Derrida1985,
	author = {{Derrida, B.}},
	title = {A generalization of the Random Energy Model which includes correlations between energies},
	DOI= "10.1051/jphyslet:01985004609040100",
	url= "https://doi.org/10.1051/jphyslet:01985004609040100",
	journal = {J. Physique Lett.},
	year = 1985,
	volume = 46,
	number = 9,
	pages = "401-407",
}

@article{GrossMezard1986,
title = {The simplest spin glass},
journal = {Nuclear Physics B},
volume = {240},
number = {4},
pages = {431-452},
year = {1984},
issn = {0550-3213},
doi = {https://doi.org/10.1016/0550-3213(84)90237-2},
url = {https://www.sciencedirect.com/science/article/pii/0550321384902372},
author = {D.J. Gross and M. Mezard}
}

@article{Gardner1985,
title = {Spin glasses with p-spin interactions},
journal = {Nuclear Physics B},
volume = {257},
pages = {747-765},
year = {1985},
issn = {0550-3213},
doi = {https://doi.org/10.1016/0550-3213(85)90374-8},
url = {https://www.sciencedirect.com/science/article/pii/0550321385903748},
author = {E. Gardner}
}

@article{GrossKanterSompolinsky1985,
  title = {Mean-field theory of the Potts glass},
  author = {Gross, D. J. and Kanter, I. and Sompolinsky, H.},
  journal = {Phys. Rev. Lett.},
  volume = {55},
  issue = {3},
  pages = {304--307},
  numpages = {0},
  year = {1985},
  month = {Jul},
  publisher = {American Physical Society},
  doi = {10.1103/PhysRevLett.55.304},
  url = {https://link.aps.org/doi/10.1103/PhysRevLett.55.304}
}

@article{KirkpatrickThirumalai1987b,
  title = {p-spin-interaction spin-glass models: Connections with the structural glass problem},
  author = {Kirkpatrick, T. R. and Thirumalai, D.},
  journal = {Phys. Rev. B},
  volume = {36},
  issue = {10},
  pages = {5388--5397},
  numpages = {0},
  year = {1987},
  month = {Oct},
  publisher = {American Physical Society},
  doi = {10.1103/PhysRevB.36.5388},
  url = {https://link.aps.org/doi/10.1103/PhysRevB.36.5388}
}

@article{KirkpatrickWolynes1987,
  title = {Stable and metastable states in mean-field Potts and structural glasses},
  author = {Kirkpatrick, T. R. and Wolynes, P. G.},
  journal = {Phys. Rev. B},
  volume = {36},
  issue = {16},
  pages = {8552--8564},
  numpages = {0},
  year = {1987},
  month = {Dec},
  publisher = {American Physical Society},
  doi = {10.1103/PhysRevB.36.8552},
  url = {https://link.aps.org/doi/10.1103/PhysRevB.36.8552}
}

@article{KanterSompolinsky1987,
  title = {Mean-field theory of spin-glasses with finite coordination number},
  author = {Kanter, I. and Sompolinsky, H.},
  journal = {Phys. Rev. Lett.},
  volume = {58},
  issue = {2},
  pages = {164--167},
  numpages = {0},
  year = {1987},
  month = {Jan},
  publisher = {American Physical Society},
  doi = {10.1103/PhysRevLett.58.164},
  url = {https://link.aps.org/doi/10.1103/PhysRevLett.58.164}
}

@article{deAlmeidaJonesKosterlitzThouless1978,
  author = {{de Almeida}, J. R. L. and Jones, R. C. and Kosterlitz, J. M. and Thouless, D. J.},
  title = {The Infinite-Ranged Spin Glass with $m$-Component Spins},
  journal = {Journal of Physics C: Solid State Physics},
  volume = {11},
  number = {21},
  pages = {L871--L875},
  year = {1978},
  doi = {10.1088/0022-3719/11/21/005}
}

@article{CrisantiSommers1994,
	author = {Crisanti, A. and Sommers, H. -J.},
	date = {1992/10/01},
	doi = {10.1007/BF01309287},
	id = {Crisanti1992},
	isbn = {1431-584X},
	journal = {Zeitschrift f{\"u}r Physik B Condensed Matter},
	number = {3},
	pages = {341--354},
	title = {The spherical $p$-spin interaction spin glass model: the statics},
	url = {https://doi.org/10.1007/BF01309287},
	volume = {87},
	year = {1992}}

@article{FranzParisi1995,
	author = {{Silvio Franz} and {Giorgio Parisi}},
	title = {Recipes for Metastable States in Spin Glasses},
	DOI= "10.1051/jp1:1995201",
	url= "https://doi.org/10.1051/jp1:1995201",
	journal = {J. Phys. I France},
	year = 1995,
	volume = 5,
	number = 11,
	pages = "1401-1415",
	month = "",
}

@article{Monasson1995,
  title = {Structural Glass Transition and the Entropy of the Metastable States},
  author = {Monasson, R\'emi},
  journal = {Phys. Rev. Lett.},
  volume = {75},
  issue = {15},
  pages = {2847--2850},
  numpages = {0},
  year = {1995},
  month = {Oct},
  publisher = {American Physical Society},
  doi = {10.1103/PhysRevLett.75.2847},
  url = {https://link.aps.org/doi/10.1103/PhysRevLett.75.2847}
}

@article{FranzParisi1997,
  title = {Phase Diagram of Coupled Glassy Systems: A Mean-Field Study},
  author = {Franz, Silvio and Parisi, Giorgio},
  journal = {Phys. Rev. Lett.},
  volume = {79},
  issue = {13},
  pages = {2486--2489},
  numpages = {0},
  year = {1997},
  month = {Sep},
  publisher = {American Physical Society},
  doi = {10.1103/PhysRevLett.79.2486},
  url = {https://link.aps.org/doi/10.1103/PhysRevLett.79.2486}
}

@article{MezardParisi2001,
   title={The Bethe lattice spin glass revisited},
   volume={20},
   ISSN={1434-6028},
   url={http://dx.doi.org/10.1007/PL00011099},
   DOI={10.1007/pl00011099},
   number={2},
   journal={The European Physical Journal B},
   publisher={Springer Science and Business Media LLC},
   author={Mézard, M. and Parisi, G.},
   year={2001},
   month=Mar, pages={217–233} }

@article{CastellaniCavagna2005,
doi = {10.1088/1742-5468/2005/05/P05012},
url = {https://doi.org/10.1088/1742-5468/2005/05/P05012},
year = {2005},
month = {may},
publisher = {},
volume = {2005},
number = {05},
pages = {P05012},
author = {Castellani, Tommaso and Cavagna, Andrea},
title = {Spin-glass theory for pedestrians},
journal = {Journal of Statistical Mechanics: Theory and Experiment}
}

@article{CrisantiLeuzzi2004,
  title = {Spherical $2+p$ Spin-Glass Model: An Exactly Solvable Model for Glass to Spin-Glass Transition},
  author = {Crisanti, A. and Leuzzi, L.},
  journal = {Phys. Rev. Lett.},
  volume = {93},
  issue = {21},
  pages = {217203},
  numpages = {4},
  year = {2004},
  month = {Nov},
  publisher = {American Physical Society},
  doi = {10.1103/PhysRevLett.93.217203},
  url = {https://link.aps.org/doi/10.1103/PhysRevLett.93.217203}
}

@article{CrisantiLeuzzi2006,
  title = {Spherical $2+p$ spin-glass model: An analytically solvable model with a glass-to-glass transition},
  author = {Crisanti, A. and Leuzzi, L.},
  journal = {Phys. Rev. B},
  volume = {73},
  issue = {1},
  pages = {014412},
  numpages = {20},
  year = {2006},
  month = {Jan},
  publisher = {American Physical Society},
  doi = {10.1103/PhysRevB.73.014412},
  url = {https://link.aps.org/doi/10.1103/PhysRevB.73.014412}
}

@article{CrisantiLeuzzi2007,
  title = {Amorphous-amorphous transition and the two-step replica symmetry breaking phase},
  author = {Crisanti, Andrea and Leuzzi, Luca},
  journal = {Phys. Rev. B},
  volume = {76},
  issue = {18},
  pages = {184417},
  numpages = {8},
  year = {2007},
  month = {Nov},
  publisher = {American Physical Society},
  doi = {10.1103/PhysRevB.76.184417},
  url = {https://link.aps.org/doi/10.1103/PhysRevB.76.184417}
}

@article{KurchanParisiZamponi2012,
   title={Exact theory of dense amorphous hard spheres in high dimension I. The free energy},
   volume={2012},
   ISSN={1742-5468},
   url={http://dx.doi.org/10.1088/1742-5468/2012/10/P10012},
   DOI={10.1088/1742-5468/2012/10/p10012},
   number={10},
   journal={Journal of Statistical Mechanics: Theory and Experiment},
   publisher={IOP Publishing},
   author={Kurchan, Jorge and Parisi, Giorgio and Zamponi, Francesco},
   year={2012},
   month=Oct, pages={P10012} }

@article{KurchanParisiUrbaniZamponi2013,
   title={Exact Theory of Dense Amorphous Hard Spheres in High Dimension. II. The High Density Regime and the Gardner Transition},
   volume={117},
   ISSN={1520-5207},
   url={http://dx.doi.org/10.1021/jp402235d},
   DOI={10.1021/jp402235d},
   number={42},
   journal={The Journal of Physical Chemistry B},
   publisher={American Chemical Society (ACS)},
   author={Kurchan, Jorge and Parisi, Giorgio and Urbani, Pierfrancesco and Zamponi, Francesco},
   year={2013},
   month=May, pages={12979–12994} }

@article{CharbonneauKurchanParisiUrbaniZamponi2014,
   title={Exact theory of dense amorphous hard spheres in high dimension. III. The full replica symmetry breaking solution},
   volume={2014},
   ISSN={1742-5468},
   url={http://dx.doi.org/10.1088/1742-5468/2014/10/P10009},
   DOI={10.1088/1742-5468/2014/10/p10009},
   number={10},
   journal={Journal of Statistical Mechanics: Theory and Experiment},
   publisher={IOP Publishing},
   author={Charbonneau, Patrick and Kurchan, Jorge and Parisi, Giorgio and Urbani, Pierfrancesco and Zamponi, Francesco},
   year={2014},
   month=Oct, pages={P10009} }

@article{FranzParisi2016,
doi = {10.1088/1751-8113/49/14/145001},
url = {https://doi.org/10.1088/1751-8113/49/14/145001},
year = {2016},
month = {feb},
publisher = {IOP Publishing},
volume = {49},
number = {14},
pages = {145001},
author = {Franz, Silvio and Parisi, Giorgio},
title = {The simplest model of jamming},
journal = {Journal of Physics A: Mathematical and Theoretical}
}

@article{Sompolinsky1981,
  title = {Time-Dependent Order Parameters in Spin-Glasses},
  author = {Sompolinsky, H.},
  journal = {Phys. Rev. Lett.},
  volume = {47},
  issue = {13},
  pages = {935--938},
  numpages = {0},
  year = {1981},
  month = {Sep},
  publisher = {American Physical Society},
  doi = {10.1103/PhysRevLett.47.935},
  url = {https://link.aps.org/doi/10.1103/PhysRevLett.47.935}
}

@article{SompolinskyZippelius1982,
  title = {Relaxational dynamics of the Edwards-Anderson model and the mean-field theory of spin-glasses},
  author = {Sompolinsky, H. and Zippelius, Annette},
  journal = {Phys. Rev. B},
  volume = {25},
  issue = {11},
  pages = {6860--6875},
  numpages = {0},
  year = {1982},
  month = {Jun},
  publisher = {American Physical Society},
  doi = {10.1103/PhysRevB.25.6860},
  url = {https://link.aps.org/doi/10.1103/PhysRevB.25.6860}
}

@article{KirkpatrickThirumalai1987,
  title = {Dynamics of the Structural Glass Transition and the $p$-Spin---Interaction Spin-Glass Model},
  author = {Kirkpatrick, T. R. and Thirumalai, D.},
  journal = {Phys. Rev. Lett.},
  volume = {58},
  issue = {20},
  pages = {2091--2094},
  numpages = {0},
  year = {1987},
  month = {May},
  publisher = {American Physical Society},
  doi = {10.1103/PhysRevLett.58.2091},
  url = {https://link.aps.org/doi/10.1103/PhysRevLett.58.2091}
}

@article{Bouchaud1992,
	author = {{J. P. Bouchaud}},
	title = {Weak ergodicity breaking and aging in disordered systems},
	DOI= "10.1051/jp1:1992238",
	url= "https://doi.org/10.1051/jp1:1992238",
	journal = {J. Phys. I France},
	year = 1992,
	volume = 2,
	number = 9,
	pages = "1705-1713",
	month = "",
}

@article{CugliandoloKurchan1993,
  title = {Analytical solution of the off-equilibrium dynamics of a long-range spin-glass model},
  author = {Cugliandolo, L. F. and Kurchan, J.},
  journal = {Phys. Rev. Lett.},
  volume = {71},
  issue = {1},
  pages = {173--176},
  numpages = {0},
  year = {1993},
  month = {Jul},
  publisher = {American Physical Society},
  doi = {10.1103/PhysRevLett.71.173},
  url = {https://link.aps.org/doi/10.1103/PhysRevLett.71.173}
}

@article{CrisantiHornerSommers1993,
	author = {Crisanti, A. and Horner, H. and Sommers, H. -J.},
	date = {1993/06/01},
	doi = {10.1007/BF01312184},
	id = {Crisanti1993},
	isbn = {1431-584X},
	journal = {Zeitschrift f{\"u}r Physik B Condensed Matter},
	number = {2},
	pages = {257--271},
	title = {The spherical $p$-spin interaction spin-glass model},
	url = {https://doi.org/10.1007/BF01312184},
	volume = {92},
	year = {1993}}

@article{CugliandoloKurchan1994,
   title={On the out-of-equilibrium relaxation of the Sherrington-Kirkpatrick model},
   volume={27},
   ISSN={1361-6447},
   url={http://dx.doi.org/10.1088/0305-4470/27/17/011},
   DOI={10.1088/0305-4470/27/17/011},
   number={17},
   journal={Journal of Physics A: Mathematical and General},
   publisher={IOP Publishing},
   author={Cugliandolo, L F and Kurchan, J},
   year={1994},
   month=Sept, pages={5749–5772} }

@article{Franz1994,
   title={On mean field glassy dynamics out of equilibrium},
   volume={210},
   ISSN={0378-4371},
   url={http://dx.doi.org/10.1016/0378-4371(94)00057-3},
   DOI={10.1016/0378-4371(94)00057-3},
   number={1-2},
   journal={Physica A: Statistical Mechanics and its Applications},
   publisher={Elsevier BV},
   author={Franz, Silvio and Mézard, Marc},
   year={1994},
   month=Sept, pages={48–72} }

@article{CugliandoloLeticiaKurchan1997,
  title = {Energy flow, partial equilibration, and effective temperatures in systems with slow dynamics},
  author = {Cugliandolo, Leticia F. and Kurchan, Jorge and Peliti, Luca},
  journal = {Phys. Rev. E},
  volume = {55},
  issue = {4},
  pages = {3898--3914},
  numpages = {0},
  year = {1997},
  month = {Apr},
  publisher = {American Physical Society},
  doi = {10.1103/PhysRevE.55.3898},
  url = {https://link.aps.org/doi/10.1103/PhysRevE.55.3898}
}

@article{FranzMezardParisiPeliti1998,
  title = {Measuring Equilibrium Properties in Aging Systems},
  author = {Franz, Silvio and M\'ezard, Marc and Parisi, Giorgio and Peliti, Luca},
  journal = {Phys. Rev. Lett.},
  volume = {81},
  issue = {9},
  pages = {1758--1761},
  numpages = {0},
  year = {1998},
  month = {Aug},
  publisher = {American Physical Society},
  doi = {10.1103/PhysRevLett.81.1758},
  url = {https://link.aps.org/doi/10.1103/PhysRevLett.81.1758}
}

@article{FranzMezardParisiPeliti1999,
   title={The Response of Glassy Systems to Random Perturbations: A Bridge Between Equilibrium and Off-Equilibrium},
   volume={97},
   ISSN={1572-9613},
   url={http://dx.doi.org/10.1023/A:1004602906332},
   DOI={10.1023/a:1004602906332},
   number={3-4},
   journal={Journal of Statistical Physics},
   publisher={Springer Science and Business Media LLC},
   author={Franz, Silvio and Mézard, Marc and Parisi, Giorgio and Peliti, Luca},
   year={1999},
   month=Nov, pages={459–488} }

@article{JohOrbachWoodHammannVincent1999,
  title = {Extraction of the Spin Glass Correlation Length},
  author = {Joh, Y. G. and Orbach, R. and Wood, G. G. and Hammann, J. and Vincent, E.},
  journal = {Phys. Rev. Lett.},
  volume = {82},
  issue = {2},
  pages = {438--441},
  numpages = {0},
  year = {1999},
  month = {Jan},
  publisher = {American Physical Society},
  doi = {10.1103/PhysRevLett.82.438},
  url = {https://link.aps.org/doi/10.1103/PhysRevLett.82.438}
}

@article{MontanariTersenghi2003,
   title={On the nature of the low-temperature phase in discontinuous mean-field spin glasses},
   volume={33},
   ISSN={1434-6036},
   url={http://dx.doi.org/10.1140/epjb/e2003-00174-7},
   DOI={10.1140/epjb/e2003-00174-7},
   number={3},
   journal={The European Physical Journal B - Condensed Matter},
   publisher={Springer Science and Business Media LLC},
   author={Montanari, A. and Ricci-Tersenghi, F.},
   year={2003},
   month=June, pages={339–346} }

@article{FolenaFranzTersenghi2020,
  title = {Rethinking Mean-Field Glassy Dynamics and Its Relation with the Energy Landscape: The Surprising Case of the Spherical Mixed $p$-Spin Model},
  author = {Folena, Giampaolo and Franz, Silvio and Ricci-Tersenghi, Federico},
  journal = {Phys. Rev. X},
  volume = {10},
  issue = {3},
  pages = {031045},
  numpages = {26},
  year = {2020},
  month = {Aug},
  publisher = {American Physical Society},
  doi = {10.1103/PhysRevX.10.031045},
  url = {https://link.aps.org/doi/10.1103/PhysRevX.10.031045}
}

@article{Mattis1976,
title = {Solvable spin systems with random interactions},
journal = {Physics Letters A},
volume = {56},
number = {5},
pages = {421-422},
year = {1976},
issn = {0375-9601},
doi = {https://doi.org/10.1016/0375-9601(76)90396-0},
url = {https://www.sciencedirect.com/science/article/pii/0375960176903960},
author = {D.C. Mattis}
}

@INCOLLECTION{Toulouse1977,
       author = {{Toulouse}, G.},
        title = "{Theory of the frustration effect in spin glasses: I}",
    booktitle = {SPIN GLASS THEORY AND BEYOND: AN INTRODUCTION TO THE REPLICA METHOD AND ITS APPLICATIONS. Edited by MEZARD M ET AL. Published by World Scientific Press},
         year = 1987,
       editor = {{Mezard}, Marc and {Parisi}, Giorgio and {Virasoro}, Miguel Angel},
        pages = {99-103},
          doi = {10.1142/9789812799371_0009},
       adsurl = {https://ui.adsabs.harvard.edu/abs/1987sgtb.book...99T}
}

@article{VannimenusToulouse1977,
doi = {10.1088/0022-3719/10/18/008},
url = {https://doi.org/10.1088/0022-3719/10/18/008},
year = {1977},
month = {sep},
publisher = {},
volume = {10},
number = {18},
pages = {L537},
author = {J Vannimenus and G Toulouse},
title = {Theory of the frustration effect. II. Ising spins on a square lattice},
journal = {Journal of Physics C: Solid State Physics}
}

@article{Kirkpatrick1977,
  title = {Frustration and ground-state degeneracy in spin glasses},
  author = {Kirkpatrick, Scott},
  journal = {Phys. Rev. B},
  volume = {16},
  issue = {10},
  pages = {4630--4641},
  numpages = {0},
  year = {1977},
  month = {Nov},
  publisher = {American Physical Society},
  doi = {10.1103/PhysRevB.16.4630},
  url = {https://link.aps.org/doi/10.1103/PhysRevB.16.4630}
}

@article{FradkinHubermanShenker1978,
  title = {Gauge symmetries in random magnetic systems},
  author = {Fradkin, Eduardo and Huberman, B. A. and Shenker, Stephen H.},
  journal = {Phys. Rev. B},
  volume = {18},
  issue = {9},
  pages = {4789--4814},
  numpages = {0},
  year = {1978},
  month = {Nov},
  publisher = {American Physical Society},
  doi = {10.1103/PhysRevB.18.4789},
  url = {https://link.aps.org/doi/10.1103/PhysRevB.18.4789}
}

@article{Hertz1978,
  title = {Gauge models for spin-glasses},
  author = {Hertz, J. A.},
  journal = {Phys. Rev. B},
  volume = {18},
  issue = {9},
  pages = {4875--4885},
  numpages = {0},
  year = {1978},
  month = {Nov},
  publisher = {American Physical Society},
  doi = {10.1103/PhysRevB.18.4875},
  url = {https://link.aps.org/doi/10.1103/PhysRevB.18.4875}
}

@article{Nishimori1980,
doi = {10.1088/0022-3719/13/21/012},
url = {https://doi.org/10.1088/0022-3719/13/21/012},
year = {1980},
month = {jul},
publisher = {},
volume = {13},
number = {21},
pages = {4071},
author = {H Nishimori},
title = {Exact results and critical properties of the Ising model with competing interactions},
journal = {Journal of Physics C: Solid State Physics}
}

@article{Nishimori1981,
    author = {Nishimori, Hidetoshi},
    title = {Internal Energy, Specific Heat and Correlation Function of the Bond-Random Ising Model},
    journal = {Progress of Theoretical Physics},
    volume = {66},
    number = {4},
    pages = {1169-1181},
    year = {1981},
    month = {10},
    issn = {0033-068X},
    doi = {10.1143/PTP.66.1169},
    url = {https://doi.org/10.1143/PTP.66.1169},
    eprint = {https://academic.oup.com/ptp/article-pdf/66/4/1169/5265369/66-4-1169.pdf},
}

@article{GeorgesHanselBouchaud1985,
	author = {{Georges, A.} and {Hansel, D.} and {Le Doussal, P.} and {Bouchaud, J.-P.}},
	title = {Exact properties of spin glasses. II. Nishimori's line : new results and physical implications},
	DOI= "10.1051/jphys:0198500460110182700",
	url= "https://doi.org/10.1051/jphys:0198500460110182700",
	journal = {J. Phys. France},
	year = 1985,
	volume = 46,
	number = 11,
	pages = "1827-1836",
}

@book{Nishimori_book,
  address = {Oxford; New York},
  author = {Nishimori, Hidetoshi},
  isbn = {0198509405 9780198509400 0198509413 9780198509417},
  publisher = {Oxford University Press},
  refid = {47063322},
  title = {Statistical Physics of Spin Glasses and Information Processing: an Introduction},
  year = 2001
}

@article{GoldbartMartinFradkin1986,
  title = {Gauge-invariant spin glasses},
  author = {Goldbart, Paul and Martin, Olivier and Fradkin, Eduardo},
  journal = {Phys. Rev. B},
  volume = {34},
  issue = {1},
  pages = {301--305},
  numpages = {0},
  year = {1986},
  month = {Jul},
  publisher = {American Physical Society},
  doi = {10.1103/PhysRevB.34.301},
  url = {https://link.aps.org/doi/10.1103/PhysRevB.34.301}
}

@article{SouthernYoung1977,
doi = {10.1088/0022-3719/10/12/023},
url = {https://doi.org/10.1088/0022-3719/10/12/023},
year = {1977},
month = {jun},
publisher = {},
volume = {10},
number = {12},
pages = {2179},
author = {B W Southern and A P Young},
title = {Real space rescaling study of spin glass behaviour in three dimensions},
journal = {Journal of Physics C: Solid State Physics}
}

@article{BrayMoore1984,
doi = {10.1088/0022-3719/17/18/004},
url = {https://doi.org/10.1088/0022-3719/17/18/004},
year = {1984},
month = {jun},
publisher = {},
volume = {17},
number = {18},
pages = {L463},
author = {A J Bray and M A Moore},
title = {Lower critical dimension of Ising spin glasses: a numerical study},
journal = {Journal of Physics C: Solid State Physics}
}

@article{McMillan1984,
doi = {10.1088/0022-3719/17/18/010},
url = {https://doi.org/10.1088/0022-3719/17/18/010},
year = {1984},
month = {jun},
publisher = {},
volume = {17},
number = {18},
pages = {3179},
author = {W L McMillan},
title = {Scaling theory of Ising spin glasses},
journal = {Journal of Physics C: Solid State Physics}
}

@article{FisherHuse1986,
  title = {Ordered Phase of Short-Range Ising Spin-Glasses},
  author = {Fisher, Daniel S. and Huse, David A.},
  journal = {Phys. Rev. Lett.},
  volume = {56},
  issue = {15},
  pages = {1601--1604},
  numpages = {0},
  year = {1986},
  month = {Apr},
  publisher = {American Physical Society},
  doi = {10.1103/PhysRevLett.56.1601},
  url = {https://link.aps.org/doi/10.1103/PhysRevLett.56.1601}
}

@article{BrayMoore1987,
  title = {Chaotic Nature of the Spin-Glass Phase},
  author = {Bray, A. J. and Moore, M. A.},
  journal = {Phys. Rev. Lett.},
  volume = {58},
  issue = {1},
  pages = {57--60},
  numpages = {0},
  year = {1987},
  month = {Jan},
  publisher = {American Physical Society},
  doi = {10.1103/PhysRevLett.58.57},
  url = {https://link.aps.org/doi/10.1103/PhysRevLett.58.57}
}

@article{FisherHuse1987,
doi = {10.1088/0305-4470/20/15/013},
url = {https://doi.org/10.1088/0305-4470/20/15/013},
year = {1987},
month = {oct},
publisher = {},
volume = {20},
number = {15},
pages = {L1005},
author = {D S Fisher and D A Huse},
title = {Absence of many states in realistic spin glasses},
journal = {Journal of Physics A: Mathematical and General}
}

@article{HuseFisher1987,
doi = {10.1088/0305-4470/20/15/012},
url = {https://doi.org/10.1088/0305-4470/20/15/012},
year = {1987},
month = {oct},
publisher = {},
volume = {20},
number = {15},
pages = {L997},
author = {D A Huse and D S Fisher},
title = {Pure states in spin glasses},
journal = {Journal of Physics A: Mathematical and General}
}

@article{FisherHuse1988,
  title = {Equilibrium behavior of the spin-glass ordered phase},
  author = {Fisher, Daniel S. and Huse, David A.},
  journal = {Phys. Rev. B},
  volume = {38},
  issue = {1},
  pages = {386--411},
  numpages = {0},
  year = {1988},
  month = {Jul},
  publisher = {American Physical Society},
  doi = {10.1103/PhysRevB.38.386},
  url = {https://link.aps.org/doi/10.1103/PhysRevB.38.386}
}

@article{NewmanStein1992,
  title = {Multiple states and thermodynamic limits in short-ranged Ising spin-glass models},
  author = {Newman, C. M. and Stein, D. L.},
  journal = {Phys. Rev. B},
  volume = {46},
  issue = {2},
  pages = {973--982},
  numpages = {0},
  year = {1992},
  month = {Jul},
  publisher = {American Physical Society},
  doi = {10.1103/PhysRevB.46.973},
  url = {https://link.aps.org/doi/10.1103/PhysRevB.46.973}
}

@article{NewmanStein1996,
  title = {Non-Mean-Field Behavior of Realistic Spin Glasses},
  author = {Newman, C. M. and Stein, D. L.},
  journal = {Phys. Rev. Lett.},
  volume = {76},
  issue = {3},
  pages = {515--518},
  numpages = {0},
  year = {1996},
  month = {Jan},
  publisher = {American Physical Society},
  doi = {10.1103/PhysRevLett.76.515},
  url = {https://link.aps.org/doi/10.1103/PhysRevLett.76.515}
}

@article{NewmanStein1997,
  title = {Metastate approach to thermodynamic chaos},
  author = {Newman, C. M. and Stein, D. L.},
  journal = {Phys. Rev. E},
  volume = {55},
  issue = {5},
  pages = {5194--5211},
  numpages = {0},
  year = {1997},
  month = {May},
  publisher = {American Physical Society},
  doi = {10.1103/PhysRevE.55.5194},
  url = {https://link.aps.org/doi/10.1103/PhysRevE.55.5194}
}

@article{Read2014,
  title = {Short-range Ising spin glasses: The metastate interpretation of replica symmetry breaking},
  author = {Read, N.},
  journal = {Phys. Rev. E},
  volume = {90},
  issue = {3},
  pages = {032142},
  numpages = {28},
  year = {2014},
  month = {Sep},
  publisher = {American Physical Society},
  doi = {10.1103/PhysRevE.90.032142},
  url = {https://link.aps.org/doi/10.1103/PhysRevE.90.032142}
}

@article{HarrisLubenskyChen1976,
  title = {Critical Properties of Spin-Glasses},
  author = {Harris, A. B. and Lubensky, T. C. and Chen, Jing-Huei},
  journal = {Phys. Rev. Lett.},
  volume = {36},
  issue = {8},
  pages = {415--418},
  numpages = {0},
  year = {1976},
  month = {Feb},
  publisher = {American Physical Society},
  doi = {10.1103/PhysRevLett.36.415},
  url = {https://link.aps.org/doi/10.1103/PhysRevLett.36.415}
}

@article{BrayRoberts1980,
doi = {10.1088/0022-3719/13/29/019},
url = {https://doi.org/10.1088/0022-3719/13/29/019},
year = {1980},
month = {oct},
publisher = {},
volume = {13},
number = {29},
pages = {5405},
author = {A J Bray and S A Roberts},
title = {Renormalisation-group approach to the spin glass transition in finite magnetic fields},
journal = {Journal of Physics C: Solid State Physics}
}

@article{FisherSompolinsky1985,
  title = {Scaling in Spin-Glasses},
  author = {Fisher, Daniel S. and Sompolinsky, H.},
  journal = {Phys. Rev. Lett.},
  volume = {54},
  issue = {10},
  pages = {1063--1066},
  numpages = {0},
  year = {1985},
  month = {Mar},
  publisher = {American Physical Society},
  doi = {10.1103/PhysRevLett.54.1063},
  url = {https://link.aps.org/doi/10.1103/PhysRevLett.54.1063}
}

@article{DominicisKondorTemesvari1993,
  title={ISING SPIN GLASS: RECENT PROGRESS IN THE FIELD THEORY APPROACH},
  author={Cyrano de Dominicis and Imre Kondor and Tamas Temesvari},
  journal={International Journal of Modern Physics B},
  year={1993},
  volume={07},
  pages={986-992},
  url={https://api.semanticscholar.org/CorpusID:121229080}
}

@article{DrosselBokilMoore2000,
  title = {Spin glasses without time-reversal symmetry and the absence of a genuine structural glass transition},
  author = {Drossel, Barbara and Bokil, Hemant and Moore, M. A.},
  journal = {Phys. Rev. E},
  volume = {62},
  issue = {6},
  pages = {7690--7699},
  numpages = {0},
  year = {2000},
  month = {Dec},
  publisher = {American Physical Society},
  doi = {10.1103/PhysRevE.62.7690},
  url = {https://link.aps.org/doi/10.1103/PhysRevE.62.7690}
}

@article{MooreDrossel2002,
  title = {$p$-Spin Model in Finite Dimensions and Its Relation to Structural Glasses},
  author = {Moore, M. A. and Drossel, Barbara},
  journal = {Phys. Rev. Lett.},
  volume = {89},
  issue = {21},
  pages = {217202},
  numpages = {4},
  year = {2002},
  month = {Nov},
  publisher = {American Physical Society},
  doi = {10.1103/PhysRevLett.89.217202},
  url = {https://link.aps.org/doi/10.1103/PhysRevLett.89.217202}
}

@article{TemesváriDominicisPimentel2002,
   title={Generic replica symmetric field-theory for short range Ising spin glasses},
   volume={25},
   ISSN={1434-6028},
   url={http://dx.doi.org/10.1140/epjb/e20020041},
   DOI={10.1140/epjb/e20020041},
   number={3},
   journal={The European Physical Journal B},
   publisher={Springer Science and Business Media LLC},
   author={Temesvári, T. and De Dominicis, C. and Pimentel, I.R.},
   year={2002},
   month=Feb, pages={361–372} }

@article{TemesvariDominicis2002,
  title = {Replica Field Theory and Renormalization Group for the Ising Spin Glass in an External Magnetic Field},
  author = {Temesv\'ari, T. and De Dominicis, C.},
  journal = {Phys. Rev. Lett.},
  volume = {89},
  issue = {9},
  pages = {097204},
  numpages = {4},
  year = {2002},
  month = {Aug},
  publisher = {American Physical Society},
  doi = {10.1103/PhysRevLett.89.097204},
  url = {https://link.aps.org/doi/10.1103/PhysRevLett.89.097204}
}

@article{Temesvari2008,
  title = {Almeida-Thouless transition below six dimensions},
  author = {Temesv\'ari, T.},
  journal = {Phys. Rev. B},
  volume = {78},
  issue = {22},
  pages = {220401(R)},
  numpages = {4},
  year = {2008},
  month = {Dec},
  publisher = {American Physical Society},
  doi = {10.1103/PhysRevB.78.220401},
  url = {https://link.aps.org/doi/10.1103/PhysRevB.78.220401}
}

@article{ParisiTemesvári2012,
   title={Replica symmetry breaking in and around six dimensions},
   volume={858},
   ISSN={0550-3213},
   url={http://dx.doi.org/10.1016/j.nuclphysb.2012.01.014},
   DOI={10.1016/j.nuclphysb.2012.01.014},
   number={2},
   journal={Nuclear Physics B},
   publisher={Elsevier BV},
   author={Parisi, G. and Temesvári, T.},
   year={2012},
   month=May, pages={293–316} }

@article{MooreBray2011,
  title = {Disappearance of the de Almeida-Thouless line in six dimensions},
  author = {Moore, M. A. and Bray, A. J.},
  journal = {Phys. Rev. B},
  volume = {83},
  issue = {22},
  pages = {224408},
  numpages = {5},
  year = {2011},
  month = {Jun},
  publisher = {American Physical Society},
  doi = {10.1103/PhysRevB.83.224408},
  url = {https://link.aps.org/doi/10.1103/PhysRevB.83.224408}
}

@article{CharbonneauYaida2017,
  title = {Nontrivial Critical Fixed Point for Replica-Symmetry-Breaking Transitions},
  author = {Charbonneau, Patrick and Yaida, Sho},
  journal = {Phys. Rev. Lett.},
  volume = {118},
  issue = {21},
  pages = {215701},
  numpages = {6},
  year = {2017},
  month = {May},
  publisher = {American Physical Society},
  doi = {10.1103/PhysRevLett.118.215701},
  url = {https://link.aps.org/doi/10.1103/PhysRevLett.118.215701}
}

@article{MooreRead2018,
  title = {Multicritical Point on the de Almeida--Thouless Line in Spin Glasses in $d>6$ Dimensions},
  author = {Moore, M. A. and Read, N.},
  journal = {Phys. Rev. Lett.},
  volume = {120},
  issue = {13},
  pages = {130602},
  numpages = {6},
  year = {2018},
  month = {Mar},
  publisher = {American Physical Society},
  doi = {10.1103/PhysRevLett.120.130602},
  url = {https://link.aps.org/doi/10.1103/PhysRevLett.120.130602}
}

@article{HollerRead2020,
   title={One-step replica-symmetry-breaking phase below the de Almeida–Thouless line in low-dimensional spin glasses},
   volume={101},
   ISSN={2470-0053},
   url={http://dx.doi.org/10.1103/PhysRevE.101.042114},
   DOI={10.1103/physreve.101.042114},
   number={4},
   journal={Physical Review E},
   publisher={American Physical Society (APS)},
   author={Höller, J. and Read, N.},
   year={2020},
   month=Apr }

@article{AngeliniLucibelloParisiPerrupatoTersenghiRizzo2022,
  title = {Unexpected Upper Critical Dimension for Spin Glass Models in a Field Predicted by the Loop Expansion around the Bethe Solution at Zero Temperature},
  author = {Angelini, Maria Chiara and Lucibello, Carlo and Parisi, Giorgio and Perrupato, Gianmarco and Ricci-Tersenghi, Federico and Rizzo, Tommaso},
  journal = {Phys. Rev. Lett.},
  volume = {128},
  issue = {7},
  pages = {075702},
  numpages = {6},
  year = {2022},
  month = {Feb},
  publisher = {American Physical Society},
  doi = {10.1103/PhysRevLett.128.075702},
  url = {https://link.aps.org/doi/10.1103/PhysRevLett.128.075702}
}

@article{AngeliniParisiTersenghi2013,
   title={Ensemble renormalization group for disordered systems},
   volume={87},
   ISSN={1550-235X},
   url={http://dx.doi.org/10.1103/PhysRevB.87.134201},
   DOI={10.1103/physrevb.87.134201},
   number={13},
   journal={Physical Review B},
   publisher={American Physical Society (APS)},
   author={Angelini, Maria Chiara and Parisi, Giorgio and Ricci-Tersenghi, Federico},
   year={2013},
   month=Apr }

@article{Castellana2025,
  title = {Order parameter for non-mean-field spin glasses},
  author = {Castellana, Michele},
  journal = {Phys. Rev. B},
  volume = {112},
  issue = {22},
  pages = {224206},
  numpages = {8},
  year = {2025},
  month = {Dec},
  publisher = {American Physical Society},
  doi = {10.1103/7lm2-ljrd},
  url = {https://link.aps.org/doi/10.1103/7lm2-ljrd}
}

@article{BhattYoung1985,
  title = {Search for a transition in the three-dimensional $\pm J$ Ising spin-glass},
  author = {Bhatt, R. N. and Young, A. P.},
  journal = {Phys. Rev. Lett.},
  volume = {54},
  issue = {9},
  pages = {924--927},
  numpages = {0},
  year = {1985},
  month = {Mar},
  publisher = {American Physical Society},
  doi = {10.1103/PhysRevLett.54.924},
  url = {https://link.aps.org/doi/10.1103/PhysRevLett.54.924}
}

@article{Ogielski1985,
  title = {Dynamics of three-dimensional Ising spin glasses in thermal equilibrium},
  author = {Ogielski, Andrew T.},
  journal = {Phys. Rev. B},
  volume = {32},
  issue = {11},
  pages = {7384--7398},
  numpages = {0},
  year = {1985},
  month = {Dec},
  publisher = {American Physical Society},
  doi = {10.1103/PhysRevB.32.7384},
  url = {https://link.aps.org/doi/10.1103/PhysRevB.32.7384}
}

@article{FranzParisiVirasoro1994,
	author = {{S. Franz} and {G. Parisi} and {M.A. Virasoro}},
	title = {Interfaces and lower critical dimension in a spin glass model},
	DOI= "10.1051/jp1:1994213",
	url= "https://doi.org/10.1051/jp1:1994213",
	journal = {J. Phys. I France},
	year = 1994,
	volume = 4,
	number = 11,
	pages = "1657-1667",
	month = "",
}

@article{JanusCollab2013,
  title = {Critical parameters of the three-dimensional Ising spin glass},
  author = {Baity-Jesi, M. and Ba\~nos, R. A. and Cruz, A. and Fernandez, L. A. and Gil-Narvion, J. M. and Gordillo-Guerrero, A. and I\~niguez, D. and Maiorano, A. and Mantovani, F. and Marinari, E. and Martin-Mayor, V. and Monforte-Garcia, J. and Sudupe, A. Mu\~noz and Navarro, D. and Parisi, G. and Perez-Gaviro, S. and Pivanti, M. and Ricci-Tersenghi, F. and Ruiz-Lorenzo, J. J. and Schifano, S. F. and Seoane, B. and Tarancon, A. and Tripiccione, R. and Yllanes, D.},
  collaboration = {Janus Collaboration},
  journal = {Phys. Rev. B},
  volume = {88},
  issue = {22},
  pages = {224416},
  numpages = {9},
  year = {2013},
  month = {Dec},
  publisher = {American Physical Society},
  doi = {10.1103/PhysRevB.88.224416},
  url = {https://link.aps.org/doi/10.1103/PhysRevB.88.224416}
}

@article{HukushimaNemoto1996,
author = {Hukushima ,Koji and Nemoto ,Koji},
title = {Exchange Monte Carlo Method and  Application to Spin Glass Simulations},
journal = {Journal of the Physical Society of Japan},
volume = {65},
number = {6},
pages = {1604-1608},
year = {1996},
doi = {10.1143/JPSJ.65.1604},
URL = { 
        https://doi.org/10.1143/JPSJ.65.1604
},
eprint = { 
        https://doi.org/10.1143/JPSJ.65.1604
    }
}

@article{KatzgraberKorner2006,
  title = {Universality in three-dimensional Ising spin glasses: A Monte Carlo study},
  author = {Katzgraber, Helmut G. and K\"orner, Mathias and Young, A. P.},
  journal = {Phys. Rev. B},
  volume = {73},
  issue = {22},
  pages = {224432},
  numpages = {11},
  year = {2006},
  month = {Jun},
  publisher = {American Physical Society},
  doi = {10.1103/PhysRevB.73.224432},
  url = {https://link.aps.org/doi/10.1103/PhysRevB.73.224432}
}

@article{HasenbuschPelissetto2008,
  title = {Critical behavior of three-dimensional Ising spin glass models},
  author = {Hasenbusch, Martin and Pelissetto, Andrea and Vicari, Ettore},
  journal = {Phys. Rev. B},
  volume = {78},
  issue = {21},
  pages = {214205},
  numpages = {23},
  year = {2008},
  month = {Dec},
  publisher = {American Physical Society},
  doi = {10.1103/PhysRevB.78.214205},
  url = {https://link.aps.org/doi/10.1103/PhysRevB.78.214205}
}

@article{WangMachtaKatzgraber2015,
  title = {Comparing Monte Carlo methods for finding ground states of Ising spin glasses: Population annealing, simulated annealing, and parallel tempering},
  author = {Wang, Wenlong and Machta, Jonathan and Katzgraber, Helmut G.},
  journal = {Phys. Rev. E},
  volume = {92},
  issue = {1},
  pages = {013303},
  numpages = {8},
  year = {2015},
  month = {Jul},
  publisher = {American Physical Society},
  doi = {10.1103/PhysRevE.92.013303},
  url = {https://link.aps.org/doi/10.1103/PhysRevE.92.013303}
}

@article{Kawamura1992,
  title = {Chiral ordering in Heisenberg spin glasses in two and three dimensions},
  author = {Kawamura, Hikaru},
  journal = {Phys. Rev. Lett.},
  volume = {68},
  issue = {25},
  pages = {3785--3788},
  numpages = {0},
  year = {1992},
  month = {Jun},
  publisher = {American Physical Society},
  doi = {10.1103/PhysRevLett.68.3785},
  url = {https://link.aps.org/doi/10.1103/PhysRevLett.68.3785}
}

@article{MarinariParisiTersenghiRuizLorenzoZuliani2000,
   title={Replica Symmetry Breaking in Short-Range Spin Glasses: Theoretical Foundations and Numerical Evidences},
   volume={98},
   ISSN={1572-9613},
   url={http://dx.doi.org/10.1023/A:1018607809852},
   DOI={10.1023/a:1018607809852},
   number={5-6},
   journal={Journal of Statistical Physics},
   publisher={Springer Science and Business Media LLC},
   author={Marinari, Enzo and Parisi, Giorgio and Ricci-Tersenghi, Federico and Ruiz-Lorenzo, Juan J. and Zuliani, Francesco},
   year={2000},
   month=Mar, pages={973–1074} }

@article{VedulaMooreSharma2025,
  title = {Nature of spin glass order in physical dimensions},
  author = {Vedula, Bharadwaj and Moore, M. A. and Sharma, Auditya},
  journal = {Phys. Rev. E},
  volume = {111},
  issue = {3},
  pages = {034102},
  numpages = {9},
  year = {2025},
  month = {Mar},
  publisher = {American Physical Society},
  doi = {10.1103/PhysRevE.111.034102},
  url = {https://link.aps.org/doi/10.1103/PhysRevE.111.034102}
}

@article{ParisiRitort1993,
  title={Replica symmetry breaking in four-dimensional spin glasses},
  author={Giorgio Parisi and Felix Ritort},
  journal={Journal of Physics A},
  year={1993},
  volume={26},
  pages={6711-6730},
  url={https://api.semanticscholar.org/CorpusID:120062910}
}

@article{MarinariParisiRuizLorenzoRitort1996,
  title = {Numerical Evidence for Spontaneously Broken Replica Symmetry in 3D Spin Glasses},
  author = {Marinari, E. and Parisi, G. and Ruiz-Lorenzo, J. and Ritort, F.},
  journal = {Phys. Rev. Lett.},
  volume = {76},
  issue = {5},
  pages = {843--846},
  numpages = {0},
  year = {1996},
  month = {Jan},
  publisher = {American Physical Society},
  doi = {10.1103/PhysRevLett.76.843},
  url = {https://link.aps.org/doi/10.1103/PhysRevLett.76.843}
}

@book{Young1997,
author = {Young, A P},
title = {Spin Glasses and Random Fields},
publisher = {WORLD SCIENTIFIC},
year = {1997},
doi = {10.1142/3517},
address = {},
edition   = {},
URL = {https://www.worldscientific.com/doi/abs/10.1142/3517},
eprint = {https://www.worldscientific.com/doi/pdf/10.1142/3517}
}

@article{MarinariParisiRuizLorenzo1998,
  author = {Marinari, Enzo and Parisi, Giorgio and Ruiz-Lorenzo, J. J.},
  title = {Phase Structure of the Three-Dimensional Edwards-Anderson Spin Glass},
  journal = {Physical Review B},
  volume = {58},
  number = {22},
  pages = {14852--14863},
  year = {1998},
  doi = {10.1103/PhysRevB.58.14852}
}

@article{KrzakalaMartin2000,
  title = {Spin and Link Overlaps in Three-Dimensional Spin Glasses},
  author = {Krzakala, F. and Martin, O. C.},
  journal = {Phys. Rev. Lett.},
  volume = {85},
  issue = {14},
  pages = {3013--3016},
  numpages = {0},
  year = {2000},
  month = {Oct},
  publisher = {American Physical Society},
  doi = {10.1103/PhysRevLett.85.3013},
  url = {https://link.aps.org/doi/10.1103/PhysRevLett.85.3013}
}

@article{PalassiniYoung2000,
  title = {Nature of the Spin Glass State},
  author = {Palassini, Matteo and Young, A. P.},
  journal = {Phys. Rev. Lett.},
  volume = {85},
  issue = {14},
  pages = {3017--3020},
  numpages = {0},
  year = {2000},
  month = {Oct},
  publisher = {American Physical Society},
  doi = {10.1103/PhysRevLett.85.3017},
  url = {https://link.aps.org/doi/10.1103/PhysRevLett.85.3017}
}

@article{HartmannYoung2001,
  title = {Lower critical dimension of Ising spin glasses},
  author = {Hartmann, A. K. and Young, A. P.},
  journal = {Phys. Rev. B},
  volume = {64},
  issue = {18},
  pages = {180404(R)},
  numpages = {4},
  year = {2001},
  month = {Oct},
  publisher = {American Physical Society},
  doi = {10.1103/PhysRevB.64.180404},
  url = {https://link.aps.org/doi/10.1103/PhysRevB.64.180404}
}

@article{AspelmeierBrayMoore2002,
  title = {Why Temperature Chaos in Spin Glasses Is Hard to Observe},
  author = {Aspelmeier, T. and Bray, A. J. and Moore, M. A.},
  journal = {Phys. Rev. Lett.},
  volume = {89},
  issue = {19},
  pages = {197202},
  numpages = {4},
  year = {2002},
  month = {Oct},
  publisher = {American Physical Society},
  doi = {10.1103/PhysRevLett.89.197202},
  url = {https://link.aps.org/doi/10.1103/PhysRevLett.89.197202}
}

@article{KatzgraberHelmutYoung2003,
  title = {Monte Carlo studies of the one-dimensional Ising spin glass with power-law interactions},
  author = {Katzgraber, Helmut G. and Young, A. P.},
  journal = {Phys. Rev. B},
  volume = {67},
  issue = {13},
  pages = {134410},
  numpages = {8},
  year = {2003},
  month = {Apr},
  publisher = {American Physical Society},
  doi = {10.1103/PhysRevB.67.134410},
  url = {https://link.aps.org/doi/10.1103/PhysRevB.67.134410}
}

@article{Boettcher2005,
  title = {Stiffness of the Edwards-Anderson Model in all Dimensions},
  author = {Boettcher, Stefan},
  journal = {Phys. Rev. Lett.},
  volume = {95},
  issue = {19},
  pages = {197205},
  numpages = {4},
  year = {2005},
  month = {Nov},
  publisher = {American Physical Society},
  doi = {10.1103/PhysRevLett.95.197205},
  url = {https://link.aps.org/doi/10.1103/PhysRevLett.95.197205}
}

@article{KatzgraberHamzZhueOchoaMunozBauza2015,
  title = {Seeking Quantum Speedup Through Spin Glasses: The Good, the Bad, and the Ugly},
  author = {Katzgraber, Helmut G. and Hamze, Firas and Zhu, Zheng and Ochoa, Andrew J. and Munoz-Bauza, H.},
  journal = {Phys. Rev. X},
  volume = {5},
  issue = {3},
  pages = {031026},
  numpages = {15},
  year = {2015},
  month = {Sep},
  publisher = {American Physical Society},
  doi = {10.1103/PhysRevX.5.031026},
  url = {https://link.aps.org/doi/10.1103/PhysRevX.5.031026}
}

@article{YucesoyKatzgraberHelmut2012,
  title = {Evidence of Non-Mean-Field-Like Low-Temperature Behavior in the Edwards-Anderson Spin-Glass Model},
  author = {Yucesoy, B. and Katzgraber, Helmut G. and Machta, J.},
  journal = {Phys. Rev. Lett.},
  volume = {109},
  issue = {17},
  pages = {177204},
  numpages = {5},
  year = {2012},
  month = {Oct},
  publisher = {American Physical Society},
  doi = {10.1103/PhysRevLett.109.177204},
  url = {https://link.aps.org/doi/10.1103/PhysRevLett.109.177204}
}

@article{WangMachtaKatzgraber2014,
  title = {Evidence against a mean-field description of short-range spin glasses revealed through thermal boundary conditions},
  author = {Wang, Wenlong and Machta, Jonathan and Katzgraber, Helmut G.},
  journal = {Phys. Rev. B},
  volume = {90},
  issue = {18},
  pages = {184412},
  numpages = {12},
  year = {2014},
  month = {Nov},
  publisher = {American Physical Society},
  doi = {10.1103/PhysRevB.90.184412},
  url = {https://link.aps.org/doi/10.1103/PhysRevB.90.184412}
}

@article{WangWallinLidmar2020,
  title = {Evidence of many thermodynamic states of the three-dimensional Ising spin glass},
  author = {Wang, Wenlong and Wallin, Mats and Lidmar, Jack},
  journal = {Phys. Rev. Res.},
  volume = {2},
  issue = {4},
  pages = {043241},
  numpages = {8},
  year = {2020},
  month = {Nov},
  publisher = {American Physical Society},
  doi = {10.1103/PhysRevResearch.2.043241},
  url = {https://link.aps.org/doi/10.1103/PhysRevResearch.2.043241}
}

@article{HedYoungDomany2004,
  title = {Lack of Ultrametricity in the Low-Temperature Phase of Three-Dimensional Ising Spin Glasses},
  author = {Hed, Guy and Young, A. P. and Domany, Eytan},
  journal = {Phys. Rev. Lett.},
  volume = {92},
  issue = {15},
  pages = {157201},
  numpages = {4},
  year = {2004},
  month = {Apr},
  publisher = {American Physical Society},
  doi = {10.1103/PhysRevLett.92.157201},
  url = {https://link.aps.org/doi/10.1103/PhysRevLett.92.157201}
}

@article{ContucciGiardinaGibertiParisiVernia2007,
  title = {Ultrametricity in the Edwards-Anderson Model},
  author = {Contucci, Pierluigi and Giardin\`a, Cristian and Giberti, Claudio and Parisi, Giorgio and Vernia, Cecilia},
  journal = {Phys. Rev. Lett.},
  volume = {99},
  issue = {5},
  pages = {057206},
  numpages = {4},
  year = {2007},
  month = {Jul},
  publisher = {American Physical Society},
  doi = {10.1103/PhysRevLett.99.057206},
  url = {https://link.aps.org/doi/10.1103/PhysRevLett.99.057206}
}

@article{KatzgraberHartmann2009,
  title = {Ultrametricity and Clustering of States in Spin Glasses: A One-Dimensional View},
  author = {Katzgraber, Helmut G. and Hartmann, Alexander K.},
  journal = {Phys. Rev. Lett.},
  volume = {102},
  issue = {3},
  pages = {037207},
  numpages = {4},
  year = {2009},
  month = {Jan},
  publisher = {American Physical Society},
  doi = {10.1103/PhysRevLett.102.037207},
  url = {https://link.aps.org/doi/10.1103/PhysRevLett.102.037207}
}

@article{LarsonKatzgraberMooreYoung2010,
  title = {Numerical studies of a one-dimensional three-spin spin-glass model with long-range interactions},
  author = {Larson, Derek and Katzgraber, Helmut G. and Moore, M. A. and Young, A. P.},
  journal = {Phys. Rev. B},
  volume = {81},
  issue = {6},
  pages = {064415},
  numpages = {8},
  year = {2010},
  month = {Feb},
  publisher = {American Physical Society},
  doi = {10.1103/PhysRevB.81.064415},
  url = {https://link.aps.org/doi/10.1103/PhysRevB.81.064415}
}

@article{YoungKatzgraber2004,
  title = {Absence of an Almeida-Thouless Line in Three-Dimensional Spin Glasses},
  author = {Young, A. P. and Katzgraber, Helmut G.},
  journal = {Phys. Rev. Lett.},
  volume = {93},
  issue = {20},
  pages = {207203},
  numpages = {4},
  year = {2004},
  month = {Nov},
  publisher = {American Physical Society},
  doi = {10.1103/PhysRevLett.93.207203},
  url = {https://link.aps.org/doi/10.1103/PhysRevLett.93.207203}
}

@article{KatzgraberLarsonYoung2009,
  title = {Study of the de Almeida--Thouless Line Using Power-Law Diluted One-Dimensional Ising Spin Glasses},
  author = {Katzgraber, Helmut G. and Larson, Derek and Young, A. P.},
  journal = {Phys. Rev. Lett.},
  volume = {102},
  issue = {17},
  pages = {177205},
  numpages = {4},
  year = {2009},
  month = {Apr},
  publisher = {American Physical Society},
  doi = {10.1103/PhysRevLett.102.177205},
  url = {https://link.aps.org/doi/10.1103/PhysRevLett.102.177205}
}

@article{LeuzziParisiTersenghiRuizLorenzo2008,
  title = {Dilute One-Dimensional Spin Glasses with Power Law Decaying Interactions},
  author = {Leuzzi, L. and Parisi, G. and Ricci-Tersenghi, F. and Ruiz-Lorenzo, J. J.},
  journal = {Phys. Rev. Lett.},
  volume = {101},
  issue = {10},
  pages = {107203},
  numpages = {4},
  year = {2008},
  month = {Sep},
  publisher = {American Physical Society},
  doi = {10.1103/PhysRevLett.101.107203},
  url = {https://link.aps.org/doi/10.1103/PhysRevLett.101.107203}
}

@article{SinghYoung2017,
   title={de Almeida–Thouless instability in short-range Ising spin glasses},
   volume={96},
   ISSN={2470-0053},
   url={http://dx.doi.org/10.1103/PhysRevE.96.012127},
   DOI={10.1103/physreve.96.012127},
   number={1},
   journal={Physical Review E},
   publisher={American Physical Society (APS)},
   author={Singh, R. R. P. and Young, A. P.},
   year={2017},
   month=July }

@article{Fernandezetal2022,
  title = {Numerical test of the replica-symmetric Hamiltonian for correlations of the critical state of spin glasses in a field},
  author = {Fernandez, L. A. and Gonzalez-Adalid Pemartin, I. and Martin-Mayor, V. and Parisi, G. and Ricci-Tersenghi, F. and Rizzo, T. and Ruiz-Lorenzo, J. J. and Veca, M.},
  journal = {Phys. Rev. E},
  volume = {105},
  issue = {5},
  pages = {054106},
  numpages = {7},
  year = {2022},
  month = {May},
  publisher = {American Physical Society},
  doi = {10.1103/PhysRevE.105.054106},
  url = {https://link.aps.org/doi/10.1103/PhysRevE.105.054106}
}

@article{Aguilaretal2024,
  title = {Evidence of a second-order phase transition in the six-dimensional Ising spin glass in a field},
  author = {Aguilar-Janita, M. and Martin-Mayor, V. and Moreno-Gordo, J. and Ruiz-Lorenzo, J. J.},
  journal = {Phys. Rev. E},
  volume = {109},
  issue = {5},
  pages = {055302},
  numpages = {15},
  year = {2024},
  month = {May},
  publisher = {American Physical Society},
  doi = {10.1103/PhysRevE.109.055302},
  url = {https://link.aps.org/doi/10.1103/PhysRevE.109.055302}
}

@article{VedulaMooreSharma2024,
  title = {Evidence that the de Almeida--Thouless transition disappears below six dimensions},
  author = {Vedula, Bharadwaj and Moore, M. A. and Sharma, Auditya},
  journal = {Phys. Rev. E},
  volume = {110},
  issue = {5},
  pages = {054131},
  numpages = {11},
  year = {2024},
  month = {Nov},
  publisher = {American Physical Society},
  doi = {10.1103/PhysRevE.110.054131},
  url = {https://link.aps.org/doi/10.1103/PhysRevE.110.054131}
}

@article{ImryMa,
  title = {Random-Field Instability of the Ordered State of Continuous Symmetry},
  author = {Imry, Yoseph and Ma, Shang-keng},
  journal = {Phys. Rev. Lett.},
  volume = {35},
  issue = {21},
  pages = {1399--1401},
  numpages = {0},
  year = {1975},
  month = {Nov},
  publisher = {American Physical Society},
  doi = {10.1103/PhysRevLett.35.1399},
  url = {https://link.aps.org/doi/10.1103/PhysRevLett.35.1399}
}

@article{PelcovitsPytte1978,
  title = {Spin-Glass and Ferromagnetic Behavior Induced by Random Uniaxial Anisotropy},
  author = {Pelcovits, Robert A. and Pytte, E. and Rudnick, Joseph},
  journal = {Phys. Rev. Lett.},
  volume = {40},
  issue = {7},
  pages = {476--479},
  numpages = {0},
  year = {1978},
  month = {Feb},
  publisher = {American Physical Society},
  doi = {10.1103/PhysRevLett.40.476},
  url = {https://link.aps.org/doi/10.1103/PhysRevLett.40.476}
}

@article{AizenmanWehr1989,
  title = {Rounding of first-order phase transitions in systems with quenched disorder},
  author = {Aizenman, Michael and Wehr, Jan},
  journal = {Phys. Rev. Lett.},
  volume = {62},
  issue = {21},
  pages = {2503--2506},
  numpages = {0},
  year = {1989},
  month = {May},
  publisher = {American Physical Society},
  doi = {10.1103/PhysRevLett.62.2503},
  url = {https://link.aps.org/doi/10.1103/PhysRevLett.62.2503}
}

@article{AizenmanWehr1990,
	author = {Aizenman, Michael and Wehr, Jan},
	date = {1990/06/01},
	doi = {10.1007/BF02096933},
	id = {Aizenman1990},
	isbn = {1432-0916},
	journal = {Communications in Mathematical Physics},
	number = {3},
	pages = {489--528},
	title = {Rounding effects of quenched randomness on first-order phase transitions},
	url = {https://doi.org/10.1007/BF02096933},
	volume = {130},
	year = {1990}}

@article{Fisher1985,
  title = {Random fields, random anisotropies, nonlinear \ensuremath{\sigma} models, and dimensional reduction},
  author = {Fisher, Daniel S.},
  journal = {Phys. Rev. B},
  volume = {31},
  issue = {11},
  pages = {7233--7251},
  numpages = {0},
  year = {1985},
  month = {Jun},
  publisher = {American Physical Society},
  doi = {10.1103/PhysRevB.31.7233},
  url = {https://link.aps.org/doi/10.1103/PhysRevB.31.7233}
}

@misc{DominicisTemesvariKondor1998,
      title={On Ward-Takahashi identities for the Parisi spin glass}, 
      author={C. De Dominicis and T. Temesvari and I. Kondor},
      year={1998},
      eprint={cond-mat/9802166},
      archivePrefix={arXiv},
      primaryClass={cond-mat.dis-nn},
      url={https://arxiv.org/abs/cond-mat/9802166}, 
}

@article{TemesváriKondorDominicis2000,
   title={Reparametrization invariance: a gauge-like symmetry of ultrametrically organised states},
   volume={18},
   ISSN={1434-6028},
   url={http://dx.doi.org/10.1007/s100510070038},
   DOI={10.1007/s100510070038},
   number={3},
   journal={The European Physical Journal B},
   publisher={Springer Science and Business Media LLC},
   author={Temesvári, T. and Kondor, I. and De Dominicis, C.},
   year={2000},
   month=Dec, pages={493–500} }

@inbook{DeDominicisGiardina2006, place={Cambridge}, title={Ward–Takahashi Identities and Goldstone modes}, booktitle={Random Fields and Spin Glasses: A Field Theory Approach}, publisher={Cambridge University Press}, author={De Dominicis, Cirano and Giardina, Irene}, year={2006}, pages={183–192}}

@article{BrayMooreYoung1986,
  author = {Bray, A. J. and Moore, M. A. and Young, A. P.},
  title = {Lower Critical Dimension of Metallic Vector Spin-Glasses},
  journal = {Physical Review Letters},
  volume = {56},
  number = {24},
  pages = {2641--2644},
  year = {1986},
  doi = {10.1103/PhysRevLett.56.2641}
}

@article{AspelmeierMoore2004,
  author = {Aspelmeier, T. and Moore, M. A.},
  title = {Generalized Bose--Einstein Phase Transition in Large-$m$ Component Spin Glasses},
  journal = {Physical Review Letters},
  volume = {92},
  number = {7},
  pages = {077201},
  year = {2004},
  doi = {10.1103/PhysRevLett.92.077201}
}

@article{LeeYoung2005,
  author = {Lee, L. W. and Young, A. P.},
  title = {Defect Energy of Infinite-Component Vector Spin Glasses},
  journal = {Physical Review E},
  volume = {72},
  number = {3},
  pages = {036124},
  year = {2005},
  doi = {10.1103/PhysRevE.72.036124}
}

@article{LeeDharYoung2005,
  author = {Lee, L. W. and Dhar, A. and Young, A. P.},
  title = {Spin Glasses in the Limit of an Infinite Number of Spin Components},
  journal = {Physical Review E},
  volume = {71},
  number = {3},
  pages = {036146},
  year = {2005},
  doi = {10.1103/PhysRevE.71.036146}
}

@article{SharmaYoung2010,
  author = {Sharma, Auditya and Young, A. P.},
  title = {de Almeida--Thouless Line in Vector Spin Glasses},
  journal = {Physical Review E},
  volume = {81},
  number = {6},
  pages = {061115},
  year = {2010},
  doi = {10.1103/PhysRevE.81.061115}
}

@article{Moore2012,
  author = {Moore, M. A.},
  title = {$1/m$ Expansion in Spin Glasses and the de Almeida--Thouless Line},
  journal = {Physical Review E},
  volume = {86},
  number = {3},
  pages = {031114},
  year = {2012},
  doi = {10.1103/PhysRevE.86.031114}
}

@article{CampelloneParisiRanieri1999,
  title = {Finite-dimensional corrections to the mean field in a short-range $p$-spin glassy model},
  author = {Campellone, Matteo and Parisi, Giorgio and Ranieri, Paola},
  journal = {Phys. Rev. B},
  volume = {59},
  issue = {2},
  pages = {1036--1045},
  numpages = {0},
  year = {1999},
  month = {Jan},
  publisher = {American Physical Society},
  doi = {10.1103/PhysRevB.59.1036},
  url = {https://link.aps.org/doi/10.1103/PhysRevB.59.1036}
}

@article{FranzToninelli2004,
  title = {Kac Limit for Finite-Range Spin Glasses},
  author = {Franz, Silvio and Toninelli, Fabio Lucio},
  journal = {Phys. Rev. Lett.},
  volume = {92},
  issue = {3},
  pages = {030602},
  numpages = {3},
  year = {2004},
  month = {Jan},
  publisher = {American Physical Society},
  doi = {10.1103/PhysRevLett.92.030602},
  url = {https://link.aps.org/doi/10.1103/PhysRevLett.92.030602}
}

@article{FranzToninelli2004b,
   title={Finite-range spin glasses in the Kac limit: free energy and local observables},
   volume={37},
   ISSN={1361-6447},
   url={http://dx.doi.org/10.1088/0305-4470/37/30/003},
   DOI={10.1088/0305-4470/37/30/003},
   number={30},
   journal={Journal of Physics A: Mathematical and General},
   publisher={IOP Publishing},
   author={Franz, Silvio and Toninelli, Fabio Lucio},
   year={2004},
   month=July, pages={7433–7446} }

@article{FranzMontanari2007,
doi = {10.1088/1751-8113/40/11/F01},
url = {https://doi.org/10.1088/1751-8113/40/11/F01},
year = {2007},
month = {feb},
publisher = {},
volume = {40},
number = {11},
pages = {F251},
author = {Franz, S and Montanari, A},
title = {Analytic determination of dynamical and mosaic length scales in a Kac glass model},
journal = {Journal of Physics A: Mathematical and Theoretical}
}

@article{MouhannaTarjus2016,
  title = {Phase diagram and criticality of the random anisotropy model in the large-$N$ limit},
  author = {Mouhanna, Dominique and Tarjus, Gilles},
  journal = {Phys. Rev. B},
  volume = {94},
  issue = {21},
  pages = {214205},
  numpages = {22},
  year = {2016},
  month = {Dec},
  publisher = {American Physical Society},
  doi = {10.1103/PhysRevB.94.214205},
  url = {https://link.aps.org/doi/10.1103/PhysRevB.94.214205}
}

@article{Ginzburg1981,
  title={Spin glasses with random anisotropic exchange},
  author={Ginzburg, SL},
  journal={Zh. Eksp. Teor. Fiz},
  volume={80},
  pages={244},
  year={1981}
}

@article{Goldschmidt1983,
title = {Magnets with random uniaxial anisotropy: Thermodynamic properties in the large-N limit},
journal = {Nuclear Physics B},
volume = {225},
number = {1},
pages = {123-144},
year = {1983},
issn = {0550-3213},
doi = {https://doi.org/10.1016/0550-3213(83)90015-9},
url = {https://www.sciencedirect.com/science/article/pii/0550321383900159},
author = {Yadin Y. Goldschmidt}
}

@article{Goldschmidt1984,
  title = {$\frac{1}{N}$ expansion in the random anisotropy model: A solution with replica-symmetry breaking},
  author = {Goldschmidt, Yadin Y.},
  journal = {Phys. Rev. B},
  volume = {30},
  issue = {3},
  pages = {1632(R)--1635(R)},
  numpages = {0},
  year = {1984},
  month = {Aug},
  publisher = {American Physical Society},
  doi = {10.1103/PhysRevB.30.1632},
  url = {https://link.aps.org/doi/10.1103/PhysRevB.30.1632}
}

@article{MezardParisi1991,
	author = {{Marc Mézard} and {Giorgio Parisi}},
	title = {Replica field theory for random manifolds},
	DOI= "10.1051/jp1:1991171",
	url= "https://doi.org/10.1051/jp1:1991171",
	journal = {J. Phys. I France},
	year = 1991,
	volume = 1,
	number = 6,
	pages = "809-836",
	month = "",
}

@article{LeDoussalWiese2003,
  title = {Functional renormalization group at large N for disordered elastic systems, and relation to replica symmetry breaking},
  author = {Doussal, Pierre Le and Wiese, Kay J\"org},
  journal = {Phys. Rev. B},
  volume = {68},
  issue = {17},
  pages = {174202},
  numpages = {35},
  year = {2003},
  month = {Nov},
  publisher = {American Physical Society},
  doi = {10.1103/PhysRevB.68.174202},
  url = {https://link.aps.org/doi/10.1103/PhysRevB.68.174202}
}

@article{DomanyHedPalassiniYoung2001,
   title={State hierarchy induced by correlated spin domains in short-range spin glasses},
   volume={64},
   ISSN={1095-3795},
   url={http://dx.doi.org/10.1103/PhysRevB.64.224406},
   DOI={10.1103/physrevb.64.224406},
   number={22},
   journal={Physical Review B},
   publisher={American Physical Society (APS)},
   author={Domany, Eytan and Hed, Guy and Palassini, Matteo and Young, A. P.},
   year={2001},
   month=Nov }

@article{OppermannSherrington2005,
   title={Scaling and Renormalization Group in Replica-Symmetry-Breaking Space: Evidence for a Simple Analytical Solution of the Sherrington-Kirkpatrick Model at Zero Temperature},
   volume={95},
   ISSN={1079-7114},
   url={http://dx.doi.org/10.1103/PhysRevLett.95.197203},
   DOI={10.1103/physrevlett.95.197203},
   number={19},
   journal={Physical Review Letters},
   publisher={American Physical Society (APS)},
   author={Oppermann, R. and Sherrington, D.},
   year={2005},
   month=Nov }

@misc{Vijay2026,
    title  = {The Gauge--Disorder Correspondence (to appear)},
    author={Vijay, Akash and Lee, Jong Yeon},
    year={2026},
    archivePrefix={arXiv},
}

@article{YeSachdevRead1993,
  title = {Solvable spin glass of quantum rotors},
  author = {Ye, J. and Sachdev, S. and Read, N.},
  journal = {Phys. Rev. Lett.},
  volume = {70},
  issue = {25},
  pages = {4011--4014},
  numpages = {0},
  year = {1993},
  month = {Jun},
  publisher = {American Physical Society},
  doi = {10.1103/PhysRevLett.70.4011},
  url = {https://link.aps.org/doi/10.1103/PhysRevLett.70.4011}
}

@article{GeorgesParcolletSachdev2001,
  title = {Quantum fluctuations of a nearly critical Heisenberg spin glass},
  author = {Georges, A. and Parcollet, O. and Sachdev, S.},
  journal = {Phys. Rev. B},
  volume = {63},
  issue = {13},
  pages = {134406},
  numpages = {17},
  year = {2001},
  month = {Mar},
  publisher = {American Physical Society},
  doi = {10.1103/PhysRevB.63.134406},
  url = {https://link.aps.org/doi/10.1103/PhysRevB.63.134406}
}

@article{CugliandoloGrempelSilva2001,
  title = {Imaginary-time replica formalism study of a quantum spherical p-spin-glass model},
  author = {Cugliandolo, Leticia F. and Grempel, D. R. and da Silva Santos, Constantino A.},
  journal = {Phys. Rev. B},
  volume = {64},
  issue = {1},
  pages = {014403},
  numpages = {26},
  year = {2001},
  month = {Jun},
  publisher = {American Physical Society},
  doi = {10.1103/PhysRevB.64.014403},
  url = {https://link.aps.org/doi/10.1103/PhysRevB.64.014403}
}

@article{ReadSachdevYe1995,
  title = {Landau theory of quantum spin glasses of rotors and Ising spins},
  author = {Read, N. and Sachdev, Subir and Ye, J.},
  journal = {Phys. Rev. B},
  volume = {52},
  issue = {1},
  pages = {384--410},
  numpages = {0},
  year = {1995},
  month = {Jul},
  publisher = {American Physical Society},
  doi = {10.1103/PhysRevB.52.384},
  url = {https://link.aps.org/doi/10.1103/PhysRevB.52.384}
}

@article{GeorgesParcolletSachdev2000,
  title = {Mean Field Theory of a Quantum Heisenberg Spin Glass},
  author = {Georges, Antoine and Parcollet, Olivier and Sachdev, Subir},
  journal = {Phys. Rev. Lett.},
  volume = {85},
  issue = {4},
  pages = {840--843},
  numpages = {0},
  year = {2000},
  month = {Jul},
  publisher = {American Physical Society},
  doi = {10.1103/PhysRevLett.85.840},
  url = {https://link.aps.org/doi/10.1103/PhysRevLett.85.840}
}

@article{Kauzmann1948,
    title={The Nature of the Glassy State and the Behavior of Liquids at Low Temperatures.},
    author={Walter Kauzmann},
    journal={Chem. Rev.},
    volume = {43},
    issue = {2},
    pages = {219-256},
    year={1948},
    month={Oct},
    doi={https://doi.org/10.1021/cr60135a002},
    url={https://pubs.acs.org/doi/10.1021/cr60135a002}
}

\appendix
\addtocontents{toc}{\protect\hideappendixsubsections}


\section{Local Overlaps and the Real-Space Parisi Hierarchy}
\label{app:pure_states}

This appendix develops the pure-state interpretation of the local replica-overlap field and the difference-field correlators introduced in the main text. We work with the lattice spin model defined
in~\eqnref{eq:Ham} and follow the pure-state conventions of~\secref{sec:replica_correlators}.

\subsection{The Thermodynamic Overlap and Difference Field Correlators}
\label{app:pure_state_decomposition}

Fix a realization of the quenched couplings,
\begin{align}
    \mathcal J \equiv \{\mathcal J_l,\mathcal J_P\}.
\end{align}
The Gibbs distribution is defined as $\cP({\cal J})[s] = \frac{1}{Z[{\cal J}]} e^{-\beta H[s;{\cal J}]}$.
In the thermodynamic limit, the corresponding Gibbs state admits an extremal decomposition into pure states labeled by $a$:
\begin{align}
    \cP({\cal J}) = \sum_a w_{a}({\cal J}) \cP_{a}(\cJ).
\end{align}
The coefficient of this formal decomposition is the pure-state weight $w_a(\mathcal J)$, satisfying $w_a(\mathcal J)\geq0$ and $\sum_a w_a(\mathcal J)=1$. Accordingly, thermal expectation values can be written as
\begin{align}
    \langle \mathcal O\rangle_{\mathcal J}
    =
    \sum_a
    w_a(\mathcal J)
    \langle \mathcal O\rangle_{a,\mathcal J}.
    \label{eq:pure_state_decomposition}
\end{align}
When the decomposition is not discrete, the sum should be understood as an integral over pure states. The states appearing in~\eqnref{eq:pure_state_decomposition} are equilibrium pure Gibbs states and should be distinguished from metastable minima of a finite-volume free-energy landscape.

For each pure state, define the normalized frozen magnetization
\begin{align}
    \mu_{a,\mathcal J}(v)
    \equiv
    \frac{1}{\sqrt N}
    \langle S(v)\rangle_{a,\mathcal J}.
    \label{eq:pure_state_magnetization}
\end{align}
The similarity between two pure states $a$ and $b$ within the same disorder realization is characterized by the local overlap
\begin{align}
    q_{ab}^{\mathcal J}(v)
    \equiv
    \mu_{a,\mathcal J}(v)^T
    \mu_{b,\mathcal J}(v).
    \label{eq:local_pure_state_overlap}
\end{align}
The corresponding thermodynamic overlap is obtained by spatially averaging the local overlap,
\begin{align}
    q_{ab}^{\mathcal J}
    \equiv
    \frac{1}{V}\sum_v q_{ab}^{\mathcal J}(v)
    =
    \frac{1}{V}\sum_v
    \mu_{a,\mathcal J}(v)^T
    \mu_{b,\mathcal J}(v).
    \label{eq:global_pure_state_overlap}
\end{align}
Thus, $q_{ab}^{\mathcal J}$ measures the global similarity between the frozen configurations of two pure states.



For a fixed disorder realization $\mathcal J$, the overlap distribution is obtained by independently sampling two pure states from the Gibbs measure and recording their mutual overlap,
\begin{align}
    P_{\mathcal J}(q)
    =
    \sum_{a,b}
    w_a(\mathcal J)w_b(\mathcal J)
    \delta\!\left(q-q_{ab}^{\mathcal J}\right).
    \label{eq:sample_overlap_distribution}
\end{align}
The quenched overlap distribution is obtained by subsequently averaging over the disorder,
\begin{align}
    P(q) =\overline{P_{\mathcal J}(q)}.
    \label{eq:overlap_distribution}
\end{align}
Thus, the pure-state decomposition is defined separately for each disorder realization, while $P(q)$ characterizes the disorder-averaged statistics of mutual overlaps between pure states.

The same distribution can be parametrized by the Parisi function $q(x)$. Defining the cumulative overlap distribution
\begin{align} 
    x(q) \equiv \int_{-\infty}^{q}dq'\,P(q') ,\label{eq:cumulative_overlap_distribution} 
\end{align}
the Parisi function is its generalized inverse,
\begin{align}
    q(x) =
    \inf\left\{
        q' : x(q')\geq x
    \right\},
    \qquad
    x\in[0,1].
    \label{eq:parisi_quantile}
\end{align}
It is therefore nondecreasing, with $P(q)=dx(q)/dq$ wherever the overlap distribution is continuous. A plateau of $q(x)$ corresponds to finite probability weight at a single overlap, while a discontinuous jump in $q(x)$ skips an interval of overlap values that is absent from the support of $P(q)$.

In this interpretation, the coordinate $x$ labels the level of the Parisi hierarchy at which two pure states separate, and $q(x)$ gives their corresponding thermodynamic overlap. To resolve how their frozen configurations differ in real space, define the difference field
\begin{align}
    \delta\mu_{ab}^{\mathcal J}(v)
    \equiv
    \mu_{a,\mathcal J}(v)-\mu_{b,\mathcal J}(v),
    \label{eq:difference_field}
\end{align}
and the overlap-conditioned difference-field correlator
\begin{align}
    D^x(r)
    \equiv
    \frac{1}{2}
    \overline{
    \left\langle
        \frac{1}{V}\sum_v
        \delta\mu_{ab}(v)^T
        \delta\mu_{ab}(v+r)
    \right\rangle_{q_{ab}=q(x)}
    }.
    \label{eq:hierarchical_difference_field_correlator}
\end{align}
Here the inner brackets denote the Gibbs-weighted average over pairs of pure states with mutual overlap $q(x)$, while the overline denotes the quenched disorder average. 

At leading order in the large-$N$ limit, the homogeneous Parisi saddles have a state-independent self-overlap $q_{aa}^{\mathcal J}=q_{\rm EA}$. Consequently,
\begin{align}
    D^x(0)
    &=
    \frac{1}{2}
    \overline{
    \left\langle
        \frac{1}{V}\sum_v
        \left|\delta\mu_{ab}(v)\right|^2
    \right\rangle_{q_{ab}=q(x)}
    }
    \nonumber\\
    &=
    q_{\rm EA}-q(x).
    \label{eq:hierarchical_difference_field_norm}
\end{align}
Thus, the overlap deficit $q_{\rm EA}-q(x)$ measures the total mean-square separation between the frozen configurations of states that branch at level $x$, while the spatial dependence of $D^x(r)$ resolves how this difference is organized in space.

In the next subsection, we show that $D^x(r)$ is encoded directly in the replica-resolved spin correlator, providing the connection between the pure-state description and the spatial hierarchy obtained from the replicated theory.

\label{app:pure_state_hierarchy}




\subsection{Replica Realization of the Overlap and Difference Field Correlators}
\label{app:replica_realization}

For a fixed disorder realization $\mathcal J$, the replicated partition function $Z_{\mathcal J}^n$ describes $n$ independent copies of the same system subject to the same quenched couplings. The local
replica-overlap field is
\begin{align}
    Q^{\alpha\beta}(v) = \frac{1}{N}
    S^\alpha(v)^T S^\beta(v).
    \label{eq:replica_local_overlap_appendix}
\end{align}
Consider two distinct replicas, $\alpha\neq\beta$, occupying pure states $a$ and $b$, respectively, within the same disorder realization. Once the pure states are fixed, the thermal fluctuations of the two replicas are independent. Their cross-correlator therefore factorizes as
\begin{align}
    \left\langle
        S^\alpha(v)^T S^\beta(v')
    \right\rangle_{a,b;\mathcal J}
    =
    \left\langle S(v)\right\rangle_{a,\mathcal J}^{T}
    \left\langle S(v')\right\rangle_{b,\mathcal J}.
    \label{eq:replicA_Pure_state_factorization}
\end{align}
At coincident points, this gives
\begin{align}
    \left\langle
        Q^{\alpha\beta}(v)
    \right\rangle_{a,b;\mathcal J}
    =
    q_{ab}^{\mathcal J}(v),
    \qquad
    \alpha\neq\beta.
    \label{eq:replica_local_overlap_pure_state}
\end{align}
Thus, an off-diagonal replica pair directly probes the local frozen overlap between the two pure states occupied by the replicas.

After averaging over the quenched disorder, the large-$N$ theory is dominated by the saddle-point overlap matrix $Q_*^{\alpha\beta}$. 
Its off-diagonal entries, together with their multiplicities in the replica limit, encode the quenched statistics of the thermodynamic overlaps between pure states. More precisely, for any suitable test function $f$,
\begin{align}
    \int dq\,P(q)f(q)
    =
    \lim_{n\rightarrow0}
    \frac{1}{n(n-1)}
    \sum_{\alpha\neq\beta}
    f\!\left(Q_*^{\alpha\beta}\right).
    \label{eq:replica_overlap_distribution}
\end{align}
The saddle $Q_*^{\alpha\beta}$ should therefore not be interpreted as the overlap matrix of a particular disorder realization. Rather, it is the replica representation of the quenched statistics of pure-state overlaps.

The spatial structure associated with these overlaps is encoded in the replica-resolved spin correlator,
\begin{align}
    G^{\alpha\beta}(r)
    \equiv
    \frac{1}{N}
    \overline{
        \left\langle
            S^\alpha(0)^T S^\beta(r)
        \right\rangle
    }
    =
    \left[\mathcal M_*^{-1}(r)\right]^{\alpha\beta}.
    \label{eq:replica_spin_correlator_appendix}
\end{align}
For notational convenience, we denote the off-diagonal replica correlator associated with the Parisi level $x$ by
\begin{align}
    G^x(r)
    \equiv
    \left.
        G^{\alpha\beta}(r)
    \right|_{Q_*^{\alpha\beta}=q(x)},
    \qquad
    \alpha\neq\beta .
\end{align}
The terminal level $x=1$, for which
$q(1)=q_{\rm EA}$, corresponds to replicas occupying the same pure state, and we denote the corresponding frozen correlator by $G^1(r)$. Since distinct replicas have independent thermal fluctuations once the pure state is fixed, $G^1(r)$ contains only the frozen contribution. Subtracting it from the diagonal correlator therefore isolates the connected thermal correlations within a pure state,
\begin{align}
    G^{\alpha\alpha}(r)-G^1(r)
    =
    \overline{
        \sum_a w_a\,
        \frac{1}{V}\sum_v
        C_a(v,v+r)
    }.
    \label{eq:fullrsb_thermal_projection}
\end{align}
The spatial average over $v$ restores translation invariance after the quenched average, so the resulting correlator depends only on the
separation $r$.

The off-diagonal correlator $G^x(r)$, by contrast, probes the frozen cross-correlation between two states whose mutual overlap is $q(x)$. Expanding the difference-field correlator defined in
Eq.~\eqref{eq:hierarchical_difference_field_correlator} then gives
\begin{align}
    D^x(r)
    =
    G^1(r)-G^x(r).
    \label{eq:difference_replicA_Projection}
\end{align}
In other words, the difference-field correlator is the loss of frozen spatial correlation incurred by moving from two replicas in the same pure state to two replicas with mutual overlap $q(x)$.
At zero separation,
\begin{align}
    G^1(0)=q_{\rm EA},
    \qquad
    G^x(0)=q(x),
\end{align}
and therefore
\begin{align}
    D^x(0)
    =
    q_{\rm EA}-q(x).
\end{align}
Thus, the usual overlap deficit is simply the zero-distance value of the spatially resolved difference-field correlator. The remaining spatial dependence of $D^x(r)$ resolves how this difference is organized in real space.

Equation~\eqref{eq:difference_replicA_Projection} provides the desired connection between the pure-state and replica descriptions. Its eigenchannel representation and long-distance behavior are derived
in the main text.

\section{$q(0)=0$ on Stable Unmagnetized RSB Branches}
\label{app:q0}

For any RSB ansatz, a stable unmagnetized branch
must have vanishing overlap at the outermost level of the Parisi hierarchy. For 1RSB this condition is $q_0=0$, whereas for FRSB and 1FRSB it is $q(0)=0$. The result follows from the stability of the
level-$0$ replicon. We present the derivation using the continuous-RSB notation appropriate to FRSB and 1FRSB; the 1RSB result follows from the replacements
$q(0)\to q_0$, $\lambda(0)\to\lambda_0$, and
$\epsilon(k,0)\to\epsilon_0(k)$.

For an unmagnetized saddle, the isolated replica-uniform component of the saddle-point equation gives
\begin{align}
    q(0)= -\lambda(0) \frac{1}{V}\sum_k\frac{1}{\epsilon(k,0)^2}.
    \label{eq:app_q0_from_iso}
\end{align}
The local saddle equation at the same level is
\begin{align}
    \lambda(0) = -2\beta^{2}\left[
        A_P q(0)^3+A_l q(0)
    \right].
\end{align}
If $q(0)\neq0$, combining these two equations gives
\begin{align}
    1 = 2\beta^{2}\left[
        A_P q(0)^2+A_l
    \right]
    \frac{1}{V}\sum_k
    \frac{1}{\epsilon(k,0)^2}.
    \label{eq:app_q0_nonzero_condition}
\end{align}

On the other hand, the level-$0$ replicon eigenvalue is
\begin{align}
    \lambda_R^{(0)} = 1- \beta^{2}\left[6A_Pq(0)^2+2A_l\right]
    \frac{1}{V}\sum_k
    \frac{1}{\epsilon(k,0)^2}.
\end{align}
Eliminating the momentum integral using
Eq.~\eqref{eq:app_q0_nonzero_condition} yields the exact identity
\begin{align}
    \lambda_R^{(0)}
    =
    -\frac{2A_Pq(0)^2}
    {A_Pq(0)^2+A_l}.
    \label{eq:app_q0_identity}
\end{align}
For any nonzero plaquette disorder, $A_P>0$, this quantity is strictly negative whenever $q(0)\neq0$. Thus every gapped unmagnetized RSB saddle with a nonzero outer overlap is unstable in the level-$0$ replicon channel. Replicon stability requires
\begin{align}
    q(0)=\lambda(0)=0.
\end{align}
For 1RSB, the same argument gives $q_0=\lambda_0=0$.

This argument does not apply to magnetized branches. In that case, the replica-uniform condensate contributes to the isolated mode and modifies Eq.~\eqref{eq:app_q0_from_iso}, so a nonzero outer overlap is not excluded by the preceding stability argument.

\section{Details of the Full-RSB Solution}
\label{app:frsb}

\subsection{Spectrum and Replica Limit of Hierarchical Matrices}
\label{app:frsb_spectrum}
We first summarize the spectral decomposition of a finite-step
hierarchical matrix and then take its replica limit. Consider a $k$-step hierarchy with block sizes $n=m_0>m_1>\cdots>m_{k+1}=1$. A hierarchical matrix with diagonal element $a_d$ and off-diagonal level values $a_i$ may be written as
\begin{align}
    \mathcal M = a_d\mathbb{I} + \sum_{i=0}^{k} a_i\left(J_i-J_{i+1}\right),
    \label{eq:app_hierarchical_matrix}
\end{align}
where $(J_i)^{\alpha\beta}=1$ whenever replicas $\alpha$ and $\beta$ belong to the same block of size $m_i$, and vanishes otherwise. Writing $J_i=m_iP_i$, the operator $P_i$ is the orthogonal projector onto the subspace of vectors that are constant within each $m_i$-block. Since the block decompositions are nested, the projectors $P_i$ commute and may be simultaneously diagonalized.

The resulting eigenspaces are organized by the hierarchy itself. Let
\begin{align}
    V_0&\equiv \mathrm{range}\,P_0 ,
    \\
    V_i&\equiv
    \mathrm{range}\,P_i\ominus\mathrm{range}\,P_{i-1},
    \qquad i=1,\ldots,k+1 .
\end{align}
The one-dimensional space $V_0$ is spanned by the replica-uniform vector $(1,\cdots,1)$. For $i\geq1$, vectors in $V_i$ are constant on each $m_i$-block but sum to zero within every parent $m_{i-1}$-block. Consequently, for $v\in V_i$,
\begin{align}
    J_jv =
    \begin{cases}
        0, & j<i,\\[2pt]
        m_j v, & j\geq i.
    \end{cases}
\end{align}
Substitution into~\eqref{eq:app_hierarchical_matrix} gives the
hierarchical eigenvalues
\begin{align}
    z_i =
    a_d
    -m_i a_{i-1}
    +
    \sum_{j=i}^{k}
    a_j\left(m_j-m_{j+1}\right),
    \label{eq:app_finite_spectrum}
\end{align}
for $i=1,\ldots,k+1$, with multiplicities
\begin{align}
    \dim V_i = n\left(
        \frac{1}{m_i} - \frac{1}{m_{i-1}}
    \right).
    \label{eq:app_finite_multiplicity}
\end{align}
The replica-uniform sector $V_0$, by contrast, carries the isolated eigenvalue
\begin{align}
    z_{\mathrm{iso}}
    =
    a_d
    +
    \sum_{j=0}^{k}
    a_j\left(m_j-m_{j+1}\right).
    \label{eq:app_finite_iso}
\end{align}
This isolated mode will play an important role in the replica limit. Although its multiplicity is only one, its eigenvalue differs from the outer edge of the hierarchical band by an amount of order $n$.

Under analytic continuation to $n\to0$, the discrete hierarchy is replaced by the continuous Parisi coordinate $x\in[0,1]$. The finite-step combinations appearing in~\eqref{eq:app_finite_spectrum} become
\begin{align}
    m_i a_{i-1}
    &\longrightarrow
    x\,a(x),
    \\
    \sum_{j=i}^{k}
    a_j\left(m_j-m_{j+1}\right)
    &\longrightarrow
    -\int_x^1 dy\,a(y).
\end{align}
For the spin kernel $\mathcal M(k)$, whose diagonal element is
$\lambda_d-\beta(1-p)K(k)$ and whose hierarchical level function is $\lambda(x)$, this gives
\begin{align}
    \epsilon(k,x)
    =
    \lambda_d-\beta(1-p)K(k)
    -x\lambda(x)
    -\int_x^1dy\,\lambda(y),
\end{align}
which is Eq.~\eqref{eq:frsb_epsilon} of the main text. The isolated replica-uniform eigenvalue takes the form
\begin{align}
    \epsilon_{\mathrm{iso}}(k)
    =
    \epsilon(k,0)+n\lambda(0)+O(n^2).
    \label{eq:app_iso_eigenvalue}
\end{align}

We now apply the same spectral decomposition to the trace-log term. At finite replica number, the normalized trace can be written as
\begin{align}
    \frac{1}{n}&\Tr\log\mathcal M(k)
    =
    \frac{1}{n}\log\epsilon_{\rm iso}(k) \nonumber \\
    &\qquad +
    \sum_{i=1}^{k+1}
    \left(
        \frac{1}{m_i}-\frac{1}{m_{i-1}}
    \right)
    \log\epsilon_i(k).
    \label{eq:app_finite_trace_log}
\end{align}
The normalized multiplicities of the hierarchical band satisfy
\begin{align}
    \sum_{i=1}^{k+1}
    \left(
        \frac{1}{m_i}-\frac{1}{m_{i-1}}
    \right)
    =
    1-\frac{1}{n}.
\end{align}
It is therefore useful to subtract the outermost band eigenvalue
$\epsilon(k,0)$ from each logarithm. Equation~\eqref{eq:app_finite_trace_log}
may then be rearranged as
\begin{align}
    \frac{1}{n}\Tr\log\mathcal M(k)
    ={}&
    \frac{1}{n}\log\epsilon_{\rm iso}(k)
    +
    \left(1-\frac{1}{n}\right)\log\epsilon(k,0)
    \nonumber\\
    &+
    \sum_{i=1}^{k+1}
    \left(
        \frac{1}{m_i}-\frac{1}{m_{i-1}}
    \right)
    \log\frac{\epsilon_i(k)}{\epsilon(k,0)} .
    \label{eq:app_finite_trace_log_ratio}
\end{align}
This form makes the replica limit transparent. From
Eq.~\eqref{eq:app_finite_multiplicity}, the hierarchical multiplicities
become
\begin{align}
    \dim V_i
    \longrightarrow
    -\,n\,\frac{dx}{x^2},
\end{align}
so the last term in~\eqref{eq:app_finite_trace_log_ratio} approaches
\begin{align}
    -\int_0^1\frac{dx}{x^2}
    \log\frac{\epsilon(k,x)}{\epsilon(k,0)} .
\end{align}
The ratio is essential for obtaining a finite continuum limit. Indeed, if one were to integrate $\log\epsilon(k,x)$ directly, the integrand would contain a singular contribution
$\log\epsilon(k,0)/x^2$ as $x\to0$. Subtracting the outer-edge value $\log\epsilon(k,0)$ removes this common contribution and leaves only the variation of the hierarchical eigenvalue away from $\epsilon(k,0)$.

The remaining contribution comes from the isolated replica-uniform mode. Its eigenvalue differs from the outer edge of the hierarchical band by $\epsilon_{\rm iso}(k) = \epsilon(k,0)+n\lambda(0)+O(n^2)$. Hence
\begin{align}
    \frac{1}{n}\log\epsilon_{\rm iso}(k)
    =
    \frac{1}{n}\log\epsilon(k,0) 
    +
    \frac{\lambda(0)}{\epsilon(k,0)}
    +O(n).
\end{align}
Combining this with the second term in
Eq.~\eqref{eq:app_finite_trace_log_ratio}, the apparent $1/n$
singularities cancel, leaving
\begin{align}
    \lim_{n\to0}\frac{1}{n}\Tr\log\mathcal M(k)
    =
    \log\epsilon(k,0)
    &-
    \int_0^1\frac{dx}{x^2}
    \log\frac{\epsilon(k,x)}{\epsilon(k,0)} \nonumber \\
    &+
    \frac{\lambda(0)}{\epsilon(k,0)} .
    \label{eq:app_Tk_derivation}
\end{align}
Thus the logarithmic ratio and the final
$\lambda(0)/\epsilon(k,0)$ term have the same origin: both arise from properly separating the isolated replica-uniform mode from the hierarchical band before taking the $n\to0$ limit.

The above eigenvalue decomposition simultaneously diagonalizes all hierarchical matrices sharing the same Parisi block structure. In the resulting eigenbasis, matrix multiplication acts pointwise on the hierarchical eigenvalues, so matrix inversion amounts simply to inverting each eigenvalue. The matrix saddle-point equation $Q=\frac{1}{V}\sum_k\mathcal M^{-1}(k)$ therefore reduces, level by level, to the scalar relation
\begin{align}
    \widehat q(x)
    =
    \frac{1}{V}\sum_k\frac{1}{\epsilon(k,x)},
\end{align}
which is precisely the full-RSB sum rule~\eqref{eq:frsb_sumrule}.

The continuum band, however, does not exhaust the saddle-point
information. At finite $n$, there remains a distinguished
replica-uniform eigenmode, whose contribution must be treated separately. The replica-uniform component of the same matrix equation reads
\begin{align}
    \widehat q(0)+nq(0) = \frac{1}{V}\sum_k
    \frac{1}{
        \epsilon(k,0)+n\lambda(0)
    } .
    \label{eq:app_iso_matching}
\end{align}
Expanding both sides in powers of $n$, the leading $O(n^0)$ term gives
\begin{align}
    \widehat q(0)
    =
    \frac{1}{V}\sum_k\frac{1}{\epsilon(k,0)},
\end{align}
which is simply the $x\to0$ limit of the continuum sum rule and hence provides no independent constraint. The additional information is contained in the $O(n)$ term. Matching at this order yields
\begin{align}
    q(0)
    =
    -\lambda(0)
    \frac{1}{V}\sum_k
    \frac{1}{\epsilon(k,0)^2}
    \label{eq:app_q0_from_iso}
\end{align}
Thus, while the $O(n^0)$ part of the isolated mode merges smoothly with the outer edge of the continuum band, its $O(n)$ splitting retains an additional saddle-point  condition. 

\subsection{No-Folding Condition}
\label{app:frsb_monotonicity}

The local FRSB equations define a formal continuous trajectory through $\Delta=\Delta(q)$ and $x=x(q)$.
For this trajectory to represent a physical Parisi profile, larger overlaps must occur at progressively deeper levels of the hierarchy.
Equivalently, $x(q)$ must be strictly increasing,
\begin{align} 
    \frac{dx}{dq}>0. \label{eq:frsb_monotonicity_condition} 
\end{align} 
This is an admissibility condition in addition to the local saddle-point equations.

Differentiating Eq.~\eqref{eq:frsb_x_of_q} gives,
\begin{align}
    \frac{1}{x(q)}\frac{dx(q)}{dq}
    &=
    \frac{1}{q}
    -\frac{18A_P q}{3A_P q^2+ A_l}
    -\frac{\partial_{\Delta}w_3(D,\Delta)}
           {w_3(D,\Delta)}
      \frac{d\Delta}{dq}.
\end{align}
Differentiating the local gap equation~\eqref{eq:frsb_gap_equation} with respect to $q$, and using
$\partial_{\Delta}w_2(D,\Delta)=-2w_3(D,\Delta)$, gives
\begin{align}
    \frac{d\Delta}{dq}
    =
    \frac{3A_P q}{3A_P q^2+A_l}
    \frac{w_2(D,\Delta)}{w_3(D,\Delta)}.
    \label{eq:frsb_Delta_dq}
\end{align}
Substituting this relation together with
$\partial_{\Delta}w_3(D,\Delta)=-3w_4(D,\Delta)$ yields
\begin{align}
    \frac{1}{x(q)}\frac{dx(q)}{dq}
    =
    \frac{1}{q}
    +
    \frac{9A_P q}{3A_P q^2 + A_l}
    \left[
        \frac{
            w_2(D,\Delta)w_4(D,\Delta)
        }{
            w_3(D,\Delta)^2
        }
        -2
    \right].
    \label{eq:frsb_monotonicity}
\end{align}
It is convenient to introduce the spectral ratio
\begin{align}
    R(\Delta)
    \equiv
    \frac{
        w_2(D,\Delta)w_4(D,\Delta)
    }{
        w_3(D,\Delta)^2
    }.
\end{align}
To establish its lower bound, write
\begin{align}
    w_2&=\int_k f(k)^2, \quad w_3=\int_k f(k)g(k), \quad w_4=\int_k g(k)^2, 
\end{align}
with
\begin{align}
    f(k)=\frac{1}{\Delta+2D-K(k)}, 
    \qquad
    g(k)=f(k)^2. 
\end{align}
The Cauchy--Schwarz inequality then gives
\begin{align}
    w_3(D,\Delta)^2
    \leq 
    w_2(D,\Delta)w_4(D,\Delta), 
\end{align}
and hence
\begin{align}
    R(\Delta)\geq1.
\end{align}
Equality holds only when $f$ and $g$ are linearly dependent almost everywhere, equivalently when the spectral measure of $\Delta+2D-K(k)$ collapses to a single point.
Finally, multiplying Eq.~\eqref{eq:frsb_monotonicity} by the positive quantity $q\,[3A_P q^2+A_l]$ gives
\begin{align}
    \operatorname{sgn}\!\left(\frac{dx}{dq}\right)
    =
    \operatorname{sgn}\!\left\{
        A_l+3A_P q^2\bigl[3R(\Delta)-5\bigr]
    \right\}.
\end{align}

Consequently, we find,
\begin{align} 
     A_l + 3 A_P q^2 \bigl[ 3R(\Delta)-5 \bigr] >0 . \label{eq:frsb_admissibility} 
\end{align}

We refer to Eq.~\eqref{eq:frsb_admissibility} as the \emph{no-folding condition}. As long as it is satisfied, $x(q)$ is monotone and can be inverted to obtain a valid nondecreasing profile $q(x)$. When the expression reaches zero, the continuous trajectory reaches the boundary of admissibility.
Continuing into a region where it is negative would give $dx/dq<0$, so $x(q)$ would fold back and its inverse would no longer define a nondecreasing Parisi function.

The roles of the link and plaquette disorders are particularly transparent. Since $R(\Delta)\geq1$,
\begin{align}
    A_l + 3A_P q^2\bigl[3R(\Delta)-5\bigr]
    \geq
    A_l - 6A_P q^2.
\end{align}
Hence $A_l > 6 A_P q^2$ is a sufficient condition for admissibility. In particular, whenever $A_l > 0$, the continuous solution is necessarily admissible at sufficiently small overlap, where the positive link-disorder term dominates.
Thus link disorder ensures that a  continuous Parisi branch is locally admissible near $q=0$, while plaquette disorder can cause it to fold at larger overlap.

Finite link disorder does not, however, prevent the trajectory from folding at larger overlap. When $R(\Delta)<5/3$, the plaquette contribution becomes negative and can eventually overwhelm $A_l$.
Saturation of Eq.~\eqref{eq:frsb_admissibility} then marks the endpoint of the admissible pure-FRSB continuation. In the parameter regime studied in this work, however, the pure-FRSB continuation is preempted by a transition to 1FRSB before the no-folding boundary is reached.

\section{Classification of Admissible Parisi Structures}
\label{app:no_other_phases}
In this section we show that, within the translation-invariant large-$N$ theory, a replicon-stable Parisi profile can contain at most one continuously breaking interval and at most one finite jump. If both are present, the continuous interval must come first. The only possible structures are therefore RS, 1RSB, FRSB, and 1FRSB, together with their magnetized counterparts where allowed.

The proof has a simple geometric interpretation. For a Parisi profile, define the integrated overlap $\chi(q)=\int_q^1 x(u)\,du$. 
Because $x(q)$ is nondecreasing, $\chi(q)$ is concave. Replicon stability requires $\chi(q)$ to lie below a marginal stability curve $h(q)$ at every occupied overlap, while a continuously breaking sector lies exactly on this curve, $\chi(q)=h(q)$. The classification then follows from two facts. First, $h(q)$ can change curvature at most once, from concave to convex. Second, across any finite jump, $\chi(q)$ is affine over the skipped overlap interval, and the saddle equation forces this line to pass above $h(q)$ somewhere between the two endpoints. This can only happen if the jump reaches the convex part of $h(q)$. Once it does, the geometry prevents the profile from returning to the stability curve. Hence every finite jump is terminal.
We assume $p<1$, $A_P>0$, and $D\ge2$. 

\subsubsection{Integrated overlap and stability.}

Let $x(q)$ denote the cumulative overlap distribution and define
\begin{align}
    \chi(q)=\int_q^1 x(u)\,du.
    \label{eq:app_integrated_overlap}
\end{align}
Since $\chi'(q) = -x(q)$, and $x(q)$ is nondecreasing, $\chi(q)$ is concave.

At an occupied overlap $q = q(x)$, $\chi(q)$ is precisely the hierarchical eigenvalue $\widehat q(x)$. To see this, note that 
\begin{align}
    \chi(q(x))
    &=\int_{q(x)}^1 x(u)\,du \nonumber\\
    &=\bigl[u\,x(u)\bigr]_{q(x)}^1
      -\int_{q(x)}^1 u\,dx(u) \nonumber\\
    &=1-xq(x)-\int_x^1 q(y)\,dy \nonumber\\
    &=\widehat q(x).
    \label{eq:app_chi_hatq}
\end{align}
The level-resolved band sum rule, Eq. \eqref{eq:frsb_sumrule_watson}, therefore gives 
\begin{align}
    \chi(q) = \frac{w_1(D,\Delta)}{\beta (1-p)},
        \label{eq:app_chi_gap}
\end{align}
at every occupied overlap.

The corresponding zero-momentum replicon eigenvalue is
\begin{align}
    \lambda_R(0;q,\Delta) &=1 - \Gamma_{q}(0)\Pi_{q}(0;\Delta)  \nonumber \\
    &=1-\frac{6A_P q^2 + 2A_l}{(1-p)^2}w_2(D,\Delta).
    \label{eq:app_general_replicon}
\end{align}
Here we have made the dependence on the gap $\Delta$ explicit. 

Since $w_2(D,\Delta)$ decreases with increasing $\Delta$, it is convenient to define, for each fixed $q$, the marginal gap $\Delta_*(q)\geq0$ at which the replicon becomes marginal. When such a nonnegative solution exists, it is determined by
\begin{align}
    2(3A_Pq^2+A_l)\,
    w_2(D,\Delta_*(q))
    =(1-p)^2.
    \label{eq:app_marginal_gap}
\end{align}
Replicon stability then requires $\Delta\geq\Delta_*(q)$. If Eq.~\eqref{eq:app_marginal_gap} has no nonnegative solution, the replicon is already stable at $\Delta=0$, and we set $\Delta_*(q)=0$.

We define the corresponding stability envelope by
\begin{align}
    h(q)
    \equiv
    \frac{w_1(D,\Delta_*(q))}
    {\beta(1-p)}.
    \label{eq:app_h_def}
\end{align}
Since $w_1(D,\Delta)$ also decreases with $\Delta$, replicon stability at an occupied overlap immediately implies
\begin{align}
    \chi(q)\leq h(q).
\end{align}
When no marginal gap exists, the same inequality holds automatically, since $\Delta_*(q)=0$ and $h(q)$ is then the largest band-sum value attainable for any $\Delta\geq0$.

On a continuously breaking interval, the saddle equations make the replicon exactly marginal. Hence
\begin{align}
    \chi(q)=h(q),
    \qquad
    x(q)=-h'(q).
    \label{eq:app_unique_arc}
\end{align}
Thus every continuously breaking sector must lie on the same curve $h(q)$.

\subsubsection{$h(q)$ changes curvature at most once.}

We next determine the shape of $h(q)$. On the branch where the marginal gap is positive, $\Delta_*(q)>0$, and for $q>0$, differentiating the marginality condition in Eq.~\eqref{eq:app_marginal_gap} and using $\partial_\Delta w_r=-r w_{r+1}$ gives
\begin{align}
    \Delta_*'(q)
    =
    \frac{3A_Pq\,w_2(D,\Delta_*(q))}
    {(3A_Pq^2+A_l)w_3(D,\Delta_*(q))}
    >0,
    \label{eq:app_Delta_star_prime}
\end{align}
Differentiating the stability envelope then gives
\begin{align}
    h'(q)
    &=
    -\frac{3A_Pq}{\beta(1-p)}
    \frac{w_2(D,\Delta_*(q))^2}
    {(3A_Pq^2+A_l)w_3(D,\Delta_*(q))} \nonumber \\
    &<0.
    \label{eq:app_h_prime}
\end{align}
It is convenient to determine the curvature by taking a logarithmic derivative of $-h'(q)$:
\begin{align}
    \frac{d}{dq}\log[-h'(q)]
    &=
    \frac{S(q)}
    {q(3A_Pq^2+A_l)},
    \\
    S(q)
    &=
    A_l+3A_Pq^2
    \left[3R(\Delta_*(q))-5\right],
    \label{eq:app_Sq}
\end{align}
where
\begin{align}
    R(\Delta)
    =
    \frac{w_2(D,\Delta)w_4(D,\Delta)}
    {w_3(D,\Delta)^2}.
    \label{eq:app_R_def}
\end{align}
Since $\frac{d}{dq}\log[-h'(q)] = \frac{h''(q)}{h'(q)}$, and $h'(q)<0$, the sign of the curvature is opposite to that of the logarithmic derivative:
\begin{align}
    \operatorname{sgn}[-h''(q)]
    =
    \operatorname{sgn}S(q).
\end{align}
Thus $h(q)$ is concave where $S(q)>0$ and convex where $S(q)<0$. When the positive-gap branch extends to $q=0$ and $A_l>0$, $S(q)=A_l+O(q^2)>0$, so $h(q)$ is initially concave. 

The only lattice-specific input required in the proof is the following inequality, valid for $\Delta>0$:
\begin{align}
    R(\Delta)\leq\frac53
    \quad\Longrightarrow\quad
    \mathcal C_D(\Delta)>0,
    \label{eq:app_spectral_lemma}
\end{align}
where 
\begin{align}
    \mathcal C_D(\Delta)
    =
    w_2^2w_5-5w_2w_3w_4+5w_3^3,
    \label{eq:app_CD_def}
\end{align}
with $w_{r} \equiv w_{r}(D,\Delta)$. The proof of Eq.~\eqref{eq:app_spectral_lemma} is given
in Sec.~\ref{app:spectral_inequality}. This is the only computer-assisted input to the classification.

For $R\geq5/3$, Eq.~\eqref{eq:app_Sq} gives $S\geq0$, so the envelope can become convex only in the region $R<5/3$. To show that its curvature can change at most once, it is therefore enough to determine the direction in which $S(q)$ crosses zero.

Note that at any zero of $S$,
\begin{align}
    R(\Delta_*(q))
    =
    \frac53-\frac{A_l}{9A_Pq^2}
    \leq\frac53.
    \label{eq:app_R_at_zero}
\end{align}
Using $\partial_\Delta w_r=-r w_{r+1}$, the derivative of $R(\Delta)$ can be written as
\begin{align}
    \frac{w_2}{w_3}R'(\Delta)
    =
    2(3R-5)(R-2)
    -4\frac{\mathcal C_D}{w_3^3}.
    \label{eq:app_R_prime_identity}
\end{align}
Combining this identity with Eq.~\eqref{eq:app_Delta_star_prime}, and then imposing $S(q)=0$, gives
\begin{align}
    \left.S'(q)\right|_{S(q)=0}
    =
    -\frac{12A_Pq}{2-R}
    \frac{\mathcal C_D}{w_3(D,\Delta_*(q))^3}.
    \label{eq:app_S_crossing}
\end{align}
At a zero of $S$, Eq.~\eqref{eq:app_R_at_zero} gives
$R\leq5/3<2$, while the spectral inequality
Eq.~\eqref{eq:app_spectral_lemma} gives $\mathcal C_D>0$. Therefore
\begin{align}
    \left.S'(q)\right|_{S(q)=0}<0.
\end{align}
Every zero of $S$ is thus a downward crossing. Once $S$ becomes negative, it cannot return to positive values, since doing so would require an upward crossing. Hence $S(q)$ can change sign at most once, from positive to negative.

It follows that $h(q)$ can change curvature at most once. We denote the possible inflection overlap by $q_f$: the envelope is concave for $q<q_f$ and convex for $q>q_f$. Either region may be absent.
If the marginal gap exists only above some minimum overlap, the extension
\begin{align}
    h(q)=\frac{w_1(D,0)}{\beta(1-p)}
\end{align}
is constant below that point and therefore introduces no additional change of curvature.

\subsubsection{A finite jump must cross the stability envelope.}

Consider a finite jump from overlap $a$ to $b>a$ at Parisi coordinate $0<m<1$. Since the interval $(a,b)$ is absent from the support of the
overlap distribution,
\begin{align}
    x(q)=m,
    \qquad
    a<q<b.
\end{align}
Equation~\eqref{eq:app_integrated_overlap} therefore gives
\begin{align}
    \chi(q)
    =
    \chi(a)-m(q-a),
    \qquad
    a\leq q\leq b.
    \label{eq:app_jump_chi}
\end{align}
Thus $\chi(q)$ is affine across the skipped interval.

Let $\Delta_a$ and $\Delta_b$ be the physical gaps at the two occupied endpoints. The undifferentiated saddle equation across the jump gives the exact matching condition
\begin{align}
    \beta(1-p)(\Delta_b-\Delta_a)
    =
    2m\beta^2
    \int_a^b
    \left(3A_Pq^2+A_l\right)dq.
    \label{eq:app_jump_gap}
\end{align}
Although the overlaps between $a$ and $b$ are unoccupied, it is useful to define an auxiliary gap $\widetilde\Delta(q)$ through
\begin{align}
    \chi(q)
    =
    \frac{w_1(D,\widetilde\Delta(q))}
    {\beta(1-p)}.
    \label{eq:app_aux_gap}
\end{align}
This gap is only a convenient interpolation across the skipped interval. At the endpoints it agrees with the physical gaps,
\begin{align}
    \widetilde\Delta(a)=\Delta_a,
    \qquad
    \widetilde\Delta(b)=\Delta_b.
\end{align}
Since $\chi'(q)=-m$, differentiating
Eq.~\eqref{eq:app_aux_gap} gives
\begin{align}
    \widetilde\Delta'(q)
    =
    \frac{m\beta(1-p)}
    {w_2(D,\widetilde\Delta(q))}.
    \label{eq:app_aux_gap_prime}
\end{align}
Integrating from $a$ to $b$ and comparing with
Eq.~\eqref{eq:app_jump_gap} yields
\begin{align}
    \int_a^b dq\,
    \left[
        \frac{(1-p)^2}
        {w_2(D,\widetilde\Delta(q))}
        -2(3A_Pq^2+A_l)
    \right]
    =0.
    \label{eq:app_jump_identity}
\end{align}
This identity is the central constraint on a finite jump. Importantly, it is an integrated saddle equation over the skipped interval; it does not require the unoccupied overlaps themselves to be stable.

The sign of the integrand has a direct geometric meaning. Whenever a marginal gap exists,
\begin{align}
    \chi(q)>h(q)
    &\quad\Longleftrightarrow\quad
    \widetilde\Delta(q)<\Delta_*(q)
    \nonumber\\
    &\quad\Longleftrightarrow\quad
    \frac{(1-p)^2}
    {w_2(D,\widetilde\Delta(q))}
    -2(3A_Pq^2+A_l)<0.
    \label{eq:app_jump_sign}
\end{align}
The first equivalence follows because $w_1$ decreases with the gap, and the second because $w_2$ does as well. If no marginal gap exists, the bracket in Eq.~\eqref{eq:app_jump_identity} is strictly positive for every nonnegative $\widetilde\Delta$.

Suppose, for contradiction, that
\begin{align}
    \chi(q)\leq h(q)
    \qquad
    \text{for all }q\in[a,b].
\end{align}
The integrand in Eq.~\eqref{eq:app_jump_identity} would then be nonnegative everywhere. Since its integral vanishes, it would have to vanish throughout the interval. This would require
\begin{align}
    \chi(q)=h(q)
\end{align}
throughout the jump. But $\chi(q)$ is a nonconstant affine function, whereas the curvature result above prevents $h(q)$ from coinciding with such a function over a finite interval. We conclude that every finite jump must satisfy
\begin{align}
    \chi(q)>h(q)
\end{align}
at some point between its endpoints.

Thus, although both occupied endpoints are replicon stable, the affine continuation of $\chi(q)$ through the skipped overlap interval must pass above the stability envelope somewhere inside the jump.

\subsubsection{Every finite jump is terminal.}

Define the separation between the stability envelope and the integrated overlap,
\begin{align}
    F(q)=h(q)-\chi(q).
    \label{eq:app_F_def}
\end{align}
Replicon stability at the two occupied endpoints gives
\begin{align}
    F(a)\geq0,
    \qquad
    F(b)\geq0,
\end{align}
while the result above implies
\begin{align}
    F(q)<0
\end{align}
somewhere inside $(a,b)$.

Suppose first that $b\leq q_f$. Then $h$ is concave throughout the jump. Since $\chi$ is affine, $F=h-\chi$ is also concave. A concave function lies above the chord joining its endpoint values, so $F(a),F(b)\geq0$ would imply $F(q)\geq0$ throughout the interval. This contradicts the existence of an interior point with $F<0$. Hence
\begin{align}
    b>q_f.
    \label{eq:app_jump_past_qf}
\end{align}
Every finite jump must therefore extend into the convex part of the
stability envelope.

Moreover, there exists a point
\begin{align}
    v\in[\max\{a,q_f\},b)
\end{align}
such that $F(v)<0$. If the negative point already lies to the right of $q_f$, this is immediate. Otherwise, let $u<q_f$ be a point with $F(u)<0$. Since $F$ is concave on $[a,q_f]$ and $F(a)\geq0$, the secant slope from $a$ to $u$ is negative,
\begin{align}
    \frac{F(u)-F(a)}{u-a}<0.
\end{align}
Concavity of $F$ on $[a,q_f]$ implies that secant slopes decrease as one moves to the right, so
\begin{align}
    \frac{F(q_f)-F(u)}{q_f-u}
    \leq
    \frac{F(u)-F(a)}{u-a}
    <0.
\end{align}
Hence $F(q_f)<F(u)<0$, and we may take $v=q_f$.

On $[v,b]$, $h$ is convex and $\chi$ is affine, so $F=h-\chi$ is convex. Since $F(v)<0$ and $F(b)\geq0$, the left derivative at $b$ must be positive:
\begin{align}
    F'_-(b)
    \geq
    \frac{F(b)-F(v)}{b-v}
    >0.
    \label{eq:app_positive_slope_after_jump}
\end{align}
For $q>b$, the envelope $h$ remains convex. The integrated overlap $\chi$ remains concave because $\chi'(q)=-x(q)$ and $x(q)$ is nondecreasing. Therefore $F=h-\chi$ remains convex beyond the jump.
Its slope can only increase from the positive value at $b$, so
\begin{align}
    F(q)
    \geq
    F(b)+F'_-(b)(q-b)
    >0,
    \qquad q>b.
    \label{eq:app_after_jump}
\end{align}

Thus, once the profile has completed a finite jump, it remains strictly below the stability envelope, $\chi(q)<h(q)$, at all larger overlaps. A later continuously breaking sector is therefore impossible, since it would require $\chi=h$. A second finite jump is also excluded: its skipped interval would satisfy $\chi<h$ throughout, making the integrand in Eq.~\eqref{eq:app_jump_identity} strictly positive and violating the jump-matching condition.

Every finite jump is therefore terminal, and a stable Parisi profile can contain at most one such jump.

\subsubsection{There is at most one continuous sector.}

It remains to exclude two continuously breaking sectors separated only by a plateau. Suppose the plateau occurs at overlap $q_*$. As either continuous sector approaches $q_*$, Eq.~\eqref{eq:app_unique_arc} fixes the Parisi coordinate to
\begin{align}
    x=-h'(q_*).
\end{align}
Both ends of the plateau therefore occur at the same value of $x$, so the plateau must have zero width.

There can thus be at most one continuously breaking sector. Since any finite jump is terminal, a continuous sector, when present, must precede the jump.

\subsubsection{$p=1$}

At $p=1$, the spin spectrum becomes dispersionless and the scaled gap used above is no longer the most convenient variable. The stability envelope can instead be obtained directly:
\begin{align}
    h(q)
    =
    \frac{1}
    {\beta\sqrt{2(3A_Pq^2+A_l)}}.
\end{align}
Its curvature changes sign only once, at
\begin{align}
    6A_Pq^2=A_l.
\end{align}
The same geometric argument therefore gives the same classification without the lattice spectral inequality.

This concludes the proof.

\subsection{Spectral Inequality for the Hypercubic Lattice}
\label{app:spectral_inequality}
In this appendix, we provide the proof for the spectral inequality used in \appref{app:no_other_phases}.
Writing $R_D(\Delta)\equiv R(\Delta)$ to make the dependence on dimension explicit, we require
\begin{align}
    R_D(\Delta)\leq\frac53
    \quad\Longrightarrow\quad
    \mathcal C_D(\Delta)>0,
    \qquad
    \Delta>0,
    \label{eq:D1_target}
\end{align}
for every integer $D\ge2$.

The proof has three parts. A simple analytic bound controls large gaps, $\Delta\ge2D$, while a comparison between dimensions handles all $D\ge24$. The remaining compact region, $2\le D\le23$ and $0<\Delta<2D$, is covered by rigorous interval bounds, supplemented by analytic estimates near $\Delta=0$.

\subsubsection{Analytic reductions.}
Cauchy--Schwarz gives $w_3w_5\ge w_4^2$ and $R_D\ge1$, so
\begin{equation}
    \frac{\mathcal C_D}{w_3^3}
    \ge R_D^2-5R_D+5.
\end{equation}
In particular,
\begin{equation}
    R_D\le\frac{11}{8}
    \quad\Longrightarrow\quad
    \mathcal C_D>0.
    \label{eq:D1_safe}
\end{equation}

This immediately controls large gaps. Let
$f(k)=[\Delta+2D-K(k)]^{-1}$ and introduce the normalized probability measure proportional to $f(k)^3$. Then $R_D=\mathbb E[f]\mathbb E[f^{-1}]$. Since $(\Delta+4D)^{-1}\le f\le\Delta^{-1}$, the Kantorovich inequality gives
\begin{equation}
    R_D(\Delta)
    \le
    \frac{(\Delta+2D)^2}{\Delta(\Delta+4D)}
    \le\frac43<\frac{11}{8},
    \qquad
    \Delta\ge2D.
    \label{eq:D1_large_gap}
\end{equation}
It therefore remains to consider $0<\Delta<2D$.

We next control all sufficiently large dimensions at once. Define
\begin{equation}
    Y_D=\frac1D\sum_{j=1}^D\cos\Theta_j,
    \qquad
    z=\left(\frac{2D}{\Delta+2D}\right)^2,
\end{equation}
where the $\Theta_j$ are independent and uniform on $[-\pi,\pi]$. Then
\begin{equation}
    (\Delta+2D)^r w_r(D,\Delta)
    =
    \mathbb E(1-\sqrt z\,Y_D)^{-r}.
    \label{eq:D1_YD}
\end{equation}
To compare dimensions, form the $D+1$ leave-one-out averages
\begin{equation}
    Y_D^{(i)}
    =
    \frac1D\sum_{j\neq i}^{D+1}\cos\Theta_j,
    \qquad i=1,\ldots,D+1.
\end{equation}
Their average is $Y_{D+1}$. Hence Jensen's inequality gives
\begin{equation}
    \mathbb E\,\varphi(Y_{D+1})
    \le
    \mathbb E\,\varphi(Y_D)
\end{equation}
for any convex $\varphi$ for which the expectation exists. Taking $\varphi(y)=(1-\sqrt z\,y)^{-r}$ shows that the right-hand side of Eq.~\eqref{eq:D1_YD} decreases with $D$.

It is also nondecreasing in $z$. Since the distribution of $Y_D$ is symmetric,
\begin{equation}
    \mathbb E(1-\sqrt z\,Y_D)^{-r}
    =
    \sum_{n=0}^{\infty}
    \binom{2n+r-1}{r-1}
    \mathbb E[Y_D^{2n}]\,z^n,
\end{equation}
with nonnegative coefficients.

We choose $D_0=24$ as a convenient reference dimension: the bounds at $D=24$, $z=1$ are already sufficient to imply $R_D<11/8$, and the two monotonicity results then extend this bound to every $D\ge24$. In particular, for all $D\geq 24$,
\begin{equation}
    \mathbb E(1-\sqrt z\,Y_D)^{-r}
    \le
    \mathbb E(1-Y_{24})^{-r},
    \qquad0\le z\le1,
\end{equation}
where the $z=1$ endpoint follows by monotone convergence whenever the moment is finite.

Rigorous evaluation at this reference point gives
\begin{equation}
    \mathbb E(1-Y_{24})^{-2}<\frac{107}{100},
    \qquad
    \mathbb E(1-Y_{24})^{-4}<\frac{317}{250}.
\end{equation}
The third moment is at least one by Jensen's inequality. Since the common factors of $(\Delta+2D)^r$ cancel in $R_D$, Eq.~\eqref{eq:D1_YD} therefore gives
\begin{equation}
    R_D < \frac{107}{100}\frac{317}{250}
    = \frac{33919}{25000} < \frac{11}{8},
    \qquad D\ge24.
\end{equation}
Equation~\eqref{eq:D1_safe} proves the desired result for all $D\ge24$.

\subsubsection{Finite-dimensional verification.}

It remains to consider $2\le D\le23$ and $0<\Delta<2D$. Using the same variable
$z=[2D/(\Delta+2D)]^2$, this corresponds to
$1/4<z<1$. Define
\begin{equation}
    W_r(z)\equiv(\Delta+2D)^r w_r(D,\Delta).
\end{equation}
The rescaling leaves the relevant ratios unchanged:
\begin{align}
    R_D&=\frac{W_2W_4}{W_3^2},
    \label{eq:D1_R_scaled}\\
    (\Delta+2D)^9\mathcal C_D
    &=
    W_2^2W_5-5W_2W_3W_4+5W_3^3.
    \label{eq:D1_C_scaled}
\end{align}

\vspace{5pt}
\noindent\textbf{Closed-walk expansion.}
The functions $W_r(z)$ have the positive expansion
\begin{equation}
    W_r(z) = \sum_{n=0}^{\infty}
    c_{r,n}p_{D,n}z^n,
    \qquad
    c_{r,n} = \binom{2n+r-1}{r-1},
    \label{eq:D1_series}
\end{equation}
where
\begin{equation}
    p_{D,n} = \frac{a_{D,n}}{(2D)^{2n}}.
\end{equation}
Here $a_{D,n}$ counts closed nearest-neighbor walks of length $2n$ on $\mathbb Z^D$, so $p_{D,n}$ is the return probability after $2n$ steps.

The walk numbers are generated exactly from
\begin{align}
    a_{D,n} &=
    \sum_{j=0}^{n}
    \binom{2n}{2j}
    \binom{2j}{j}
    a_{D-1,n-j},
    \nonumber\\
    a_{0,n}
    &=\delta_{n0}.
    \label{eq:D1_walk_recursion}
\end{align}
Indeed, if $2j$ steps lie along the last coordinate, the two binomial factors choose their positions and assign equal numbers of positive and negative steps; the remaining steps form a closed walk in $\mathbb Z^{D-1}$. Thus all coefficients in Eq.~\eqref{eq:D1_series} are obtained exactly.

\vspace{5pt}
\noindent\textbf{Rigorous truncation bounds.}
Because all coefficients in Eq.~\eqref{eq:D1_series} are nonnegative, truncation after $n=M$ gives a lower bound on $W_r(z)$. An upper bound follows by controlling the omitted tail.

For $z<1$, the return probabilities $p_{D,n}$ and the ratios $c_{r,n+1}/c_{r,n}$ are nonincreasing in $n$. Hence, whenever
\begin{equation}
    z\,\frac{c_{r,M+2}}{c_{r,M+1}}<1,
\end{equation}
the remainder obeys
\begin{equation}
    \sum_{n=M+1}^{\infty}
    c_{r,n}p_{D,n}z^n
    \le
    p_{D,M}
    \frac{c_{r,M+1}z^{M+1}}
    {1-zc_{r,M+2}/c_{r,M+1}}.
    \label{eq:D1_geometric_tail}
\end{equation}

Near $z=1$, we instead use the heat-kernel representation
\begin{equation}
    w_r(D,\Delta)
    =
    \frac1{\Gamma(r)}
    \int_0^\infty
    t^{r-1}e^{-\Delta t}
    \big[e^{-2t}I_0(2t)\big]^D\,dt,
    \label{eq:D1_heat}
\end{equation}
where $I_0$ is the modified Bessel function. The bound
\begin{equation}
    e^{-2t}I_0(2t)
    \le
    \frac{2}{7\sqrt t},
    \qquad t\ge10,
\end{equation}
implies
\begin{equation}
    p_{D,n}
    \le
    2\left(\frac27\right)^D
    D^{D/2}n^{-D/2},
    \qquad n\ge10D.
    \label{eq:D1_return_bound}
\end{equation}
Since $c_{r,n}\sim n^{r-1}$, this gives a convergent power-law bound on the tail when $D>2r$.

This endpoint estimate is needed only for $9\le D\le23$, where we prove the stronger condition $R_D<11/8$ using $r=2,3,4$. For $2\le D\le8$, the small-gap region is treated analytically below, so the interval calculation remains at $z<1$ and the geometric bound~\eqref{eq:D1_geometric_tail} suffices.

\vspace{5pt}
\noindent\textbf{Certified interval bounds.}
The finite calculation uses integer arithmetic with directed rounding. We multiply the series coefficients by $S=2^{96}$ and round downward; Horner evaluation with the same rounding then gives a rigorous lower bound. Adding an upward-rounded bound on the accumulated rounding error and the omitted tail gives a rigorous upper bound.

For an interval $z\in[z_-,z_+]$, positivity of the coefficients implies monotonicity in $z$. Evaluating the lower bound at $z_-$ and the upper bound at $z_+$ therefore gives
\begin{equation}
    L_r \le  W_r(z)  \le U_r,
    \qquad
    z_-\le z\le z_+.
    \label{eq:D1_interval_bounds}
\end{equation}
Thus each calculation certifies an entire interval, rather than a discrete set of points.

Using Eqs.~\eqref{eq:D1_R_scaled} and \eqref{eq:D1_C_scaled}, an interval is accepted whenever
\begin{align}
    3L_2L_4&>5U_3^2,
    \label{eq:D1_test_outside}\\
    L_2^2L_5-5U_2U_3U_4+5L_3^3&>0.
    \label{eq:D1_test_C}
\end{align}
The first proves $R_D>5/3$, so the premise of Eq.~\eqref{eq:D1_target} is absent; the second proves $\mathcal C_D>0$ directly. We also use
\begin{equation}
    8U_2U_4<11L_3^2,
    \label{eq:D1_test_safe}
\end{equation}
which implies $R_D<11/8$ and hence $\mathcal C_D>0$ by Eq.~\eqref{eq:D1_safe}.

Intervals that fail these tests are bisected and checked again. The accepted intervals are finally verified to meet without gaps, giving complete coverage of the required range.

\vspace{5pt}
\noindent\textbf{Small-gap endpoints.}
The limit $\Delta\to0$ requires separate treatment in the lowest dimensions. For $D=3,\ldots,8$, elementary upper and lower bounds on the heat kernel in Eq.~\eqref{eq:D1_heat} give
\begin{align}
    D=3:\quad&
    R_D>\frac74,
    &&0<\Delta\le\frac1{200},
    \nonumber\\
    D=4:\quad&
    R_D>2,
    &&0<\Delta\le\frac1{200},
    \nonumber\\
    D=5:\quad&
    R_D>2,
    &&0<\Delta\le\frac1{125},
    \nonumber\\
    D=6,7,8:\quad&
    \mathcal C_D>0,
    &&0<\Delta\le\frac1{200}.
    \label{eq:D1_small_gap}
\end{align}
Thus, for $D=3,4,5$, the condition $R_D\le5/3$ is excluded in these intervals, while for $D=6,7,8$ the desired conclusion follows directly.

For $D=2$, the lattice Green function can be expressed in terms of complete elliptic integrals. Using this representation and the recurrence between successive Watson integrals gives
\begin{equation}
    \mathcal C_2(\Delta)>0,
    \qquad
    0<\Delta\le\frac1{50}.
    \label{eq:D1_D2}
\end{equation}

The remaining compact ranges are covered by the interval procedure above. For $D=2,\ldots,8$, we verify the range from the corresponding small-gap cutoff to $\Delta=2D$. For $9\le D\le23$, the full interval $0\le\Delta\le2D$ is checked directly. Altogether, the finite verification consists of $1280$ certified intervals.

Combining these checks with the small-gap estimates, the large-gap bound~\eqref{eq:D1_large_gap}, and the $D\ge24$ argument proves Eq.~\eqref{eq:D1_target} for every integer $D\ge2$ and every $\Delta>0$.

\end{document}